\documentclass[
reprint,
nofootinbib,
 amsmath,amssymb,
 aps,
 pra,
 superscriptaddress,
]{revtex4-2}
\usepackage{listings}

\usepackage{todonotes}
\usepackage{graphicx} 
\usepackage{tikz}
\usetikzlibrary{quantikz2}

\usepackage{xcolor}
\usepackage{graphicx}
\usepackage{dcolumn}
\usepackage{bm}
\usepackage{bbm}
\usepackage{amsmath}
\usepackage{amsthm}
\usepackage{orcidlink}
\usepackage{physics}
\usepackage{natbib}
\usepackage[version=4]{mhchem}
\usepackage{gensymb}
\makeatletter
\let\old@makecaption=\@makecaption
\usepackage{subcaption}
\let\@makecaption=\old@makecaption
\makeatother
\usepackage{siunitx}
\usepackage{graphicx,multirow}
\usepackage{array}
\usepackage{hyperref}
\hypersetup{
    colorlinks,
    linkcolor={red!80!black},
    citecolor={blue!80!black},
    urlcolor={blue!80!black}
}
\usepackage[capitalize]{cleveref}
\usepackage[super]{nth}
\usepackage{lipsum}

\usepackage{chemformula}

\usepackage{dsfont}

\usepackage{enumitem}

\newcommand{\angstrom}{\text{\normalfont\AA}}

\begin{document}
\title{Contextual Subspace Auxiliary-Field Quantum Monte Carlo: \\Improved bias with reduced quantum resources}
\date{\today}
\author{Matthew Kiser\orcidlink{0000-0002-9357-7583}}
\email{matthew.kiser@volkswagen.de}
\affiliation{Volkswagen AG, Wolfsburg, Germany}
\affiliation{TUM School of Natural Sciences, Technical University of Munich, Garching, Germany}

\author{Matthias Beuerle}
\affiliation{IQM Quantum Computers, Georg-Brauchle-Ring 23-25, 80992, Munich, Germany}
\author{Fedor Šimkovic\orcidlink{0000-0003-0637-5244}}
\email{fedor.simkovic@meetiqm.com}
\affiliation{IQM Quantum Computers, Georg-Brauchle-Ring 23-25, 80992, Munich, Germany}

\begin{abstract}
Using trial wavefunctions prepared on quantum devices to reduce the bias of auxiliary-field quantum Monte Carlo (QC-AFQMC) has established itself as a promising hybrid approach to the simulation of strongly correlated many body systems. Here, we further reduce the required quantum resources by decomposing the trial wavefunction into classical and quantum parts, respectively treated by classical and quantum devices, within the contextual subspace projection formalism. Importantly, we show that our algorithm is compatible with the recently developed matchgate shadow protocol for efficient overlap calculation in QC-AFQMC. Investigating the nitrogen dimer and the reductive decomposition of ethylene carbonate in lithium-based batteries, we observe that our method outperforms a number of established algorithm for ground state energy computations, while reaching chemical accuracy with less than half of the original number of qubits.
\end{abstract}

\maketitle

\section{Introduction}

Understanding collective quantum phenomena in strongly-correlated fermionic systems is of crucial importance in the fields of quantum chemistry, solid state and high-energy physics \cite{cao2019quantum, mcardle2020quantum, dalzell2023quantum, di2023quantum}. They are equally important for industrial applications such as designing batteries, high-temperature superconductors, solar cells and light-emitting diodes \cite{ho2018promise, gao2021applications, rice2021quantum, delgado2022simulating, kim2022fault, kim2023design, fomichev2024simulating, paudel2022quantum}. Notwithstanding decades of steady algorithmic development and a steep increase in available classical computational resources, the progress in understanding such problems has been only incremental. This disconnect is a consequence of the fundamental limitations of classical numerical methods due to the exponential growth of computational spaces with system size and the infamous sign problem related to the fermionic nature of particles involved. Quantum algorithms, on the other hand, were shown to scale advantageously for at least a subset of computational tasks, such as the real-time evolution of quantum systems~\cite{dalzell2023quantum}. The simulation of strongly correlated fermionic systems has thus become one of the central applications motivating the development of quantum computers.  

Despite the promise of quantum computing, currently available \emph{noisy, intermediate-scale quantum} (NISQ)~\cite{preskill2018nisq} devices are severely limited in their qubit numbers and gate fidelities. For this reason, a number of hybrid strategies have emerged which, instead of treating the full many-body problem at the quantum level, aim at leveraging the strengths of both classical techniques and quantum information processing by splitting the simulation workload between classical and quantum devices. Prominent examples include externally optimising parameterised quantum circuits \cite{cerezo2021variational, bharti2022noisy, tilly2022variational} or using the quantum computer as a solver within approaches based on embedding and active space methods \cite{jamet2021krylov, li2022toward, izsak2023quantum, iijima2023towards, rossmannek2023quantum}. 

One approach that has been proposed~\cite{kirby2019contextuality,kirby2021contextualsubspace, weaving2023stabilizer, ralli2023unitary} splits the original problem into contextual (quantum) and noncontextual (classical) parts, which leads to a customizable and often substantial reduction in the required number of qubits. This \emph{contextual subspace} formalism has been successfully applied to the simulation of small molecules on real quantum hardware~\cite{weaving2023contextual, weaving2023benchmarking, liang2023spacepulse}. 

Yet another class of hybrid algorithms was proposed in the context of auxiliary-field quantum Monte Carlo (AFQMC) \cite{zhang2003quantum, motta2018ab, lee2022twenty}, a universal workhorse of condensed matter physics and quantum chemistry. The success of AFQMC lies in mitigating the fermionic sign problem, conventionally found in quantum Monte Carlo, at the price of introducing a bias. The severity of this bias is then related to the quality of an externally acquired trial wavefunction which guides the calculation, i.e from Hartree-Fock (HF) or more complex methods \cite{landinez2019non, amsler2023quantum,  pham2024scalable, huang2024gpu}. Recently, the quantum-classical AFQMC (QC-AFQMC) approach was proposed, where a quantum computer is used to prepare trial states of superior quality to their classical alternatives \cite{huggin2022unbiasing}. Inspired by this work, a number of other hybrid algorithms based on trial states have since been presented for related methods such as Green's function Monte Carlo (GFMC), variational Monte Carlo (VMC) and full configuration interaction quantum Monte Carlo (FCIQMC) \cite{sinibaldi2023unbiasing, montanaro2023accelerating, xu2023quantum, kanno2024quantum}.

The usefulness of the originally proposed QC-AFQMC algorithm \cite{huggin2022unbiasing} within the NISQ era was put in question \cite{huggin2022unbiasing, mazzola2022exponential, lee2022response} due to an exponentially growing classical overhead in the computation of overlaps between a shadow tomographic representation \cite{aaronson2017shadow, huang2020predicting} of the quantum trial and classical Slater-determinant-type Monte Carlo walkers. It was later shown that this complexity can be reduced to polynomial time through the use of classical shadow protocols based on matchgate circuits \cite{wan2022matchgate, low2022classical, wu2024error, zhao2024group}. The actual scaling was then improved from high-polynomial \cite{kiser2024classical} to essentially equivalent to the classical implementation of AFQMC \cite{jiang2024unbiasing}. This lead to a reinvigorated interest in the further development of this approach \cite{huang2024evaluating}. 

In this paper, we further reduce the resource requirements of QC-AFQMC by using quantum trial wavefunctions computed within a relatively small contextual subspace (CS) to guide AFQMC operating on the large original Hilbert space. Our technique, which we name contextual subspace AFQMC (CS-AFQMC), allows for the CS to encompass an arbitrary number of qubits, which can be tailored to the available quantum resources and which will define the quality of the overall result. The algorithm's performance interpolates between the standard AFQMC with a Hartree-Fock trial wavefunction (HF-AFQMC) in the limit of zero qubits in the CS and the exact result when the CS is equivalent to the full Hilbert space. We show that the quantum trial wavefunctions obtained within CS-AFQMC can be combined with the matchgate shadow protocol of Ref.~\cite{wan2022matchgate} for the efficient classical computation of wavefunction overlaps.

Studying the dissociation curve of the nitrogen dimer in the STO-3G basis, we find that the CS-AFQMC approach requires only an 8-qubit CS to reach chemical accuracy for most inter-atomic distances and significantly outperforms both standard HF-AFQMC, the original contextual subspace approximation (CSA) \cite{kirby2021contextualsubspace}, as well as CCSD and the ``gold standard'' CCSD(T) methods \cite{bartlett2007coupled}. Next, we focus on a more realistic industry application of quantum chemistry with relevance to the design of lithium-based batteries, which are omnipresent in portable electronic devices and play a key role in automotive electrification and grid energy storage \cite{vanderven2020batterytheory}. Specifically, we investigate the stability of ethylene carbonate (EC) with respect to a reductive decomposition by \ce{Li}, a process which occurs at the anode-electrolyte interface of batteries \cite{wang2001ecdecomposition, debnath2023afqmcec}. We use CS-AFQMC to compute the reaction energy for one step of the reaction path describing this decomposition process within a 32-qubit active space (AS) of the original Hamiltonian in the double-zeta basis (cc-pVDZ). Using a CS with half the size of this active space our method significantly outperforms all aforementioned alternatives for the energies of individual configurations while yielding reaction energies well within chemical accuracy.

The paper is structured as follows: in \cref{sec:theory} we give a concise introduction to the main algorithms used in this work: the contextual subspace approximation (CSA) in \cref{subsec:csa} and the auxiliary-field quantum Monte Carlo (AFQMC) in \cref{subsec:afqmc}. We then outline the combined CS-AFQMC algorithm in \cref{subsec:csafqmc} and describe the procedure for efficient overlap calculation within CS-AFQMC in \cref{subsec:overlaps}. In \cref{sec:results} we benchmark CS-AFQMC against other state-of-the-art algorithms by applying it to two systems from quantum chemistry, the nitrogen dimer in \cref{subsec:nitrogen} and the reductive decomposition of EC by \ce{Li} in \cref{subsec:ec}. \cref{sec:conclusions} offers a discussion of obtained results and highlights further research directions.

\begin{figure*}
    \centering
    \includegraphics[width=0.9\textwidth]{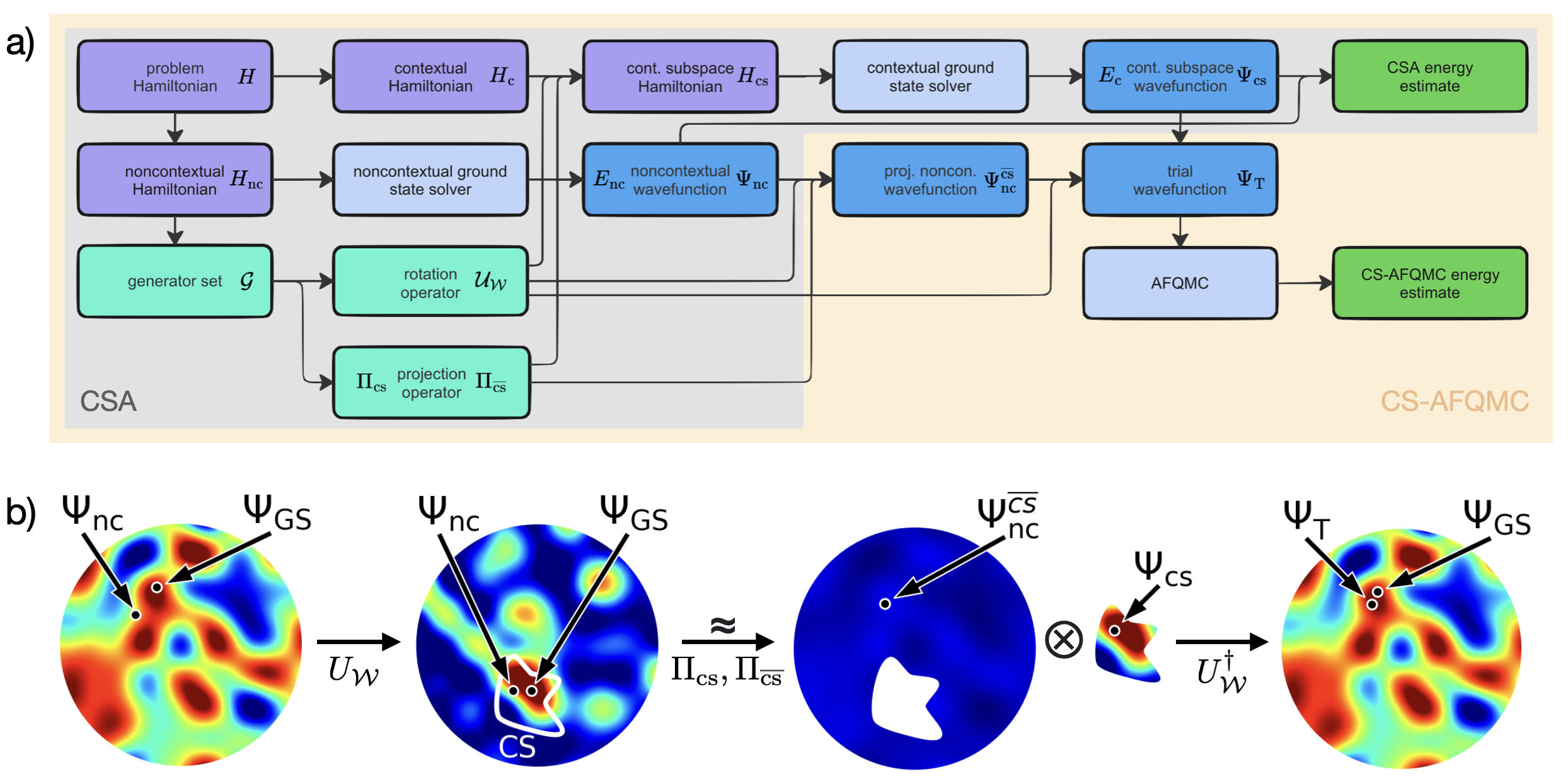}
    \caption{\textbf{a)} Workflow of CSA (grey area) and CS-AFQMC (light orange area) algorithms, as described in \cref{subsec:csa} and \cref{subsec:csafqmc}, respectively. \textbf{b)} Visual representation of the procedure of applying the rotation operator $U_{\mathcal{W}}$ and projection operator $\Pi_{\mathcal{W}}$ in these two algorithms. The color gradient between blue and red colors indicates an arbitrary abstract metric of quantumness of the underlying subspaces. First, through a unitary rotation, the quantumness is concentrated into the CS, which ideally contains the true ground state ($\ket{\Psi_{\text{GS}}}$). Then the projection is performed and a quantum device is used to find the ground state within the CS ($\ket{\Psi_\text{cs}}$), while a classical device computes the noncontextual ground state ($\ket{\Psi_\text{nc}}$), which is projected to the complement of the CS ($\ket{\Psi_\text{nc}^{\overline{\text{cs}}}}$). This product state of $\ket{\Psi_\text{cs}}$ and $\ket{\Psi_\text{nc}^{\overline{\text{cs}}}}$ is finally rotated back to the full space and the resulting state ($\ket{\Psi_{\text{T}}}$) is used as the trial wavefunction for QC-AFQMC.}
    \label{fig:algorithm_workflow}
\end{figure*}

\section{Theoretical background}\label{sec:theory}

\subsection{Contextual Subspace} \label{subsec:csa}
In this work, we utilize a stabilizer subspace projection framework known as the contextual subspace (CS) projection \cite{kirby2021contextualsubspace, weaving2023stabilizer, ralli2023unitary}. This method leverages artificial symmetries of a spin Hamiltonian to reduce the number of qubits needed for its simulation on a quantum computer. Throughout this work we will use the standard Jordan-Wigner transformation (JWT) as our fermion-to-qubit mapping. In comparison to other qubit reduction methods such as qubit tapering, which is limited to the number of available symmetries \cite{bravyi2017tapering}, the symmetries in this formalism do not have to be satisfied by the model and an arbitrary number of them can be chosen, which corresponds to removing an arbitrary number of qubits from the quantum simulation. This, however, comes at the cost of progressively decreasing the accuracy of the simulation, which interpolates between the mean-field and exact results for the cases when all or no qubits are removed, respectively. In this framework, the terms from the Hamiltonian of interest $H$ acting on $N$ qubits are split into two sub-Hamiltonians, the noncontextual part $H_{\text{nc}}$ and the contextual part $H_{\text{c}}$, where:
\begin{align}
    H = H_{\text{nc}} + {H_{\text{c}}}
\end{align}
and such that the corresponding sets of Pauli terms $\mathcal{S}$, $\mathcal{S}_{\text{nc}}$ and $\mathcal{S}_{\text{c}}$ satisfy $\mathcal{S} = \mathcal{S}_{\text{nc}} \cup \mathcal{S}_{\text{c}}$. Noncontextuality can be defined as the contradiction-free simultaneous value assignment to a given set of observables.  The opposite case where no such assignment exists is called contextuality, which can also be seen as a proxy for the ``quantumness" \cite{kirby2019contextuality, raussendorf2020phasespace, mermin1993hidden, spekkens2007evidence, spekkens2008negativity}. An approximate ground state energy $E_{\text{CSA}}$ of the Hamiltonian $H$ can be computed as the sum of the energies of the contextual and noncontextual parts:
\begin{align}
    E_{\text{CSA}} = E_{\text{nc}} + E_{\text{c}}.
\end{align}
This involves finding the relative ground states of both $H_{\text{nc}}$ and $H_{\text{c}}$. Using this framework, the main strategy to find the ground state of $H$ is to obtain the former of the two classically and the latter using a quantum device \cite{kirby2019contextuality,kirby2021contextualsubspace, weaving2023stabilizer, ralli2023unitary, liang2023spacepulse}. More concretely, one could use the variational quantum eigensolver (VQE) to solve the contextual subspace ground state \cite{kirby2021contextualsubspace}, however, the formalism works independently of the choice of ground state solver and can be equally applied in the NISQ as well as fault-tolerant settings. Below, we explain in more detail the necessary steps involved in this computation.  

For a noncontextual Hamiltonian, the corresponding set of terms, $\mathcal{S}_{\text{nc}}$, can be written as
\begin{align}
    \mathcal{S}_{\text{nc}} = \mathcal{Z} \cup  \mathcal{T},
\end{align}
where all elements of $\mathcal{Z}$ commute with all elements of $\mathcal{S}_{\text{nc}}$, while $\mathcal{T}$ can be split into $T$ cliques, $\mathcal{T} = \mathcal{C}_1\cup\mathcal{C}_2\cup\dots\cup\mathcal{C}_T$, where two given elements from cliques $\mathcal{C}_i$ and $\mathcal{C}_j$ pairwise commute if $i=j$ and anticommute otherwise. The choice of how to pick the noncontextual set of operators $\mathcal{S}_{\text{nc}}$ from the terms in $\mathcal{S}$ is not unique and the task of finding optimal decompositions with respect to some metric (such as ground state energy) is generally NP-complete \cite{ralli2023unitary}. 

For chemistry problems it is nevertheless possible to find efficient heuristics based on the fact that the energy contributions of diagonal Hamiltonian terms, made out of tensor products of Pauli $Z$ and $I$ matrices, generally dominate. Previous works ~\cite{kirby2021contextualsubspace, kirby2020classical, ralli2023unitary} selected noncontextual sets that are mainly composed of diagonal terms, but also allowed for the inclusion of off-diagonal operators, which populate the $\mathcal{T}$ part of $\mathcal{S}_{\text{nc}}$. However, these off-diagonal terms lead to added complexity of the formalism, and notably increase at least quadratically the total number of Hamiltonian terms after projection into the contextual subspace \cite{ralli2023unitary}. For this reason, we exclude $\mathcal{T}$, and therefore any cliques $\mathcal{C}_i$, from our ansatz for the noncontextual subset. More concretely, we have $\mathcal{S}_{\text{nc}} = \mathcal{Z}_{D}$, where only terms $\vec{Z}_{\boldsymbol{n}} \in \mathcal{Z}_{D}$ are included that act on $|\boldsymbol{n}|\leq N$ qubits and are of the form:  
\begin{align}
    \label{eq:only_z_terms}
    \vec{Z}_{\boldsymbol{n}} = \bigotimes_{i=1}^{|\boldsymbol{n}|} Z_{n_i},
\end{align}
where $Z_{n_i}$ is a Pauli $Z$ acting on the qubit indexed by $n_i$. While this heuristic is not always an optimal decomposition, we find that in practice it performs very well for both estimating ground state energies and wavefunctions. 
For such a noncontextual Hamiltonian, an independent set of its generators $\mathcal{G} = \{G_1 \dots G_N\}$ must be identified. Then, noncontextual states are quantum states that satisfy the value assignment of $\left< G_i \right> \equiv q_i = \pm 1$, with $\vec{q} = (q_1, \dots, q_N)$. The ground state of the noncontextual Hamiltonian can be found using polynomially scaling classical optimisation techniques which minimize the energy of $H_{\text{nc}}$ in the parameter space of $\vec{q}$. In this work we use the quadratic unconstrained signed optimisation (QUSO) \cite{qubovert} with the mean-field solution as its starting point. This defines the noncontextual ground state energy $E_{\text{nc}}$ and the corresponding wavefunction $\ket{\Psi_{\text{nc}}} \equiv \Psi_{\text{nc}}(\vec{q})$.

Next, a set of $N_\text{nc} = N - N_{\text{cs}}$ stabilizers $\mathcal{W}$ are chosen, corresponding to the number of qubits one is interested in removing from the quantum part of the simulation. They will stabilize the contextual subspace in correspondence with the noncontextual solution, and should be generated by $\mathcal{G}$. The choice of stabilizer operators is non-trivial, since there are a combinatorial number of possible options. To facilitate this step it is possible to use a heuristic based on an auxiliary operator, such as the UCCSD operator. Then, stabilizers are selected based on the weighted sum of auxiliary operator terms they preserve (commute with). Once chosen, each of these stabilizers is mapped to a single-qubit Pauli matrix by a unitary transformation and the product of all such transformations defines the rotation operator $U_\mathcal{W}$. Due to our method of selecting the noncontextual Hamiltonian, the stabilizers $\mathcal{W}$ will have the form given by \cref{eq:only_z_terms}. Consequently, the stabilizer rotations in $U_\mathcal{W}$ will be exponentials of Pauli terms of either
\begin{equation}
    \label{eq:stab_rotations_paulis}
    P_k = \bigotimes_{i=1}^{|k|} Z_{n_i} Y_j\quad \quad \text{or} \quad \quad\tilde P_k = Y_j\,,
\end{equation}
where $j$ is the index of the single qubit that the stabilizer is rotated onto, $k$ is an index for the rotations, $Y$ and $Z$ are the Pauli operators and the rotation angle is $\pi/2$. This can be clearly seen by examining the selection of these rotation operators. To rotate the stabilizers onto single qubit $Z$ operators, the stabilizers are first rotated to single qubit $X$ operators by specifying a Pauli rotation with one less $Z$ than in the stabilizer and a $Y$ on the remaining qubit index. This constructs the set $\{P_k\}$. Once these rotations have been applied, all of the stabilizers are rotated to single qubit $X$ Pauli operators. These are then rotated by applying a single $Y$ rotation in the location of the $X$ term, which forms the set $\{\tilde P_k\}$. Then, the single stabilizer rotations are given by $R_k = \{P_k\}\bigcup \{\tilde P_k\}$ and the contextual subspace rotation is $U_\mathcal{W}=\prod_k R_k$. The eigenvalues of the rotated stabilizers are fixed in accordance with the values of $\left< G_i \right>$, such that a subspace consistent with the noncontextual state is defined. 
Any state can then be projected to the reduced, contextual subspace by the non-unitary projection operator $\Pi_{\text{cs}}$ \cite{kirby2021contextualsubspace, ralli2023unitary}. Similarly, $\Pi_{\overline{\text{cs}}}$ is the projection operator into the complement of the contextual subspace. For a more detailed derivation of $U_\mathcal{W}$ and $\Pi_{\text{cs}}$ in the general setting we refer to the supplementary materials of Ref.~\cite{ralli2023unitary}. 

To obtain the ground state of the contextual Hamiltonian, we constrain it within the contextual subspace
\begin{align}
    H_{\text{c}} \to H_{\text{cs}} \equiv 
 \Pi_{\text{cs}}^{\dagger} U_\mathcal{W}^{\dagger} H_{\text{c}} U_\mathcal{W} \Pi_{\text{cs}}
\end{align}
on $N_{\text{cs}}$ qubits and the ground state contextual subspace correction energy is given by
\begin{align}
   E_{\text{c}} = \frac{\bra{\Psi_{\text{cs}}} \Pi_{\text{cs}}^{\dagger} U_\mathcal{W}^{\dagger} H_{\text{c}} U_\mathcal{W} \Pi_{\text{cs}} \ket{\Psi_{\text{cs}}}}{\bra{\Psi_{\text{cs}}} \Pi_{\text{cs}}^{\dagger} \Pi_{\text{cs}}\ket{\Psi_{\text{cs}}}}\,.
\end{align}
where $\ket{\Psi_{\text{cs}}}$ is the contextual subspace ground state wavefunction. In previous works \cite{kirby2019contextuality, ralli2023unitary, weaving2023stabilizer, liang2023spacepulse} the focus was on computing approximate ground state energies. Here, we are instead interested in obtaining a trial wavefunction $\ket{\Psi_{\text{T}}}$ with as high overlap with the original ground state wavefunction $\ket{\Psi_{\text{GS}}}$ as possible. For this, it is necessary to combine the noncontextual wavefunction $\ket{\Psi_{\text{nc}}}$ with the contextual one $\ket{\Psi_{\text{cs}}}$, which is achieved by first project the noncontextual ground state $\ket{\Psi_{\text{nc}}}$ onto the complement of the contextual subspace:
\begin{align}    \ket{\Psi_{\text{nc}}^{\overline{\text{cs}}}} \equiv  
 \Pi_{\overline{\text{cs}}}U_{\mathcal{W}}\ket{\Psi_{\text{nc}}}
\end{align}
and then $\ket{\Psi_{\text{T}}}$ is obtained as the reverse rotation of the tensor product of the two wavefunctions:
\begin{align}
    \label{eq:trial_wavefunction}
    |\Psi_{\text{T}}\rangle \equiv U_{\mathcal{W}}^{\dagger} \left[ \ket{\Psi_{\text{nc}}^{\overline{\text{cs}}}} \otimes \ket{\Psi_{\text{cs}}} \right]\,.
\end{align}

\begin{figure}
    \centering
    \includegraphics[width=0.96\columnwidth]{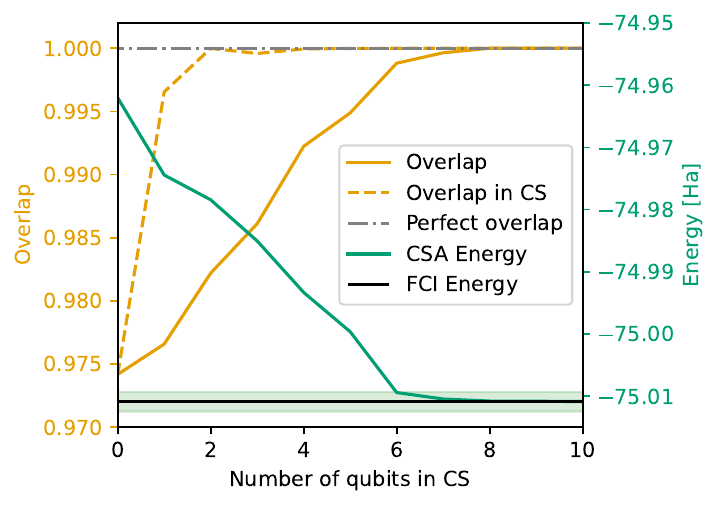}
    \caption{$\ch{H2O}$ molecule computed in the STO-3G basis with a total number of 14 qubits. The CSA energy (green curve) and the CSA wavefunction overlap with the ground state (solid, orange curve) and the overlap computed within the CS (dashed, orange curve) are shown as a function of the number of qubits in the contextual subspace. Black horizontal line indicates the FCI energy with the green shaded region signifying chemical accuracy to FCI, dot-dashed line indicates perfect overlap with the ground state. The Hartree Fock energy corresponds to the zero-qubit CSA.}
    \label{fig:csvqe_overlap}
\end{figure}

In \cref{fig:csvqe_overlap}, we use the $\ch{H2O}$ molecule in the STO-3G basis as an example to show that increasing the qubit count in the CS not only improves the ground state energy estimate, but also increases the overlap with the ground state wavefunction. While in the limit of a large CS, the overlap should approach one, it is not a priori guaranteed that this happens monotonically. We also show the overlap between the true ground state projected into the CS and the actual CS ground state (dashed yellow curve). We see that the projection loses information about the quantum state and the overlap within the CS converges faster than the overlap in the full space. This information would not be recovered upon back-projection into the original Hilbert space.

\subsection{QC-AFQMC}\label{subsec:afqmc}

We now provide a high-level description of the phaseless Auxiliary-Field Quantum Monte Carlo (AFQMC) method, and refer to  Refs.~\cite{zhang2003quantum, zhang2013auxiliary,motta2018ab, malone2022ipie} for more detailed discussions of the algorithm. The general (spin-less) electronic Hamiltonian in second quantization can be written as 
\begin{align}
    H&=H_1+H_2\nonumber\\
    &=\sum_{pq}^{[K]}h_{pq}a_p^\dagger a_q + \frac{1}{2}\sum_{pqrs}^{[K]}v_{pqrs}a_p^\dagger a_ra_q^\dagger a_s\,,
\end{align}
where $h_{pq}$ and $v_{pqrs}$ are the one-body and two-body interactions of the Hamiltonian, $a_p^\dagger$ and $a_p$ are fermionic creation and annihilation operators of the orbital $p$, and $[K]=\{1,\dots,K\}$, where $K$ is the number of spin orbitals in the given system. 

One can obtain the ground state of $H$ $\ket{\Psi_{\text{GS}}}$ through imaginary time propagation of some initial state $\ket{\Psi_I}$
\begin{equation}\label{eq:imag_evolution}
    \ket{\Psi_{\text{GS}}}=\lim_{n\rightarrow\infty}\frac{[e^{-\Delta\tau(H-E_{\text{GS}})}]^n\ket{\Psi_I}}{\bra{\Psi_I}e^{-2n\Delta\tau(H-E_{\text{GS}})}\ket{\Psi_I}}\,,
\end{equation}
where $\Delta\tau$ is a small imaginary time step and $E_{\text{GS}}$ is the energy of the ground state. For the purposes of AFQMC, we assume that the initial state has at least some overlap with the true ground state $\braket{\Psi_{\text{GS}}}{\Psi_I}\not=0$ and $\ket{\Psi_I}$ is a Slater determinant $\ket{\varphi}$ defined as
\begin{equation}
    \label{eq:slater_def}
    \ket{\varphi} = \tilde a_1^\dagger\dots \tilde a_d^\dagger\ket{\boldsymbol{0}}\,, \quad \tilde a_j = \sum_{k=1}^MV_{jk}a_k \quad\forall j \in [M]\,,
\end{equation}
where $d$ is the number of fermions, $M$ is the number of qubits, $V$ is a $M\times M$ unitary matrix and $\ket{\boldsymbol{0}}$ is the vacuum state.

In AFQMC, one decomposes the sum over two-body interactions into a sum over squares of one-body operators $v_\gamma$ through a Cholesky decomposition. Thus, the Hamiltonian is transformed into
\begin{equation}\label{eq:quad_ham}
    H=H_1 + \frac{1}{2}\sum_{\gamma=1}^{K_\gamma}v_\gamma^2\,,
\end{equation}
where $v_\gamma = i\sum_{pq}L_{pq}^\gamma a^\dagger_p a_q$. The matrix terms $L_{pq}^\gamma$ are calculated through the decomposition of the two-body terms $v_{pqrs}=\sum_{\gamma=1}^{K_\gamma} L_{pr}^\gamma L_{qs}^\gamma$.
Inserting \cref{eq:quad_ham} into \cref{eq:imag_evolution}, one can efficiently compute the exponential using the Hubbard-Stratonovich transformation \cite{motta2018ab, negele2018quantum}
\begin{equation}
    e^{-\frac{\Delta\tau}{2}\sum_\gamma v_\gamma^2}=\int \text{d}\mathbf{x} \,p(\mathbf{x})e^{-i\sqrt{\Delta\tau}\sum_\gamma x_\gamma v_\gamma^2+\mathcal{O}(\Delta\tau^2)}\,,
\end{equation}
where $p(\mathbf{x})$ is the standard Gaussian distribution and $\mathbf{x}\in\mathbb{R}^{K_\gamma}$ are auxiliary fields. 

This transformation guarantees that a Slater determinant will remain a Slater determinant at each time evolution step, allowing for the stochastic evolution of an ensemble of many Slater determinants, known as Monte Carlo walkers, $\{|\phi_z^{(n)}\rangle; z=1,\dots,N_w; n=1,\dots,N_{\Delta\tau}\}$ where $N_w$ is the total number of walkers in AFQMC and $N_{\Delta\tau}$ is the total number of imaginary time steps. These walkers each accumulate weights $\{W_{n,z}\}$ and phases $\{e^{i\theta_{n,z}}\}$. Their evolution is guided by an estimate of the ground state known as the trial wavefunction $\ket{\Psi_{\text{T}}}$.

One drawback of a naive implementation of the stochastic evolution is the phase problem, which is the complex, and more severe, analogue of the sign problem \cite{zhang2003quantum}. This effect can be mitigated using the phaseless approximation that, using the trial state, projects the walkers to the real axis in order to avoid the accumulation of a phase. The update rule for the walker is
\begin{align}
    |\phi_z^{(n+1)}\rangle=&B(\mathbf{x}_{n,z}-\bar{\mathbf{x}}_{n,z})|\phi_z^{(n)}\rangle\\
    W_{n+1,z}=&I(\mathbf{x}_{n,z}, \bar{\mathbf{x}}_{n,z}; \phi_z^{(n+1)})W_{n,z}
\end{align}
where the evolution operator
\begin{equation}
{B}({\bf x}) = \exp[-\Delta\tau\,\left(-E_{\text{GS}}+{H}_1\right)+ \sqrt{\Delta\tau}\sum_{\gamma}x_{\gamma}{v}_{\gamma}]\,,
\end{equation}
the vector $\bar{\mathbf{x}}_{n,z}$ known as the force bias is calculated as
\begin{equation}
    \left(\bar{\mathbf{x}}_{n,z}\right)_\gamma = -\sqrt{\Delta\tau}\frac{\bra{\Psi_{\text{T}}}v_\gamma|\phi_z^{(n)}\rangle}{\langle\Psi_{\text{T}}|\phi_z^{(n)}\rangle}\,,
\end{equation}
the importance function $I$ is defined as
\begin{equation}
\begin{split}
    I(\mathbf{x}_{n,z}, \bar{\mathbf{x}}_{n,z}; \phi_z^{(n+1)}) = \left|O_{n,z} e^{\mathbf{x}_{n,z} \dot{\bar{\mathbf{x}}}_{n,z} - \bar{\mathbf{x}}_{n,z} \dot{\bar{\mathbf{x}}}_{n,z}/2}\right|\\ \times \text{max}(0,\cos(\theta_{n,z}))\,,
\end{split}
\end{equation}
where the overlap ratio is 
\begin{equation}
    O_{n,z}=\frac{\langle\Psi_{\text{T}}|\phi_z^{(n+1)}\rangle}{\langle{\Psi_{\text{T}}}|{\phi_z^{(n)}}\rangle}\,,
\end{equation}
and the phase is the argument of the overlap ratio $\theta_{n,z}=\arg(O_{n,z})$.

The total energy is calculated as 
\begin{equation}
    \mathcal{E}^{(n)}=\frac{\sum_{z=1}^{N_w}W_{n,z}e^{i\theta_{n,z}}\mathcal{E}_{\text{loc}}(\phi_z^{(n)})}{\sum_{z=1}^{N_w}W_{n,z}e^{i\theta_{n,z}}}
\end{equation}
where the local energy $\mathcal{E}_{\text{loc}}(\phi_z^{(n)})$ is a mixed estimator of $H$ defined as:
\begin{equation}
    \mathcal{E}_{\text{loc}}(\phi_z^{(n)})=\frac{\bra{\Psi_{\text{T}}}H|\phi_z^{(n)}\rangle}{\langle{\Psi_{\text{T}}}|{\phi_z^{(n)}}\rangle}\,.
\end{equation}

A trial wavefunction that is more accurate or has a larger overlap with the true ground state $\braket{\Psi_{\text{T}}}{\Psi_{\text{GS}}}$ results in a more accurate ground state energy estimate using AFQMC. The potential of trial wavefunction obtained by different numerical methods has been recently explored in a number or scientific publications, including those crossing into the quantum computing field where the trial wavefunction is prepared on a quantum device known as QC-AFQMC \cite{huggins2022unbiasing, amsler2023classical, huang2024evaluating,  jiang2024unbiasing}. For such quantum-generated wavefunctions, one can use the Hadamard test to calculate the overlaps between the walkers and the trial state on the quantum device \cite{cleve1998quantum, aharonov2005a, kitaev1995quantum} or alternatively it is possible to do so classically, using matchgate shadows \cite{zhao2020fermionic, wan2022matchgate}, which we will explain in more detail in \cref{subsec:overlaps}.

\subsection{CS-AFQMC}\label{subsec:csafqmc}

As depicted in \cref{fig:algorithm_workflow}, we propose to incorporate the use of a ground state found in the contextual subspace as the trial wavefunction of AFQMC. In order to do this it is necessary to follow the below steps:
\begin{enumerate}
    \item Solve the noncontextual Hamiltonian using the heuristic described in \cref{subsec:csa}.
    \item Project the contextual Hamiltonian to a contextual subspace with the chosen or available number of qubits.
    \item Obtain an approximate ground state wavefunction in the contextual subspace using an appropriate quantum algorithm.
    \item Reverse the projection of the ground state found in the contextual subspace into the full space to obtain the trial wavefunction.
    \item Run QC-AFQMC guided by overlaps between the trial wavefunction and Monte Carlo walkers.\label{afqmc_step}
\end{enumerate}

The general idea and advantage of this method is that one can make a good enough trial state using a smaller quantum processor compared to what the full problem would require. Thus, we benefit from the quantum trial wavefunction as originally proposed in Ref.~\cite{huggin2022unbiasing} while using a more reasonable sized quantum device.

\subsection{Calculating overlaps} \label{subsec:overlaps}

In Step~\ref{afqmc_step} of the workflow presented in the previous section, one has to calculate the overlaps between Monte Carlo walkers and the trial state. This can be done via the Hadamard test or by using matchgate shadows \cite{zhao2020fermionic, wan2022matchgate}. The former method computes overlaps on a quantum computer and has quantum resource requirements that would be a severe bottleneck for NISQ quantum computers. The latter approach, on the other hand, allows for overlaps to be computed classically after the trial state has been measured and without the need for fast communication between classical and quantum processors. The variance of matchgate shadows scales as $\mathcal{O}(\sqrt{N}\log N)$ with system size, thus requires only a polynomial number of samples to be measured. The classical algorithm for overlap calculation scales as $\mathcal{O}(N^4)$. This means that matchgate shadows is an efficient formalism with respect to both classical and quantum computational resources.
Next, we outline the algorithm for computing the overlap between a matchgate shadow and a Slater determinant type walker and then illustrate how it would be incorporated into CS-AFQMC. 

\subsubsection{Matchgate shadows}
When calculating the overlap using shadow tomography between an arbitrary wave function, or in our case a trial wavefunction $\ket{\Psi_{\text{T}}}$, and a Slater determinant $\ket{\varphi}$ defined in \cref{eq:slater_def}, one prepares the state 
\begin{equation}\label{eq:tau_state}
    \ket{\boldsymbol{\tau}} = \frac{\ket{\Psi_{\text{T}}} + \ket{\boldsymbol{0}}}{\sqrt{2}}
\end{equation}
on a quantum computer. Then a random matchgate circuit $U_Q$ is appended to the state preparation. Here, $U_Q$ are fermionic Gaussian unitaries (FGU) described by a matrix in the real orthogonal group $Q\in O(2N)$ where $N$ is the number of qubits in the circuit. The unitary satisfies \begin{equation}
    \label{eq:majorana_basischange}
    U_Q^\dagger\gamma_\mu U_Q=\sum_{\nu=1}^{2N}Q_{\mu\nu}\gamma_\nu~,
\end{equation}
where $\gamma_\mu$ are Majorana operators, $\gamma_{2j-1}=a_j+a_j^\dagger$ and $\gamma_{2j}=-i(a_j-a_j^\dagger)$ for $j\in[N]$. Another way to describe matchgate circuits is as a product of rotations generated from the gate set 
\begin{equation}
    \label{eq:matchgates}
    M_n=\left\{\exp(i\theta X_j\otimes X_{j+1}),\exp(i\theta Z_j), X_N \right\}\,,
\end{equation}
acting on $N$ qubits arranged in a line, where $X_i$, $Y_i$ and $Z_i$ are the Pauli operators acting on the $i$th qubit.

Taking a measurement of the resulting state provides a snapshot $\ket{b}$ that can be classically post-processed to calculate the expectation value of this $\ket{\tau}$ with respect to the observable $\ket{\varphi}\bra{\boldsymbol{0}}$, which reduces to the overlap between the trial state $\ket{\Psi_\text{T}}$ and the arbitrary Slater determinant $\ket{\phi}$, defined by the unitary matrix $V$ (see \cref{eq:slater_def}).
The classical post-processing begins with calculating the coefficients $c_\ell$ in front of the monomial $z^\ell$ in the polynomial
\begin{equation}\label{eq:matchgateinversechannel}
    q(z)=\alpha_{\eta, N}\text{Pf}\left[\left.\left(C_{\boldsymbol{0}}+zW^*\tilde Q Q^\text{T}C_{b}Q\tilde Q^\text{T}W^\dagger\right)\right|_{\overline{S}_{\eta}}\right]~,
\end{equation}
where $\eta$ is the number of fermions in the system, $N$ is the number of qubits or spin orbitals in the system, $\alpha_{\eta, N}=i^{\eta/2}/2^{N-\eta/2}$, Pf$(A)$ is the Pfaffian of a real,  anti-symmetric matrix $A=-A^\text{T}$, $C_\textbf{i}$ is the covariance matrix of the binary string $\textbf{i}$, $W$ is a change of basis from the set $\{\sqrt{2}a_1^\dagger,\sqrt{2}a_1,\dots,\sqrt{2}a_\eta^\dagger,\sqrt{2}a_\eta,\gamma_{2\eta+1},\dots,\gamma_{2N}\}$ to the set of Majorana operators $\{\gamma_\mu\}_{\mu\in[2N]}$ \cite{wan2022matchgate}, $\tilde Q$ specifies the fermionic Gaussian unitary that generates the Slater determinant defined by the unitary matrix $V$, $Q$ is the orthogonal rotation matrix for a given snapshot and ${\overline{S}_{\eta}}=[2N]\setminus \{1,3,\dots,2\eta-1\}$. The unbiased estimate $o^{(j)}$  ($j \in [n_s]$ with $n_s$ number of shadows) of the overlap of an arbitrary state with a Slater determinant is found by calculating
\begin{equation}
    \label{eq:kianna_estimation}
    o^{(j)}=2\sum_{\ell=0}^N\binom{2N}{2\ell}\binom{N}{\ell}^{-1}c_\ell~.
\end{equation}
Either the average or a median-of-means estimate~\cite{wan2022matchgate} over the $n_s$ many unbiased estimates results in the final estimate of the overlap $\braket{\varphi}{\Psi_\mathrm{T}}$. We note that some care needs to be taken for calculations when neither $\ket{\Psi_\mathrm{T}}$ not the measured observables are of even size. In such cases a slight modification of the presented formalism can be used (see Appendix A of Ref.~\cite{wan2022matchgate}).

\subsubsection{Incorporation into CS-AFQMC}
The desired trial wavefunction $\ket{\Psi_{\text{T}}}$ can be written as an $n$-qubit density matrix $\rho$ that has been rotated by an operator $U_\mathcal{W}$. Let $\rho_{\text{unrot}}=\rho_\text{cs}\otimes\rho_{\overline{\text{cs}}}$ be the unrotated state which has the form of a  product state of the contextual ground state $\rho_\text{cs}$ on $n_\text{cs}$ qubits and the state \begin{equation}
    \rho_{\overline{\text{cs}}}=\ket{\text{nc}'}\bra{\text{nc}'}\,,\quad\text{ where }\ket{\text{nc}'}=\Pi_{\overline{\text{cs}}}U_\mathcal{W}\ket{\text{nc}},
\end{equation} 
where the state $\ket{\text{nc}}$ is the noncontextual ground state on $n$ qubits, $\text{nc}'\in\{0,1\}^{n_{\overline{\text{cs}}}}$, where $n_{\overline{\text{cs}}}$ is the number of qubits in the complement of the contextual subspace, and $\Pi_{\textbf{i}}$ is the projection operator onto the space $\textbf{i}$. The value of $\text{nc}'$ is known once the noncontextual ground state and the rotation $U_\mathcal{W}$ are calculated. Thus, we can describe the trial wavefunction as
\begin{align}
    \rho &= U_\mathcal{W}^\dagger \rho_{\text{unrot}} U_\mathcal{W}\\
         &= U_\mathcal{W}^\dagger \left( \rho_\text{cs}\otimes\rho_{\overline{\text{cs}}} \right) U_\mathcal{W}\\
         &= U_\mathcal{W}^\dagger \left[\rho_\text{cs}\otimes\left[\Pi_{\overline{\text{cs}}}U_\mathcal{W}\ket{\text{nc}}\bra{\text{nc}}U_\mathcal{W}^\dagger\Pi_{\overline{\text{cs}}}\right]\right]U_\mathcal{W}\,.
\end{align} 

In order to calculate the overlaps, we want to use the following process: 
\begin{enumerate}
    \item Sample a shadow unitary $U$ on $n_\text{cs}$ qubits from a distribution of unitaries $D$.
    \item Apply the sampled $U$ onto a state within the contextual subspace $\rho_\text{cs}$ on a quantum device.
    \item Measure the bitstring $\ket{b_\text{cs}}$ on a quantum device.
    \item Classically combine $\ket{b_\text{cs}}$ with the known bitstring $\ket{b_{\overline{\text{cs}}}}$ for the complement of the contextual subspace, which depends on $b_\text{cs}$.
    \item Classically post-process the combined bitstring to get the expectation value of an observable $O$ with respect to $\rho$.
\end{enumerate}

To start, we write the generic channel $\mathcal{M}$ for the process described above as
\begin{align}
\begin{split}\label{eq:gen_channel}
  \mathcal{M} (\rho)    & = \underset{U\sim D}{\mathbb{E}}\sum_{b\in\{0,1\}^n}\bra{b}\left(U\otimes \mathbb{I}_{\overline{\text{cs}}}\right)\rho\left(U^\dagger\otimes \mathbb{I}_{\overline{\text{cs}}}\right)\ket{b}\times \\
    &\qquad\qquad\qquad\quad\left(U^\dagger\otimes \mathbb{I}_{\overline{\text{cs}}}\right)\ket{b}\bra{b}\left(U\otimes \mathbb{I}_{\overline{\text{cs}}}\right)
\end{split}
\\[1ex]
\begin{split}\label{eq:derived_channel}
  & = \underset{U\sim D}{\mathbb{E}}\sum_{b_{\text{cs}}\in\{0,1\}^{n_{\text{cs}}}}\bra{b_\text{cs}}U\rho_\text{cs}U^\dagger\ket{b_\text{cs}}\\
    &\qquad\qquad\qquad\quad\left[U^\dagger\ket{b_\text{cs}}\bra{b_\text{cs}}U\right]\otimes\\
    &\qquad\qquad\qquad\quad\ket{b_{\overline{\text{cs}}};\text{nc}'; b_\text{cs}}\bra{b_{\overline{\text{cs}}};\text{nc}'; b_\text{cs}} 
\end{split}
\end{align}
where $b_{\textbf{i}}$ is the bitstring restricted to the subspace $\textbf{i}$, $|b'\rangle$ is the computational basis state $|b\rangle$ rotated by $\tilde{U}_\mathcal{W}$, which corresponds to the contextual subspace rotation $U_\mathcal{W}$ modified based on the commutation with the shadow unitary $U$ (see \cref{app:cont_rot_commutation}). We provide a detailed derivation of \cref{eq:derived_channel} in \cref{app:derivation}. Since we assume a distribution of random matchgate circuits, we can always find a $\tilde{U}_\mathcal{W}$ that commutes with the circuit $U$.
The state $\ket{b_{\overline{\text{cs}}};\text{nc}'; b_\text{cs}}$ is the state $\ket{b_{\overline{\text{cs}}}}$ given $\ket{b_\text{cs}}$ such that $\tilde{U}_\mathcal{W}\left[\ket{b_\text{cs}}\otimes\ket{b_{\overline{\text{cs}}}}\right]= \ket{b_\text{cs}}\otimes\ket{\text{nc}'}$. Further, in the contextual subspace $\ket{b_\text{cs}}=\Pi_\text{cs}\tilde{U}_\mathcal{W}\left[\ket{b_\text{cs}}\otimes\ket{b_{\overline{\text{cs}}}}\right]=\pm\ket{b_\text{cs}'}$ and the potential phase is cancelled when applying the unitary twice. 
This means that our procedure collapses the state $U_\mathcal{W}^\dagger\rho \,U_\mathcal{W}$ to $U^\dagger|\hat{b}\rangle\langle\hat{b}|U\otimes\ket{b_{\overline{\text{cs}}};\text{nc}'; b_\text{cs}}\bra{b_{\overline{\text{cs}}};\text{nc}'; b_\text{cs}}$ with probability $\langle b_\text{cs}| U\rho_\text{cs}U^\dagger|b_\text{cs}\rangle$. Since the probability distribution is only defined on the contextual subspace measurements, we can safely separate the two processes and classically add the bit string for the complementary space in the post-processing step.

We define the random operator $\hat\rho$ by
\begin{equation}
    \hat\rho \equiv \mathcal{M}^{-1}\left(\hat{U}^\dagger|\hat{b}_{\text{cs}}\rangle\langle\hat{b}_{\text{cs}}|\hat{U}\otimes| b_{\overline{\text{cs}}};\text{nc}'; \hat{b}_\text{cs}\rangle\langle b_{\overline{\text{cs}}};\text{nc}'; \hat{b}_\text{cs}|\right),
\end{equation}
where the hat symbol signifies that the quantity is sampled and the probability of measuring bitstring $b$ given the unitary $U$ is $\Pr[|\hat{b}\rangle=\ket{b}|\hat{U}=U]=\bra{b_\text{cs}}U\rho_\text{cs}U^\dagger\ket{b_\text{cs}}$, whereas the state $| b_{\overline{\text{cs}}};\text{nc}'; b_\text{cs}\rangle\langle b_{\overline{\text{cs}}};\text{nc}'; b_\text{cs}|$ is computed classically based on the measurement from the contextual subspace, $\ket{b_\text{cs}}$. By construction we have that the expectation value of the sampled state $\hat{\rho}$ is given by:
\begin{equation}
    \mathbb{E}[\hat{\rho}]=\rho=U_\mathcal{W}^\dagger \left( \rho_\text{cs}\otimes\rho_{\overline{\text{cs}}} \right) U_\mathcal{W}\,.
\end{equation}

Now, we want to estimate the expectation values of $M$ different observables in the space of linear operators of the $n$-qubit Hilbert space $O_i\in\mathcal{L}(\mathcal{H}_n)$ for $i\in[M]$. For a particular observable $O$, we can use estimators of the form:
\begin{equation}
\begin{split}
    \hat o \equiv \Tr[O\hat{\rho}]
    =\Tr\Bigl[&O\mathcal{M}^{-1}\Bigl(\hat{U}^\dagger|\hat{b}_\text{cs}\rangle\langle\hat{b}_\text{cs}|\hat{U}\otimes\\
    &| b_{\overline{\text{cs}}};\text{nc}'; b_\text{cs}\rangle\langle b_{\overline{\text{cs}}};\text{nc}'; b_\text{cs}|\Bigr)\Bigr]\,,
\end{split}
\end{equation}
which follows from $\mathbb{E}[\hat o] = \Tr[O\rho]$. We can compute this quantity by taking measurements $\hat{\rho}$ which contain contextual subspace information from the quantum computer, determining classically the bits in the complement of the contextual subspace, and forming $C_{b}$, which is used as input for the post-processing, from the combined bitstring $b = b_{\text{cs}} \cup b_{\overline{\text{cs}}}$, as if the entire state was obtained from a quantum computer.

When evaluating the variance of this estimator it is clear that the resulting expression will be upper-bounded by the variance bounds for the full system, which we study further in \cref{app:variance}. However, given the deterministic nature of the complement of the contextual subspace, we expect the actual variance to be bounded based solely on the size of the contextual subspace rather than the full system. The number of samples should consequently scale based on the number of fermionic modes in the contextual subspace alone.

In order to estimate the overlap of the trial state $\ket{\Psi_{\text{T}}}$ with a Slater determinant $\ket{\phi}$, one needs to express this overlap in the form of an observable. One way of achieving this is by performing measurements with respect to the observable $O=\ket{\perp}\bra{\phi}$, where $\ket{\perp}$ is a state that is orthogonal to both the Slater determinant and the state $\rho$. One then prepares a superposition of the desired state $\rho$ and the perpendicular state $\ket{\perp}\bra{\perp}$ on a quantum device. This is equally true for our case, given some careful consideration for the nature of the orthogonal state $\ket{\perp}$, since in small contextual subspaces an orthogonal state might not exist. In this case, enlarging the contextual subspace by a few auxiliary qubits---such that the even subspace requirement of the matchgate shadow protocol is satisfied---can be used as a flag to identify and generate such orthogonal states. 

\section{Results}\label{sec:results}

\begin{figure*}
    \centering    \includegraphics[width=\textwidth]{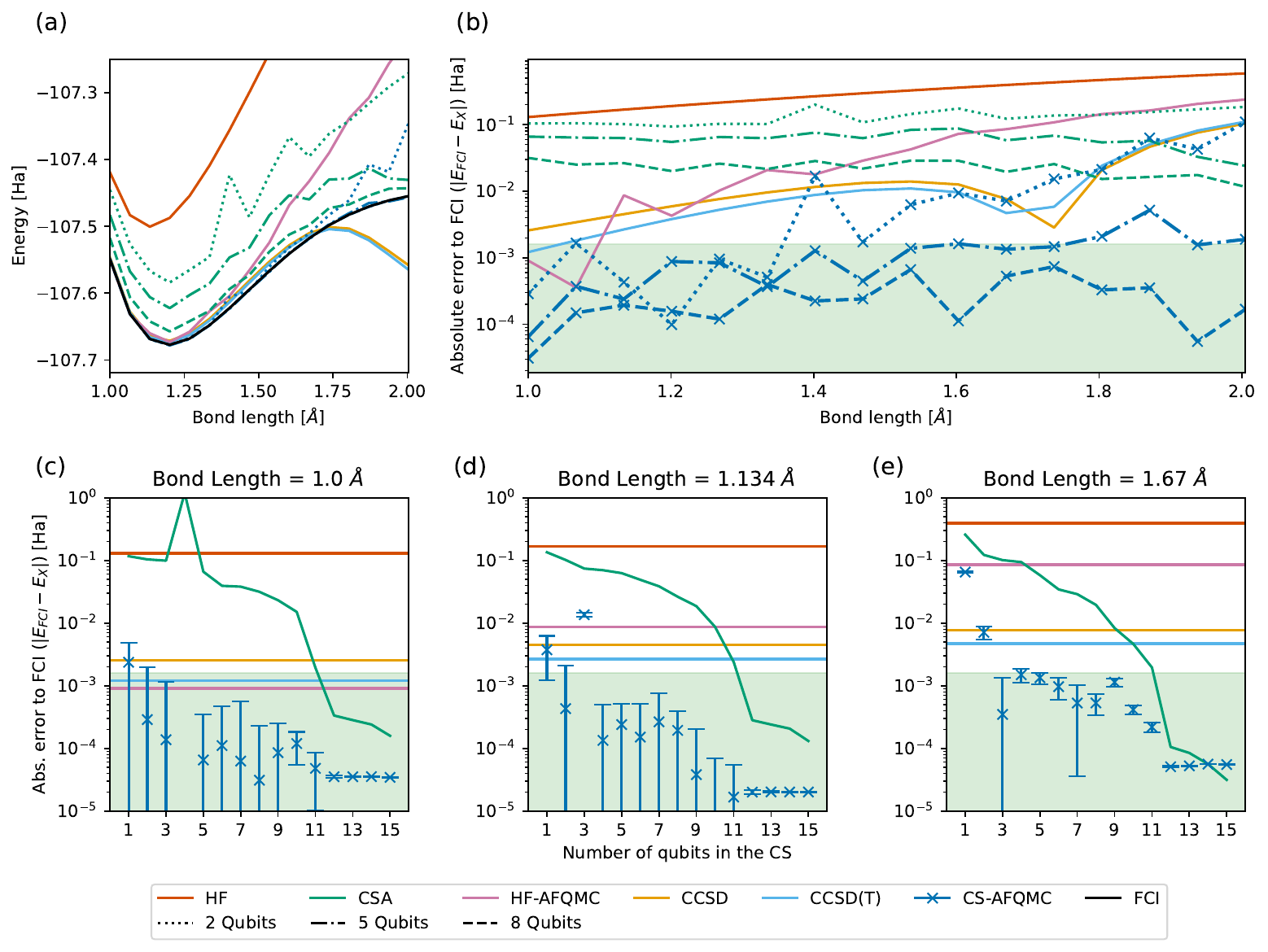}
    \caption{Energy calculation of the nitrogen dimer in the STO-3G basis set (with a total number of 20 spin orbitals). Compares HF, CCSD, CCSD(T), HF-AFQMC, CSA and CS-AFQMC for increasing size of the contextual subspace and green shading representing chemical accuracy (1 kcal/mole). The error bars are calculated based on the mean squared error (MSE) of multiple CS-AFQMC runs at each CS level for a single bond length. \emph{top:} Dissociation curve for the nitrogen dimer with \textbf{(a)} the absolute energy values and \textbf{(b)} error to the FCI calculation. \emph{bottom:} Energy comparison to the FCI solution for three bond lengths: 1.0, 1.134 (equilibrium) and 1.67.}
    \label{fig:n2_triptych}
\end{figure*}

\subsection{Nitrogen Dimer}\label{subsec:nitrogen}

We first test the CS-AFQMC method on the nitrogen dimer, $\ce{N}_2$, composed of two nitrogen atoms. We use the STO-3G basis set, in which this model corresponds to 14 electrons in 10 molecular orbitals (or 20 spin orbitals).
This system is commonly used as a toy-problem for benchmarking numerical methods due to the strongly correlated nature of the electrons, particularly as the inter-atomic separation increases beyond the equilibrium distance. At large inter-atomic distances the ground state of the nitrogen dimer is no longer well described by a single Slater determinant due to the domination of static correlations. In each AFQMC simulation, regardless of the trial wavefunction used, we propagate 600 walkers over 600 time-blocks each consisting of 25 time-steps of $\Delta \tau=0.005$ while performing population control and numerical stabilization after every 25 time-step block. 
 
We see a comparison of a number of numerical methods for $\ce{N}_2$ at different inter-atomic distances in \cref{fig:n2_triptych}. Specifically, we compare the Hartree Fock (HF), coupled cluster singles and doubles (CCSD), CCSD with perturbative triples (CCSD(T)), the ground state of the contextual subspace ansatz (CSA), AFQMC with a HF trial state (HF-AFQMC), and contextual subspace trial states (CS-AFQMC) against the full configuration interaction (FCI) serving as the exact solution for benchmarking against \cite{helgaker2013molecular}. For CSA and CS-AFQMC we perform computations with a varying size of the contextual subspace $1\leq N_{\text{cs}} \leq 15$ and use the exact ground state computed within the CS. Note that HF-AFQMC corresponds to CS-AFQMC with $N_{\text{cs}} = 0$.

In the \textbf{(a)} and \textbf{(b)} panels of \cref{fig:n2_triptych} we
only show CSA and CS-AFQMC for $N_{\text{cs}} \in \{2,5,8\}$. We observe that all methods but CS-AFQMC with 5 and 8 qubits in the CS fail to capture the dissociation curve correctly, while all of CCSD, CCSD(T) and HF-AFQMC start to diverge from the correct solution as the bond length is increased. For CSA we see that the overall shape of the curve is correct, in particular for 8 qubits in the CS, but differs by a constant energy shift. This hints that the physics has been correctly captured at this stage, which translates to a CS-AFQMC result within chemical accuracy for the same $N_{\text{cs}}$.

We proceed to inspect three particular choices for the bond length $D \in \{1.0 \angstrom, 1.134 \angstrom, 1.67 \angstrom\}$ more closely in panels \textbf{(c)}, \textbf{(d)} and \textbf{(e)}. For the CSA we see an expected monotonic decrease in energy, with the exception of the $N_{\text{cs}}=4$ value at $D=1.0\angstrom$. This is an example of occasional method failures which we have observed for very small contextual subspaces ($N_{\text{cs}} < 5$). They are related to the ground state of the contextual subspace being in the wrong particle sector when projected back to the full Hilbert space. This usually leads to completely wrong results which are easily identifiable. We see that for all three bond lengths the CSA error with respect to the exact result eventually converges below chemical accuracy at around 12 qubits in the CS. 

In contrast to CSA, the CS-AFQMC energy does not behave monotonically even if it converges extremely quickly with the size of the CS. Notably, only 5 CS qubits are needed to reaching chemical accuracy for most bond length studied and outperform all other numerical methods considered here. The relative error quickly falls to values below $10^{-3}$, where it is dominated by the stochastic noise of the AFQMC, the value of which we extract by performing multiple calculations per CS. We also observe that, remarkably, the CS-AFQMC error to FCI improves upon the CSA by up to three orders of magnitude, thus illustrating the potential of this combined approach.

As noted earlier, AFQMC is not variational and has stochastic noise. This means each time a calculation is performed the resulting estimate will be different. We found that as the size of the CS is increased in CS-AFQMC, the mean squared error (MSE) of these different runs decreases. This is expected because the trial wavefunction increases in accuracy and subsequently the evolution of the walkers are guided more accurately.
This can be understood as follows. When using an approximate trial wavefunction, there will be instances of walkers being eliminated when they cross the nodal plane of the trial wavefunction, but would not have crossed the nodal plane of the true ground state, and vice versa. This contributes to the bias, but also to the variance of the estimate of different AFQMC calculations, which is reduced as the quality of the trial wavefunction is improved.

\begin{figure}
    \centering
\includegraphics[width=0.96\columnwidth]{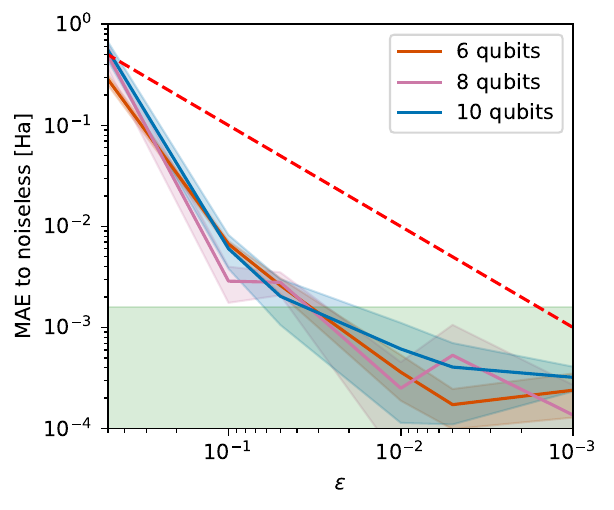}
    \caption{Mean absolute error and the corresponding variance computer from multiple CS-AFQMC runs, are shown for the  nitrogen dimer at bond length of $D = 1.737 \angstrom$ using a noisy wavefunctions as described in \cref{eq:noise_wfn}. Results are shown for  6, 8 and 10 qubits in the contextual subspace. The green shaded area signifies chemical accuracy (1 kcal/mole) and the red dashed line shows where the $x$- and $y$-axes have equal values.}
    \label{fig:toy_system_noise}
\end{figure}

We now investigate the impact of noise in the trial wavefunction $\ket{\Psi_{\text{T}}}$ on the accuracy of CS-AFQMC. To this end, the noisy trial wavefunction is perturbed according to 
\begin{equation}\label{eq:noise_wfn}
    \ket{\Psi_{\text{T}}^\varepsilon}=\frac{(1-\varepsilon)\ket{\Psi_{\text{T}}}+\varepsilon\sum_\alpha\ket{c_\alpha}}{\lVert\ket{\Psi_{\text{T}}^\varepsilon}\rVert}\,,
\end{equation}
where the sum over $\ket{c_\alpha}$ are the computational basis states with the correct fermion- and spin-numbers and $\varepsilon>0$ is an error parameter. The noise results in disproportionate error in the overlap with the noiseless wavefunction due to the increased amplitudes of basis states which originally have only a small or zero contribution to the true trial state. In \cref{fig:toy_system_noise}, we plot the mean absolute error (MAE) of the CS-AFQMC energy for the nitrogen dimer at $D = 1.737 \angstrom$. We show curves for CS wavefunctions of 6, 8 and 10 qubits and as a function of the magnitude of error parameter $\varepsilon$.

We find that the noise in the resulting CS-AFQMC energy is suppressed compared to the original perturbation $\varepsilon$ value for each size of the contextual subspace, with the relative difference reaching more than one order of magnitude. This demonstrates that the noise in the trial wavefunction is mitigated by the CS-AFQMC method, which is promising for implementations on NISQ quantum hardware. We further observe that the error rates are largely independent of the contextual subspace sizes. This demonstrates that one can, in principle, choose arbitrary-sized subspaces without concern that errors on the quantum device will severely impact the CS-AFQMC method beyond the noise already present in the quantum trial wavefunction.

\subsection{Reductive decomposition of ethylene
carbonate}\label{subsec:ec}

\begin{figure}
    \centering
\includegraphics[width=0.96\columnwidth]{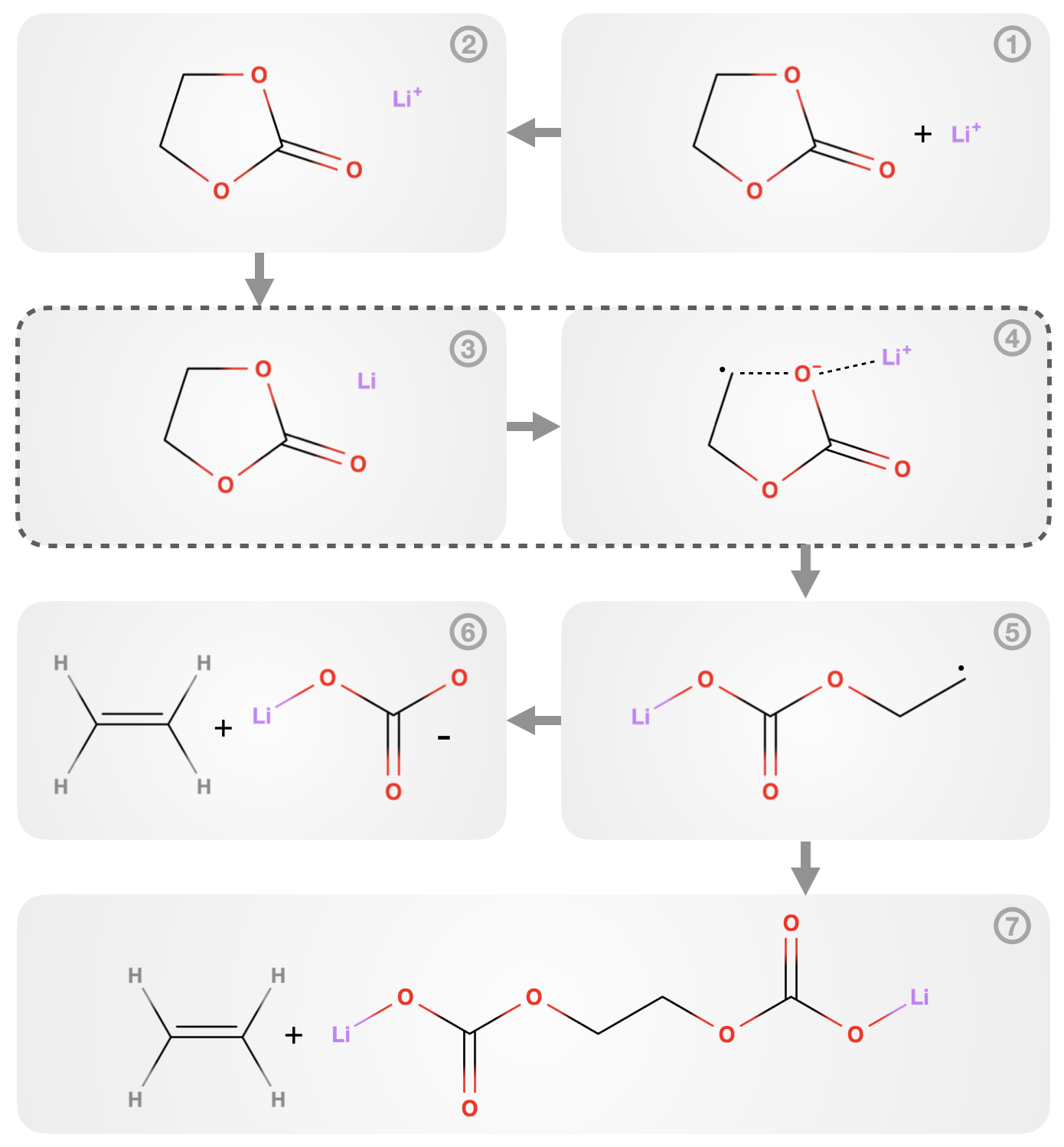}
    \caption{Reductive decomposition of EC in the presence of a lithium ion. The reaction path corresponds to that of Ref.~\cite{debnath2023afqmcec}.  Configurations (6) and (7) represent two possible termination pathways of the reaction path. Dashed line indicates the reaction step $(3)\to(4)$ investigated in this work. Commentary on the individual reaction steps can is provided in \cref{sec:results}.}
    \label{fig:electrolyte detail}
\end{figure}

Next, we consider a more realistic example that is directly relevant to applications in battery design. We investigate the stability of ethylene carbonate (EC) against the reductive decomposition by lithium (\ce{Li}) occurring at the anode electrolyte interface. EC is one of the most commonly found solvents for electrolytes used in lithium-ion batteries. While EC is not stable against reductive decomposition by \ce{Li}, the formation of a solid-electrolyte interface (SEI) alleviates this problem to some extent by passivating the electrode and thereby suppressing the reaction \cite{peled2017sei}. Understanding the reductive decomposition of EC and the role of the SEI in this process is a key factor in understanding battery safety and performance thus enabling the development of better performing batteries. The reductive decomposition steps of ethylene carbonate (\ch{C3H4O3}) in the presence of lithium ions (\ch{Li+}) is shown in \cref{fig:electrolyte detail} as adapted from Ref.~\cite{wang2001ecdecomposition} and Ref.~\cite{debnath2023afqmcec}. The process involves:
\begin{itemize}[leftmargin=2.1cm]
    \item [$(1)\to(2)$: \hspace{0.1cm}] The strong binding of an EC molecule to a lithium ion (1) to form an ion-pair intermediate (2).
    \item [$(2)\to(3)$: \hspace{0.1cm}] The intermediate (2) is reduced to form a neutral adduct (3). 
    \item [$(3)\to(4)$: \hspace{0.1cm}] The adduct (3) undergoes homolytic ring cleavage. First it goes through a transition state (4). 
    \item [$(4)\to(5)$: \hspace{0.1cm}] From the transition state (4) an intermediate radical (5) is formed. From this radical, several possible termination pathways exist. 
    \item [$(5)\to(6)$: \hspace{0.1cm}] The first option is a reductive decomposition which forms a lithium carbonate anion together with ethylene (6).
    \item [$(5)\to(7)$: \hspace{0.1cm}] The second is a dimerization which forms lithium ethylene dicarbonate together with ethylene (7). 
\end{itemize}
Both final products of lithium carbonate and lithium ethylene dicarbonate have been associated with the formation of the SEI layer. There have been multiple experimental \cite{metzger2016h2evolution, shkrob2013eprec} and theoretical \cite{wang2001ecdecomposition, debnath2023afqmcec} studies on this topic, which is still an active area of research. 

\begin{figure}
    \centering    \includegraphics[width=0.9\columnwidth]{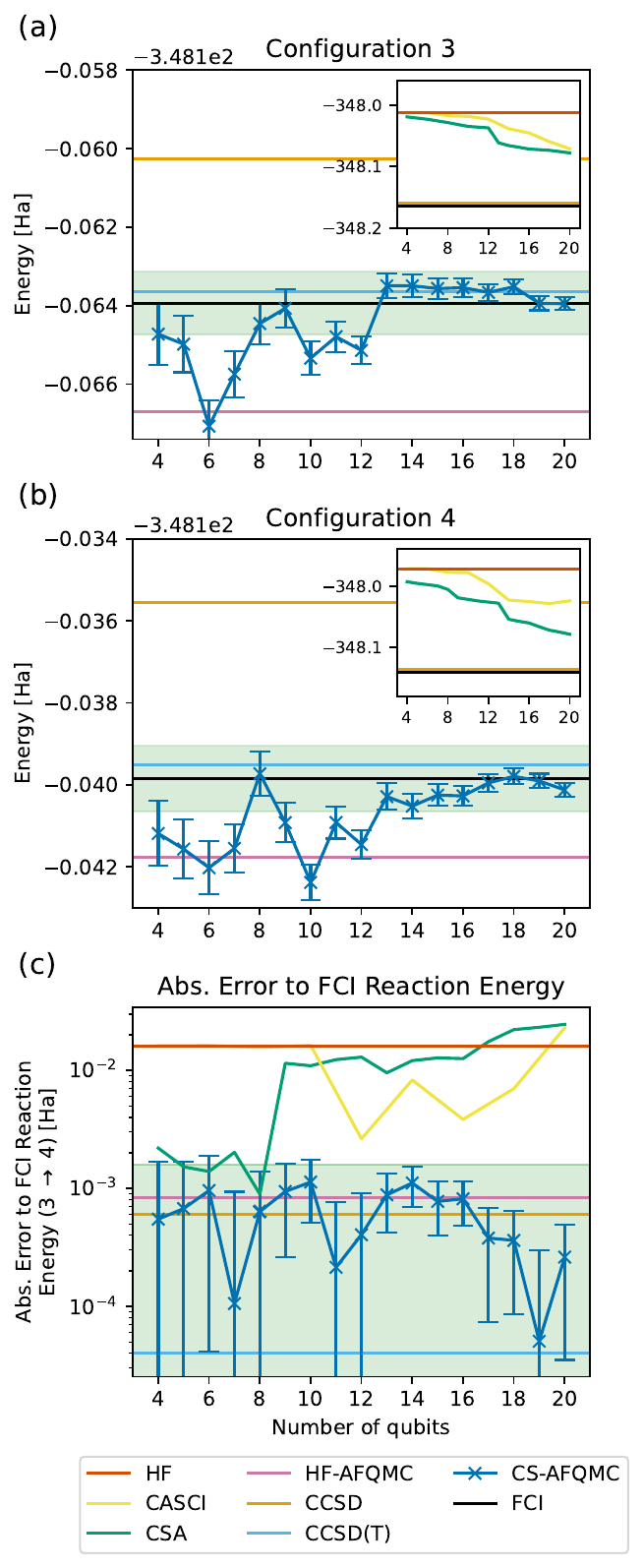}
    \caption{Third step $(3)\to(4)$ in the reductive decomposition process of EC by lithium shown in \cref{fig:electrolyte detail}. An active space of 16 molecular orbitals (32 qubits) is chosen from the original 118 orbitals in the cc-pVDZ basis set. Green shaded areas signify chemical accuracy (1 kcal/mole). \textbf{(a)} and \textbf{(b)} Energy value calculations for various methods for increasing the number of qubits up to 20 qubits in the CS for configurations 3 and 4, respectively. \textbf{(c)} Error of the reaction energy calculations to the CASCI solution on 32 spin orbitals, labeled as FCI, for up to 20 qubits in the CS. The error bars are calculated based on interpolating the mean squared error (MSE) of multiple CS-AFQMC runs for 0, 8 and 16 qubit CS for configuration 3.}
\label{fig:electrolyte_results_abs}
\end{figure}

Here, we investigate the rate determining step $(3)\to(4)$ of this reductive decomposition process, as shown in \cref{fig:electrolyte detail}. We consider the \ce{Li}-EC complex in the cc-pVDZ \cite{dunning1989ccpvnz, prascher2011ccpvnz} basis set which corresponds to 49 electrons in 118 molecular orbitals. This system size is too large for the computation of an exact FCI ground state energy  benchmark. Therefore, we perform an active-space selection procedure prior to the CS calculations which leads to an active space of 19 electrons in 16 MP2 natural orbitals~\cite{hqs_asf}. These 16 molecular orbitals correspond to 32 spin-orbitals and hence could, in principle, be simulated with 32 qubits under the JWT. In Ref.~\cite{huggin2022unbiasing} is was shown how to use wavefunctions defined within active spaces to guide AFQMC calculations on the full Hilbert space and the procedure is compatible with the CSA trial states obtained in this work.

Our results are summarised in \cref{fig:electrolyte_results_abs}.
We use CASCI energies \cite{helgaker2013molecular} for the full (32-spin-orbital) active space as a proxy for the exact FCI solutions, and label them as such in \cref{fig:electrolyte_results_abs}. CASCI results for smaller active spaces of size $4 \leq N_{\text{AS}} \leq 20$ are also computed as a function of the  number of qubits equivalent to the number of active space spin-orbitals under the JWT (x-axis in \cref{fig:electrolyte_results_abs}).
As in the previous example, we compute the HF, CCSD, CCSD(T), HF-AFQMC and CSA energies to benchmark against CS-AFQMC. For all CS-AFQMC calculations, we use 600 walkers, 1,000 time-blocks consisting of 25 time-steps with $\Delta \tau=0.005$, regardless of the size of the trial wavefunction used, while performing population control and numerical stabilization every 25 time-steps. To reduce the computational complexity, we remove the lowest amplitude terms in the trial wavefunction such that either the reduced wavefunction has an error  below $5\times10^{-5}$ compared to the true trial state, or the final wavefunction contains less than 100 computational basis state terms. 

In \cref{fig:electrolyte_results_abs} \textbf{(a)} and \textbf{(b)}, we show the absolute energies for the third and fourth configurations of \cref{fig:electrolyte detail}, respectively. Since we are mainly interested in the relative energy between these two configurations, we shrink the window for chemical accuracy by half. For both configurations we find HF, CCSD and HF-AFQMC to be outside chemical accuracy, while the CCSD(T) energies are well within. The convergence of CS-AFQMC is non-monotonic and approaches the exact value from opposite directions for the two configurations. At a CS size of about half of the active space, chemical accuracy is reached and results remain within this bound as the subspace size is further increased. For $N_{\text{cs}}\geq 19$ both CS-AFQMC energies are closer to the FCI result than CCSD(T).

Next, in \cref{fig:electrolyte_results_abs} \textbf{(c)} we focus on the reaction energy equal to the energy difference between configurations (3) and (4). Investigating the error of the energy difference with respect to FCI, we see that 
CS-AFQMC is within chemical accuracy for any number of CS qubits, including the HF-AFQMC case equivalent to $N_{\text{cs}}=0$. The same is also true for CCSD and especially CCSD(T), the accuracy of which is well below $10^{-4} $Ha, lower than any of the CS-AFQMC results with $N_{\text{cs}}\leq 20$. A partial explanation of such remarkable accuracy is that the correlations missing in CCSD(T) are of the same character for both configurations and for the most part cancel out in the difference. For CS-AFQMC, on the other hand, we found the respective contextual subspaces for the two configurations to be generally different, meaning that resulting CS-AFQMC energies approach the exact solution at different rates. We note that imposing the CS of one configuration onto the other leads to significantly inferior energy values and energy differences.

To further investigate the relationship between contextual and active spaces we focus on the CSA and CASCI energies as a function of subspace size (as shown in the insets of \cref{fig:electrolyte_results_abs} \textbf{(a)} and \textbf{(b)} as well as  \cref{fig:electrolyte_results_abs} \textbf{(c)}). One central question is whether there is any advantage from using contextual subspaces as an alternative to the established active space approach. In terms of energies, we find that both methods under-perform for these systems all the way up to 20 qubits, despite being guaranteed to converge to exact FCI results as their respective subspaces grow towards the full 32 qubits. This performance is to some degree expected, since the selection of 32 active space spin-orbitals from the original 236-orbital system in the cc-pVDZ basis has already identified the most relevant degrees of freedom responsible for non-trivial correlations. Any subspaces below this size are thus losing critical information about the system, which must be recovered in some way, for example by using AFQMC.

Comparing the two methods against each other, we see that CSA consistently yields lower energies than CASCI for all subspace sizes. Intuitively, this can be understood from regarding the active space as a special case of a contextual subspace. Specifically, the AS corresponds to a CS with all stabilizers being single-qubit Pauli $Z$ operators, for which the unitary rotation reduces to an identity. The difference between the CS identified here and a normal AS is due to the nature of the unitary rotation and projection operators, $U_{\mathcal{W}}$ and $\Pi_{\mathcal{W}}$. One consequence is that the particle number is conserved for basis states within active spaces, while the same is not true for contextual subspaces, where it is only recovered upon back-projection into the full space. Thus, more information about the original system is contained within a CS compared to an AS, leading to better quality approximations.

\section{Conclusion \& Outlook}\label{sec:conclusions}

Previous works using the contextual subspace methodology have uniquely focused on the estimation of the ground state energy. Even though this approximation must ultimately converge to the exact ground state with increasing qubit count in the CS, it was not a priori clear whether low energies translate into high wavefunction overlaps, which reduce the bias of AFQMC. We found this relationship to be the case, even if the convergence of CS-AFQMC is not guaranteed to be monotonic. 

We have successfully combined the QC-AFQMC algorithm with the contextual subspace approximation which prepares trial states with fewer quantum resources. Notably, only an extremely limited number of qubits was sufficient to push the resulting CS-AFQMC energies below chemical accuracy. Looking at the two applications investigated here, CS-AFQMC with around half of the original number of qubits in the contextual subspace has outperformed all other benchmarked methods for energy evaluations. The difference in performance was particularly stark for the nitrogen dimer, an example of a strongly correlated system where standard methods such CCSD, CCSD(T) and HF-AFQMC are known to perform poorly as the bond length is stretched. In contrast, the energies of the \ce{Li}-EC configurations (3) and (4) could be computed to within chemical accuracy using CCSD(T) and all three aforementioned methods produced very precise reaction energies. This is likely due to the fact that in these methods, similar correlations were left out for both configurations and these missing contributions effectively cancelled out in the calculation of the reaction energy. For such tasks, AFQMC has the disadvantage that energy estimates are not bound from below and the exact result can be approached from opposite directions for different configurations. This results in the compounding of biases instead of their cancellation.

One advantage of our heuristic approach to CSA is that is it straightforward and efficient to split the Hamiltonian into contextual and noncontextual parts, to solve for the noncontextual ground state and to identify and project into the contextual subspace. As a consequence, we have simulated an unprecedentedly large system and contextual subspaces thereof.
One potential bottleneck is the use of the auxiliary UCCSD operator for identifying stabilizers, as described in \cref{subsec:csa}.  Our method should thus be equally applicable to even larger systems as long as a suitable auxiliary operator can be obtained. To this end, it should also be noted that, unlike CSA, the CS-AFQMC algorithm was found to outperform CCSD (i.e. the non-unitary UCCSD equivalent with similar performance) in all examples that we have explored. We have also shown that using our heuristic, CS-AFQMC can be combined with the matchgate shadow protocol developed in Ref.~\cite{wan2022matchgate}, but it is not clear whether this result can be  generalised to contextual subspaces generated by off-diagonal terms from $\mathcal{T}$. 

Throughout this work we took an agnostic approach towards the contextual ground state solver method. Being able to reliably prepare approximations to such states on the quantum computer is an obligatory preliminary step, which would enable this hybrid method to outperform its classical alternatives. The tunability of the qubit number used in CS-AFQMC means that the break-even-point could be reached as soon as the quantum state preparation outperforms all classical methods for a given system of interest. It is not necessary to prepare the CS state with very high fidelity, as we found that AFQMC has a good degree of noise resilience built in. This can be seen from \cref{fig:toy_system_noise}) as is in accordance with the findings of Ref.~\cite{kiser2024classical} and Ref.~\cite{huang2024evaluating}. 

An open question is whether this algorithm can find universal application also in the quantum-inspired classical setting, for example by using classical methods as the contextual ground state solver. Here, we have used FCI calculations to obtain the CSA, and our results are encouraging in this regard, particularly since we found that CSA outperformed CASCI for any number of qubits in the CS and active spaces. Another open question is whether CSA and CS-AFQMC can be equally successfully applied to periodic lattice systems found in condensed matter physics.

\section*{Acknowledgements}

To obtain the results in this work, we utilized: \textsc{PySCF} \cite{PySCF:2018} to obtain the electron integrals for Hamiltonians and to perform benchmark calculations for the HF, CCSD, CCSD(T), CASCI and FCI methods; \textsc{Symmer} \cite{symmer} and \textsc{Qubovert} \cite{qubovert} for calculations related to the contextual subspace formalism; active spaces were identified using the \textsc{Active Space Finder} (\textsc{ASF}) \cite{hqs_asf} software package; for AFQMC calculations \textsc{ipie} \cite{malone2022ipie} was used. The authors would like to thank Peter V. Coveney, François Jamet, Martin Leib, Francesco Nappi, P.V. Sriluckshmy, Kianna Wan and Tim Weaving for insightful discussions.

\bibliography{main}

\begin{thebibliography}{80}%
\makeatletter
\providecommand \@ifxundefined [1]{%
 \@ifx{#1\undefined}
}%
\providecommand \@ifnum [1]{%
 \ifnum #1\expandafter \@firstoftwo
 \else \expandafter \@secondoftwo
 \fi
}%
\providecommand \@ifx [1]{%
 \ifx #1\expandafter \@firstoftwo
 \else \expandafter \@secondoftwo
 \fi
}%
\providecommand \natexlab [1]{#1}%
\providecommand \enquote  [1]{``#1''}%
\providecommand \bibnamefont  [1]{#1}%
\providecommand \bibfnamefont [1]{#1}%
\providecommand \citenamefont [1]{#1}%
\providecommand \href@noop [0]{\@secondoftwo}%
\providecommand \href [0]{\begingroup \@sanitize@url \@href}%
\providecommand \@href[1]{\@@startlink{#1}\@@href}%
\providecommand \@@href[1]{\endgroup#1\@@endlink}%
\providecommand \@sanitize@url [0]{\catcode `\\12\catcode `\$12\catcode
  `\&12\catcode `\#12\catcode `\^12\catcode `\_12\catcode `\%12\relax}%
\providecommand \@@startlink[1]{}%
\providecommand \@@endlink[0]{}%
\providecommand \url  [0]{\begingroup\@sanitize@url \@url }%
\providecommand \@url [1]{\endgroup\@href {#1}{\urlprefix }}%
\providecommand \urlprefix  [0]{URL }%
\providecommand \Eprint [0]{\href }%
\providecommand \doibase [0]{https://doi.org/}%
\providecommand \selectlanguage [0]{\@gobble}%
\providecommand \bibinfo  [0]{\@secondoftwo}%
\providecommand \bibfield  [0]{\@secondoftwo}%
\providecommand \translation [1]{[#1]}%
\providecommand \BibitemOpen [0]{}%
\providecommand \bibitemStop [0]{}%
\providecommand \bibitemNoStop [0]{.\EOS\space}%
\providecommand \EOS [0]{\spacefactor3000\relax}%
\providecommand \BibitemShut  [1]{\csname bibitem#1\endcsname}%
\let\auto@bib@innerbib\@empty
\bibitem [{\citenamefont {Cao}\ \emph {et~al.}(2019)\citenamefont {Cao},
  \citenamefont {Romero}, \citenamefont {Olson}, \citenamefont {Degroote},
  \citenamefont {Johnson}, \citenamefont {Kieferov{\'a}}, \citenamefont
  {Kivlichan}, \citenamefont {Menke}, \citenamefont {Peropadre}, \citenamefont
  {Sawaya} \emph {et~al.}}]{cao2019quantum}%
  \BibitemOpen
  \bibfield  {author} {\bibinfo {author} {\bibfnamefont {Y.}~\bibnamefont
  {Cao}}, \bibinfo {author} {\bibfnamefont {J.}~\bibnamefont {Romero}},
  \bibinfo {author} {\bibfnamefont {J.~P.}\ \bibnamefont {Olson}}, \bibinfo
  {author} {\bibfnamefont {M.}~\bibnamefont {Degroote}}, \bibinfo {author}
  {\bibfnamefont {P.~D.}\ \bibnamefont {Johnson}}, \bibinfo {author}
  {\bibfnamefont {M.}~\bibnamefont {Kieferov{\'a}}}, \bibinfo {author}
  {\bibfnamefont {I.~D.}\ \bibnamefont {Kivlichan}}, \bibinfo {author}
  {\bibfnamefont {T.}~\bibnamefont {Menke}}, \bibinfo {author} {\bibfnamefont
  {B.}~\bibnamefont {Peropadre}}, \bibinfo {author} {\bibfnamefont {N.~P.}\
  \bibnamefont {Sawaya}}, \emph {et~al.},\ }\bibfield  {title} {\bibinfo
  {title} {Quantum chemistry in the age of quantum computing},\ }\href
  {https://pubs.acs.org/doi/10.1021/acs.chemrev.8b00803} {\bibfield  {journal}
  {\bibinfo  {journal} {Chemical reviews}\ }\textbf {\bibinfo {volume} {119}},\
  \bibinfo {pages} {10856} (\bibinfo {year} {2019})}\BibitemShut {NoStop}%
\bibitem [{\citenamefont {McArdle}\ \emph {et~al.}(2020)\citenamefont
  {McArdle}, \citenamefont {Endo}, \citenamefont {Aspuru-Guzik}, \citenamefont
  {Benjamin},\ and\ \citenamefont {Yuan}}]{mcardle2020quantum}%
  \BibitemOpen
  \bibfield  {author} {\bibinfo {author} {\bibfnamefont {S.}~\bibnamefont
  {McArdle}}, \bibinfo {author} {\bibfnamefont {S.}~\bibnamefont {Endo}},
  \bibinfo {author} {\bibfnamefont {A.}~\bibnamefont {Aspuru-Guzik}}, \bibinfo
  {author} {\bibfnamefont {S.~C.}\ \bibnamefont {Benjamin}},\ and\ \bibinfo
  {author} {\bibfnamefont {X.}~\bibnamefont {Yuan}},\ }\bibfield  {title}
  {\bibinfo {title} {Quantum computational chemistry},\ }\href
  {https://journals.aps.org/rmp/abstract/10.1103/RevModPhys.92.015003}
  {\bibfield  {journal} {\bibinfo  {journal} {Reviews of Modern Physics}\
  }\textbf {\bibinfo {volume} {92}},\ \bibinfo {pages} {015003} (\bibinfo
  {year} {2020})}\BibitemShut {NoStop}%
\bibitem [{\citenamefont {Dalzell}\ \emph {et~al.}(2023)\citenamefont
  {Dalzell}, \citenamefont {McArdle}, \citenamefont {Berta}, \citenamefont
  {Bienias}, \citenamefont {Chen}, \citenamefont {Gily{\'e}n}, \citenamefont
  {Hann}, \citenamefont {Kastoryano}, \citenamefont {Khabiboulline},
  \citenamefont {Kubica} \emph {et~al.}}]{dalzell2023quantum}%
  \BibitemOpen
  \bibfield  {author} {\bibinfo {author} {\bibfnamefont {A.~M.}\ \bibnamefont
  {Dalzell}}, \bibinfo {author} {\bibfnamefont {S.}~\bibnamefont {McArdle}},
  \bibinfo {author} {\bibfnamefont {M.}~\bibnamefont {Berta}}, \bibinfo
  {author} {\bibfnamefont {P.}~\bibnamefont {Bienias}}, \bibinfo {author}
  {\bibfnamefont {C.-F.}\ \bibnamefont {Chen}}, \bibinfo {author}
  {\bibfnamefont {A.}~\bibnamefont {Gily{\'e}n}}, \bibinfo {author}
  {\bibfnamefont {C.~T.}\ \bibnamefont {Hann}}, \bibinfo {author}
  {\bibfnamefont {M.~J.}\ \bibnamefont {Kastoryano}}, \bibinfo {author}
  {\bibfnamefont {E.~T.}\ \bibnamefont {Khabiboulline}}, \bibinfo {author}
  {\bibfnamefont {A.}~\bibnamefont {Kubica}}, \emph {et~al.},\ }\bibfield
  {title} {\bibinfo {title} {Quantum algorithms: A survey of applications and
  end-to-end complexities},\ }\href {https://arxiv.org/abs/2310.03011}
  {\bibfield  {journal} {\bibinfo  {journal} {arXiv preprint arXiv:2310.03011}\
  } (\bibinfo {year} {2023})}\BibitemShut {NoStop}%
\bibitem [{\citenamefont {Di~Meglio}\ \emph {et~al.}(2024)\citenamefont
  {Di~Meglio}, \citenamefont {Jansen}, \citenamefont {Tavernelli},
  \citenamefont {Alexandrou}, \citenamefont {Arunachalam}, \citenamefont
  {Bauer}, \citenamefont {Borras}, \citenamefont {Carrazza}, \citenamefont
  {Crippa}, \citenamefont {Croft}, \citenamefont {de~Putter}, \citenamefont
  {Delgado}, \citenamefont {Dunjko}, \citenamefont {Egger}, \citenamefont
  {Fern\'andez-Combarro}, \citenamefont {Fuchs}, \citenamefont {Funcke},
  \citenamefont {Gonz\'alez-Cuadra}, \citenamefont {Grossi}, \citenamefont
  {Halimeh}, \citenamefont {Holmes}, \citenamefont {K\"uhn}, \citenamefont
  {Lacroix}, \citenamefont {Lewis}, \citenamefont {Lucchesi}, \citenamefont
  {Martinez}, \citenamefont {Meloni}, \citenamefont {Mezzacapo}, \citenamefont
  {Montangero}, \citenamefont {Nagano}, \citenamefont {Pascuzzi}, \citenamefont
  {Radescu}, \citenamefont {Ortega}, \citenamefont {Roggero}, \citenamefont
  {Schuhmacher}, \citenamefont {Seixas}, \citenamefont {Silvi}, \citenamefont
  {Spentzouris}, \citenamefont {Tacchino}, \citenamefont {Temme}, \citenamefont
  {Terashi}, \citenamefont {Tura}, \citenamefont {T\"uys\"uz}, \citenamefont
  {Vallecorsa}, \citenamefont {Wiese}, \citenamefont {Yoo},\ and\ \citenamefont
  {Zhang}}]{di2023quantum}%
  \BibitemOpen
  \bibfield  {author} {\bibinfo {author} {\bibfnamefont {A.}~\bibnamefont
  {Di~Meglio}}, \bibinfo {author} {\bibfnamefont {K.}~\bibnamefont {Jansen}},
  \bibinfo {author} {\bibfnamefont {I.}~\bibnamefont {Tavernelli}}, \bibinfo
  {author} {\bibfnamefont {C.}~\bibnamefont {Alexandrou}}, \bibinfo {author}
  {\bibfnamefont {S.}~\bibnamefont {Arunachalam}}, \bibinfo {author}
  {\bibfnamefont {C.~W.}\ \bibnamefont {Bauer}}, \bibinfo {author}
  {\bibfnamefont {K.}~\bibnamefont {Borras}}, \bibinfo {author} {\bibfnamefont
  {S.}~\bibnamefont {Carrazza}}, \bibinfo {author} {\bibfnamefont
  {A.}~\bibnamefont {Crippa}}, \bibinfo {author} {\bibfnamefont
  {V.}~\bibnamefont {Croft}}, \bibinfo {author} {\bibfnamefont
  {R.}~\bibnamefont {de~Putter}}, \bibinfo {author} {\bibfnamefont
  {A.}~\bibnamefont {Delgado}}, \bibinfo {author} {\bibfnamefont
  {V.}~\bibnamefont {Dunjko}}, \bibinfo {author} {\bibfnamefont {D.~J.}\
  \bibnamefont {Egger}}, \bibinfo {author} {\bibfnamefont {E.}~\bibnamefont
  {Fern\'andez-Combarro}}, \bibinfo {author} {\bibfnamefont {E.}~\bibnamefont
  {Fuchs}}, \bibinfo {author} {\bibfnamefont {L.}~\bibnamefont {Funcke}},
  \bibinfo {author} {\bibfnamefont {D.}~\bibnamefont {Gonz\'alez-Cuadra}},
  \bibinfo {author} {\bibfnamefont {M.}~\bibnamefont {Grossi}}, \bibinfo
  {author} {\bibfnamefont {J.~C.}\ \bibnamefont {Halimeh}}, \bibinfo {author}
  {\bibfnamefont {Z.}~\bibnamefont {Holmes}}, \bibinfo {author} {\bibfnamefont
  {S.}~\bibnamefont {K\"uhn}}, \bibinfo {author} {\bibfnamefont
  {D.}~\bibnamefont {Lacroix}}, \bibinfo {author} {\bibfnamefont
  {R.}~\bibnamefont {Lewis}}, \bibinfo {author} {\bibfnamefont
  {D.}~\bibnamefont {Lucchesi}}, \bibinfo {author} {\bibfnamefont {M.~L.}\
  \bibnamefont {Martinez}}, \bibinfo {author} {\bibfnamefont {F.}~\bibnamefont
  {Meloni}}, \bibinfo {author} {\bibfnamefont {A.}~\bibnamefont {Mezzacapo}},
  \bibinfo {author} {\bibfnamefont {S.}~\bibnamefont {Montangero}}, \bibinfo
  {author} {\bibfnamefont {L.}~\bibnamefont {Nagano}}, \bibinfo {author}
  {\bibfnamefont {V.~R.}\ \bibnamefont {Pascuzzi}}, \bibinfo {author}
  {\bibfnamefont {V.}~\bibnamefont {Radescu}}, \bibinfo {author} {\bibfnamefont
  {E.~R.}\ \bibnamefont {Ortega}}, \bibinfo {author} {\bibfnamefont
  {A.}~\bibnamefont {Roggero}}, \bibinfo {author} {\bibfnamefont
  {J.}~\bibnamefont {Schuhmacher}}, \bibinfo {author} {\bibfnamefont
  {J.}~\bibnamefont {Seixas}}, \bibinfo {author} {\bibfnamefont
  {P.}~\bibnamefont {Silvi}}, \bibinfo {author} {\bibfnamefont
  {P.}~\bibnamefont {Spentzouris}}, \bibinfo {author} {\bibfnamefont
  {F.}~\bibnamefont {Tacchino}}, \bibinfo {author} {\bibfnamefont
  {K.}~\bibnamefont {Temme}}, \bibinfo {author} {\bibfnamefont
  {K.}~\bibnamefont {Terashi}}, \bibinfo {author} {\bibfnamefont
  {J.}~\bibnamefont {Tura}}, \bibinfo {author} {\bibfnamefont {C.}~\bibnamefont
  {T\"uys\"uz}}, \bibinfo {author} {\bibfnamefont {S.}~\bibnamefont
  {Vallecorsa}}, \bibinfo {author} {\bibfnamefont {U.-J.}\ \bibnamefont
  {Wiese}}, \bibinfo {author} {\bibfnamefont {S.}~\bibnamefont {Yoo}},\ and\
  \bibinfo {author} {\bibfnamefont {J.}~\bibnamefont {Zhang}},\ }\bibfield
  {title} {\bibinfo {title} {Quantum computing for high-energy physics: State
  of the art and challenges},\ }\href
  {https://doi.org/10.1103/PRXQuantum.5.037001} {\bibfield  {journal} {\bibinfo
   {journal} {PRX Quantum}\ }\textbf {\bibinfo {volume} {5}},\ \bibinfo {pages}
  {037001} (\bibinfo {year} {2024})}\BibitemShut {NoStop}%
\bibitem [{\citenamefont {Ho}\ \emph {et~al.}(2018)\citenamefont {Ho},
  \citenamefont {McClean},\ and\ \citenamefont {Ong}}]{ho2018promise}%
  \BibitemOpen
  \bibfield  {author} {\bibinfo {author} {\bibfnamefont {A.}~\bibnamefont
  {Ho}}, \bibinfo {author} {\bibfnamefont {J.}~\bibnamefont {McClean}},\ and\
  \bibinfo {author} {\bibfnamefont {S.~P.}\ \bibnamefont {Ong}},\ }\bibfield
  {title} {\bibinfo {title} {The promise and challenges of quantum computing
  for energy storage},\ }\href
  {https://www.sciencedirect.com/science/article/pii/S254243511830182X}
  {\bibfield  {journal} {\bibinfo  {journal} {Joule}\ }\textbf {\bibinfo
  {volume} {2}},\ \bibinfo {pages} {810} (\bibinfo {year} {2018})}\BibitemShut
  {NoStop}%
\bibitem [{\citenamefont {Gao}\ \emph {et~al.}(2021)\citenamefont {Gao},
  \citenamefont {Jones}, \citenamefont {Motta}, \citenamefont {Sugawara},
  \citenamefont {Watanabe}, \citenamefont {Kobayashi}, \citenamefont
  {Watanabe}, \citenamefont {Ohnishi}, \citenamefont {Nakamura},\ and\
  \citenamefont {Yamamoto}}]{gao2021applications}%
  \BibitemOpen
  \bibfield  {author} {\bibinfo {author} {\bibfnamefont {Q.}~\bibnamefont
  {Gao}}, \bibinfo {author} {\bibfnamefont {G.~O.}\ \bibnamefont {Jones}},
  \bibinfo {author} {\bibfnamefont {M.}~\bibnamefont {Motta}}, \bibinfo
  {author} {\bibfnamefont {M.}~\bibnamefont {Sugawara}}, \bibinfo {author}
  {\bibfnamefont {H.~C.}\ \bibnamefont {Watanabe}}, \bibinfo {author}
  {\bibfnamefont {T.}~\bibnamefont {Kobayashi}}, \bibinfo {author}
  {\bibfnamefont {E.}~\bibnamefont {Watanabe}}, \bibinfo {author}
  {\bibfnamefont {Y.-y.}\ \bibnamefont {Ohnishi}}, \bibinfo {author}
  {\bibfnamefont {H.}~\bibnamefont {Nakamura}},\ and\ \bibinfo {author}
  {\bibfnamefont {N.}~\bibnamefont {Yamamoto}},\ }\bibfield  {title} {\bibinfo
  {title} {Applications of quantum computing for investigations of electronic
  transitions in phenylsulfonyl-carbazole tadf emitters},\ }\href
  {https://www.nature.com/articles/s41524-021-00540-6} {\bibfield  {journal}
  {\bibinfo  {journal} {npj Computational Materials}\ }\textbf {\bibinfo
  {volume} {7}},\ \bibinfo {pages} {70} (\bibinfo {year} {2021})}\BibitemShut
  {NoStop}%
\bibitem [{\citenamefont {Rice}\ \emph {et~al.}(2021)\citenamefont {Rice},
  \citenamefont {Gujarati}, \citenamefont {Motta}, \citenamefont {Takeshita},
  \citenamefont {Lee}, \citenamefont {Latone},\ and\ \citenamefont
  {Garcia}}]{rice2021quantum}%
  \BibitemOpen
  \bibfield  {author} {\bibinfo {author} {\bibfnamefont {J.~E.}\ \bibnamefont
  {Rice}}, \bibinfo {author} {\bibfnamefont {T.~P.}\ \bibnamefont {Gujarati}},
  \bibinfo {author} {\bibfnamefont {M.}~\bibnamefont {Motta}}, \bibinfo
  {author} {\bibfnamefont {T.~Y.}\ \bibnamefont {Takeshita}}, \bibinfo {author}
  {\bibfnamefont {E.}~\bibnamefont {Lee}}, \bibinfo {author} {\bibfnamefont
  {J.~A.}\ \bibnamefont {Latone}},\ and\ \bibinfo {author} {\bibfnamefont
  {J.~M.}\ \bibnamefont {Garcia}},\ }\bibfield  {title} {\bibinfo {title}
  {Quantum computation of dominant products in lithium--sulfur batteries},\
  }\href
  {https://pubs.aip.org/aip/jcp/article-abstract/154/13/134115/1065544/Quantum-computation-of-dominant-products-in?redirectedFrom=fulltext}
  {\bibfield  {journal} {\bibinfo  {journal} {The Journal of Chemical Physics}\
  }\textbf {\bibinfo {volume} {154}} (\bibinfo {year} {2021})}\BibitemShut
  {NoStop}%
\bibitem [{\citenamefont {Delgado}\ \emph {et~al.}(2022)\citenamefont
  {Delgado}, \citenamefont {Casares}, \citenamefont {dos Reis}, \citenamefont
  {Zini}, \citenamefont {Campos}, \citenamefont {Cruz-Hern\'andez},
  \citenamefont {Voigt}, \citenamefont {Lowe}, \citenamefont {Jahangiri},
  \citenamefont {Martin-Delgado}, \citenamefont {Mueller},\ and\ \citenamefont
  {Arrazola}}]{delgado2022simulating}%
  \BibitemOpen
  \bibfield  {author} {\bibinfo {author} {\bibfnamefont {A.}~\bibnamefont
  {Delgado}}, \bibinfo {author} {\bibfnamefont {P.~A.~M.}\ \bibnamefont
  {Casares}}, \bibinfo {author} {\bibfnamefont {R.}~\bibnamefont {dos Reis}},
  \bibinfo {author} {\bibfnamefont {M.~S.}\ \bibnamefont {Zini}}, \bibinfo
  {author} {\bibfnamefont {R.}~\bibnamefont {Campos}}, \bibinfo {author}
  {\bibfnamefont {N.}~\bibnamefont {Cruz-Hern\'andez}}, \bibinfo {author}
  {\bibfnamefont {A.-C.}\ \bibnamefont {Voigt}}, \bibinfo {author}
  {\bibfnamefont {A.}~\bibnamefont {Lowe}}, \bibinfo {author} {\bibfnamefont
  {S.}~\bibnamefont {Jahangiri}}, \bibinfo {author} {\bibfnamefont {M.~A.}\
  \bibnamefont {Martin-Delgado}}, \bibinfo {author} {\bibfnamefont {J.~E.}\
  \bibnamefont {Mueller}},\ and\ \bibinfo {author} {\bibfnamefont {J.~M.}\
  \bibnamefont {Arrazola}},\ }\bibfield  {title} {\bibinfo {title} {Simulating
  key properties of lithium-ion batteries with a fault-tolerant quantum
  computer},\ }\href {https://doi.org/10.1103/PhysRevA.106.032428} {\bibfield
  {journal} {\bibinfo  {journal} {Phys. Rev. A}\ }\textbf {\bibinfo {volume}
  {106}},\ \bibinfo {pages} {032428} (\bibinfo {year} {2022})}\BibitemShut
  {NoStop}%
\bibitem [{\citenamefont {Kim}\ \emph {et~al.}(2022)\citenamefont {Kim},
  \citenamefont {Liu}, \citenamefont {Pallister}, \citenamefont {Pol},
  \citenamefont {Roberts},\ and\ \citenamefont {Lee}}]{kim2022fault}%
  \BibitemOpen
  \bibfield  {author} {\bibinfo {author} {\bibfnamefont {I.~H.}\ \bibnamefont
  {Kim}}, \bibinfo {author} {\bibfnamefont {Y.-H.}\ \bibnamefont {Liu}},
  \bibinfo {author} {\bibfnamefont {S.}~\bibnamefont {Pallister}}, \bibinfo
  {author} {\bibfnamefont {W.}~\bibnamefont {Pol}}, \bibinfo {author}
  {\bibfnamefont {S.}~\bibnamefont {Roberts}},\ and\ \bibinfo {author}
  {\bibfnamefont {E.}~\bibnamefont {Lee}},\ }\bibfield  {title} {\bibinfo
  {title} {Fault-tolerant resource estimate for quantum chemical simulations:
  Case study on li-ion battery electrolyte molecules},\ }\href
  {https://doi.org/10.1103/PhysRevResearch.4.023019} {\bibfield  {journal}
  {\bibinfo  {journal} {Phys. Rev. Res.}\ }\textbf {\bibinfo {volume} {4}},\
  \bibinfo {pages} {023019} (\bibinfo {year} {2022})}\BibitemShut {NoStop}%
\bibitem [{\citenamefont {Kim}\ \emph {et~al.}(2023)\citenamefont {Kim},
  \citenamefont {Wu}, \citenamefont {Jian}, \citenamefont {Xiong},\ and\
  \citenamefont {Luo}}]{kim2023design}%
  \BibitemOpen
  \bibfield  {author} {\bibinfo {author} {\bibfnamefont {S.}~\bibnamefont
  {Kim}}, \bibinfo {author} {\bibfnamefont {S.}~\bibnamefont {Wu}}, \bibinfo
  {author} {\bibfnamefont {R.}~\bibnamefont {Jian}}, \bibinfo {author}
  {\bibfnamefont {G.}~\bibnamefont {Xiong}},\ and\ \bibinfo {author}
  {\bibfnamefont {T.}~\bibnamefont {Luo}},\ }\bibfield  {title} {\bibinfo
  {title} {Design of a high-performance titanium nitride metastructure-based
  solar absorber using quantum computing-assisted optimization},\ }\href
  {https://pubs.acs.org/doi/10.1021/acsami.3c08214} {\bibfield  {journal}
  {\bibinfo  {journal} {ACS Applied Materials \& Interfaces}\ }\textbf
  {\bibinfo {volume} {15}},\ \bibinfo {pages} {40606} (\bibinfo {year}
  {2023})}\BibitemShut {NoStop}%
\bibitem [{\citenamefont {Fomichev}\ \emph {et~al.}(2024)\citenamefont
  {Fomichev}, \citenamefont {Hejazi}, \citenamefont {Loaiza}, \citenamefont
  {Zini}, \citenamefont {Delgado}, \citenamefont {Voigt}, \citenamefont
  {Mueller},\ and\ \citenamefont {Arrazola}}]{fomichev2024simulating}%
  \BibitemOpen
  \bibfield  {author} {\bibinfo {author} {\bibfnamefont {S.}~\bibnamefont
  {Fomichev}}, \bibinfo {author} {\bibfnamefont {K.}~\bibnamefont {Hejazi}},
  \bibinfo {author} {\bibfnamefont {I.}~\bibnamefont {Loaiza}}, \bibinfo
  {author} {\bibfnamefont {M.~S.}\ \bibnamefont {Zini}}, \bibinfo {author}
  {\bibfnamefont {A.}~\bibnamefont {Delgado}}, \bibinfo {author} {\bibfnamefont
  {A.-C.}\ \bibnamefont {Voigt}}, \bibinfo {author} {\bibfnamefont {J.~E.}\
  \bibnamefont {Mueller}},\ and\ \bibinfo {author} {\bibfnamefont {J.~M.}\
  \bibnamefont {Arrazola}},\ }\bibfield  {title} {\bibinfo {title} {Simulating
  x-ray absorption spectroscopy of battery materials on a quantum computer},\
  }\href {https://arxiv.org/abs/2405.11015} {\bibfield  {journal} {\bibinfo
  {journal} {arXiv preprint arXiv:2405.11015}\ } (\bibinfo {year}
  {2024})}\BibitemShut {NoStop}%
\bibitem [{\citenamefont {Paudel}\ \emph {et~al.}(2022)\citenamefont {Paudel},
  \citenamefont {Syamlal}, \citenamefont {Crawford}, \citenamefont {Lee},
  \citenamefont {Shugayev}, \citenamefont {Lu}, \citenamefont {Ohodnicki},
  \citenamefont {Mollot},\ and\ \citenamefont {Duan}}]{paudel2022quantum}%
  \BibitemOpen
  \bibfield  {author} {\bibinfo {author} {\bibfnamefont {H.~P.}\ \bibnamefont
  {Paudel}}, \bibinfo {author} {\bibfnamefont {M.}~\bibnamefont {Syamlal}},
  \bibinfo {author} {\bibfnamefont {S.~E.}\ \bibnamefont {Crawford}}, \bibinfo
  {author} {\bibfnamefont {Y.-L.}\ \bibnamefont {Lee}}, \bibinfo {author}
  {\bibfnamefont {R.~A.}\ \bibnamefont {Shugayev}}, \bibinfo {author}
  {\bibfnamefont {P.}~\bibnamefont {Lu}}, \bibinfo {author} {\bibfnamefont
  {P.~R.}\ \bibnamefont {Ohodnicki}}, \bibinfo {author} {\bibfnamefont
  {D.}~\bibnamefont {Mollot}},\ and\ \bibinfo {author} {\bibfnamefont
  {Y.}~\bibnamefont {Duan}},\ }\bibfield  {title} {\bibinfo {title} {Quantum
  computing and simulations for energy applications: Review and perspective},\
  }\href {https://pubs.acs.org/doi/10.1021/acsengineeringau.1c00033} {\bibfield
   {journal} {\bibinfo  {journal} {ACS Engineering Au}\ }\textbf {\bibinfo
  {volume} {2}},\ \bibinfo {pages} {151} (\bibinfo {year} {2022})}\BibitemShut
  {NoStop}%
\bibitem [{\citenamefont {Preskill}(2018)}]{preskill2018nisq}%
  \BibitemOpen
  \bibfield  {author} {\bibinfo {author} {\bibfnamefont {J.}~\bibnamefont
  {Preskill}},\ }\bibfield  {title} {\bibinfo {title} {Quantum computing in the
  nisq era and beyond},\ }\href {https://doi.org/10.22331/q-2018-08-06-79}
  {\bibfield  {journal} {\bibinfo  {journal} {Quantum}\ }\textbf {\bibinfo
  {volume} {2}},\ \bibinfo {pages} {79} (\bibinfo {year} {2018})}\BibitemShut
  {NoStop}%
\bibitem [{\citenamefont {Cerezo}\ \emph {et~al.}(2021)\citenamefont {Cerezo},
  \citenamefont {Arrasmith}, \citenamefont {Babbush}, \citenamefont {Benjamin},
  \citenamefont {Endo}, \citenamefont {Fujii}, \citenamefont {McClean},
  \citenamefont {Mitarai}, \citenamefont {Yuan}, \citenamefont {Cincio} \emph
  {et~al.}}]{cerezo2021variational}%
  \BibitemOpen
  \bibfield  {author} {\bibinfo {author} {\bibfnamefont {M.}~\bibnamefont
  {Cerezo}}, \bibinfo {author} {\bibfnamefont {A.}~\bibnamefont {Arrasmith}},
  \bibinfo {author} {\bibfnamefont {R.}~\bibnamefont {Babbush}}, \bibinfo
  {author} {\bibfnamefont {S.~C.}\ \bibnamefont {Benjamin}}, \bibinfo {author}
  {\bibfnamefont {S.}~\bibnamefont {Endo}}, \bibinfo {author} {\bibfnamefont
  {K.}~\bibnamefont {Fujii}}, \bibinfo {author} {\bibfnamefont {J.~R.}\
  \bibnamefont {McClean}}, \bibinfo {author} {\bibfnamefont {K.}~\bibnamefont
  {Mitarai}}, \bibinfo {author} {\bibfnamefont {X.}~\bibnamefont {Yuan}},
  \bibinfo {author} {\bibfnamefont {L.}~\bibnamefont {Cincio}}, \emph
  {et~al.},\ }\bibfield  {title} {\bibinfo {title} {Variational quantum
  algorithms},\ }\href {https://www.nature.com/articles/s42254-021-00348-9}
  {\bibfield  {journal} {\bibinfo  {journal} {Nature Reviews Physics}\ }\textbf
  {\bibinfo {volume} {3}},\ \bibinfo {pages} {625} (\bibinfo {year}
  {2021})}\BibitemShut {NoStop}%
\bibitem [{\citenamefont {Bharti}\ \emph {et~al.}(2022)\citenamefont {Bharti},
  \citenamefont {Cervera-Lierta}, \citenamefont {Kyaw}, \citenamefont {Haug},
  \citenamefont {Alperin-Lea}, \citenamefont {Anand}, \citenamefont {Degroote},
  \citenamefont {Heimonen}, \citenamefont {Kottmann}, \citenamefont {Menke}
  \emph {et~al.}}]{bharti2022noisy}%
  \BibitemOpen
  \bibfield  {author} {\bibinfo {author} {\bibfnamefont {K.}~\bibnamefont
  {Bharti}}, \bibinfo {author} {\bibfnamefont {A.}~\bibnamefont
  {Cervera-Lierta}}, \bibinfo {author} {\bibfnamefont {T.~H.}\ \bibnamefont
  {Kyaw}}, \bibinfo {author} {\bibfnamefont {T.}~\bibnamefont {Haug}}, \bibinfo
  {author} {\bibfnamefont {S.}~\bibnamefont {Alperin-Lea}}, \bibinfo {author}
  {\bibfnamefont {A.}~\bibnamefont {Anand}}, \bibinfo {author} {\bibfnamefont
  {M.}~\bibnamefont {Degroote}}, \bibinfo {author} {\bibfnamefont
  {H.}~\bibnamefont {Heimonen}}, \bibinfo {author} {\bibfnamefont {J.~S.}\
  \bibnamefont {Kottmann}}, \bibinfo {author} {\bibfnamefont {T.}~\bibnamefont
  {Menke}}, \emph {et~al.},\ }\bibfield  {title} {\bibinfo {title} {Noisy
  intermediate-scale quantum algorithms},\ }\href
  {https://journals.aps.org/rmp/abstract/10.1103/RevModPhys.94.015004}
  {\bibfield  {journal} {\bibinfo  {journal} {Reviews of Modern Physics}\
  }\textbf {\bibinfo {volume} {94}},\ \bibinfo {pages} {015004} (\bibinfo
  {year} {2022})}\BibitemShut {NoStop}%
\bibitem [{\citenamefont {Tilly}\ \emph {et~al.}(2022)\citenamefont {Tilly},
  \citenamefont {Chen}, \citenamefont {Cao}, \citenamefont {Picozzi},
  \citenamefont {Setia}, \citenamefont {Li}, \citenamefont {Grant},
  \citenamefont {Wossnig}, \citenamefont {Rungger}, \citenamefont {Booth} \emph
  {et~al.}}]{tilly2022variational}%
  \BibitemOpen
  \bibfield  {author} {\bibinfo {author} {\bibfnamefont {J.}~\bibnamefont
  {Tilly}}, \bibinfo {author} {\bibfnamefont {H.}~\bibnamefont {Chen}},
  \bibinfo {author} {\bibfnamefont {S.}~\bibnamefont {Cao}}, \bibinfo {author}
  {\bibfnamefont {D.}~\bibnamefont {Picozzi}}, \bibinfo {author} {\bibfnamefont
  {K.}~\bibnamefont {Setia}}, \bibinfo {author} {\bibfnamefont
  {Y.}~\bibnamefont {Li}}, \bibinfo {author} {\bibfnamefont {E.}~\bibnamefont
  {Grant}}, \bibinfo {author} {\bibfnamefont {L.}~\bibnamefont {Wossnig}},
  \bibinfo {author} {\bibfnamefont {I.}~\bibnamefont {Rungger}}, \bibinfo
  {author} {\bibfnamefont {G.~H.}\ \bibnamefont {Booth}}, \emph {et~al.},\
  }\bibfield  {title} {\bibinfo {title} {The variational quantum eigensolver: a
  review of methods and best practices},\ }\href
  {https://www.sciencedirect.com/science/article/pii/S0370157322003118}
  {\bibfield  {journal} {\bibinfo  {journal} {Physics Reports}\ }\textbf
  {\bibinfo {volume} {986}},\ \bibinfo {pages} {1} (\bibinfo {year}
  {2022})}\BibitemShut {NoStop}%
\bibitem [{\citenamefont {Jamet}\ \emph {et~al.}(2021)\citenamefont {Jamet},
  \citenamefont {Agarwal}, \citenamefont {Lupo}, \citenamefont {Browne},
  \citenamefont {Weber},\ and\ \citenamefont {Rungger}}]{jamet2021krylov}%
  \BibitemOpen
  \bibfield  {author} {\bibinfo {author} {\bibfnamefont {F.}~\bibnamefont
  {Jamet}}, \bibinfo {author} {\bibfnamefont {A.}~\bibnamefont {Agarwal}},
  \bibinfo {author} {\bibfnamefont {C.}~\bibnamefont {Lupo}}, \bibinfo {author}
  {\bibfnamefont {D.~E.}\ \bibnamefont {Browne}}, \bibinfo {author}
  {\bibfnamefont {C.}~\bibnamefont {Weber}},\ and\ \bibinfo {author}
  {\bibfnamefont {I.}~\bibnamefont {Rungger}},\ }\bibfield  {title} {\bibinfo
  {title} {Krylov variational quantum algorithm for first principles materials
  simulations},\ }\href {https://arxiv.org/abs/2105.13298} {\bibfield
  {journal} {\bibinfo  {journal} {arXiv preprint arXiv:2105.13298}\ } (\bibinfo
  {year} {2021})}\BibitemShut {NoStop}%
\bibitem [{\citenamefont {Li}\ \emph {et~al.}(2022)\citenamefont {Li},
  \citenamefont {Huang}, \citenamefont {Cao}, \citenamefont {Huang},
  \citenamefont {Shuai}, \citenamefont {Sun}, \citenamefont {Sun},
  \citenamefont {Yuan},\ and\ \citenamefont {Lv}}]{li2022toward}%
  \BibitemOpen
  \bibfield  {author} {\bibinfo {author} {\bibfnamefont {W.}~\bibnamefont
  {Li}}, \bibinfo {author} {\bibfnamefont {Z.}~\bibnamefont {Huang}}, \bibinfo
  {author} {\bibfnamefont {C.}~\bibnamefont {Cao}}, \bibinfo {author}
  {\bibfnamefont {Y.}~\bibnamefont {Huang}}, \bibinfo {author} {\bibfnamefont
  {Z.}~\bibnamefont {Shuai}}, \bibinfo {author} {\bibfnamefont
  {X.}~\bibnamefont {Sun}}, \bibinfo {author} {\bibfnamefont {J.}~\bibnamefont
  {Sun}}, \bibinfo {author} {\bibfnamefont {X.}~\bibnamefont {Yuan}},\ and\
  \bibinfo {author} {\bibfnamefont {D.}~\bibnamefont {Lv}},\ }\bibfield
  {title} {\bibinfo {title} {Toward practical quantum embedding simulation of
  realistic chemical systems on near-term quantum computers},\ }\href
  {https://pubs.rsc.org/en/content/articlelanding/2022/sc/d2sc01492k}
  {\bibfield  {journal} {\bibinfo  {journal} {Chemical science}\ }\textbf
  {\bibinfo {volume} {13}},\ \bibinfo {pages} {8953} (\bibinfo {year}
  {2022})}\BibitemShut {NoStop}%
\bibitem [{\citenamefont {Izs{\'a}k}\ \emph {et~al.}(2023)\citenamefont
  {Izs{\'a}k}, \citenamefont {Riplinger}, \citenamefont {Blunt}, \citenamefont
  {de~Souza}, \citenamefont {Holzmann}, \citenamefont {Crawford}, \citenamefont
  {Camps}, \citenamefont {Neese},\ and\ \citenamefont
  {Schopf}}]{izsak2023quantum}%
  \BibitemOpen
  \bibfield  {author} {\bibinfo {author} {\bibfnamefont {R.}~\bibnamefont
  {Izs{\'a}k}}, \bibinfo {author} {\bibfnamefont {C.}~\bibnamefont
  {Riplinger}}, \bibinfo {author} {\bibfnamefont {N.~S.}\ \bibnamefont
  {Blunt}}, \bibinfo {author} {\bibfnamefont {B.}~\bibnamefont {de~Souza}},
  \bibinfo {author} {\bibfnamefont {N.}~\bibnamefont {Holzmann}}, \bibinfo
  {author} {\bibfnamefont {O.}~\bibnamefont {Crawford}}, \bibinfo {author}
  {\bibfnamefont {J.}~\bibnamefont {Camps}}, \bibinfo {author} {\bibfnamefont
  {F.}~\bibnamefont {Neese}},\ and\ \bibinfo {author} {\bibfnamefont
  {P.}~\bibnamefont {Schopf}},\ }\bibfield  {title} {\bibinfo {title} {Quantum
  computing in pharma: A multilayer embedding approach for near future
  applications},\ }\href {https://arxiv.org/abs/2202.04460} {\bibfield
  {journal} {\bibinfo  {journal} {Journal of Computational Chemistry}\ }\textbf
  {\bibinfo {volume} {44}},\ \bibinfo {pages} {406} (\bibinfo {year}
  {2023})}\BibitemShut {NoStop}%
\bibitem [{\citenamefont {Iijima}\ \emph {et~al.}(2023)\citenamefont {Iijima},
  \citenamefont {Imamura}, \citenamefont {Morita}, \citenamefont {Takemori},
  \citenamefont {Kasagi}, \citenamefont {Umeda},\ and\ \citenamefont
  {Yoshida}}]{iijima2023towards}%
  \BibitemOpen
  \bibfield  {author} {\bibinfo {author} {\bibfnamefont {N.}~\bibnamefont
  {Iijima}}, \bibinfo {author} {\bibfnamefont {S.}~\bibnamefont {Imamura}},
  \bibinfo {author} {\bibfnamefont {M.}~\bibnamefont {Morita}}, \bibinfo
  {author} {\bibfnamefont {S.}~\bibnamefont {Takemori}}, \bibinfo {author}
  {\bibfnamefont {A.}~\bibnamefont {Kasagi}}, \bibinfo {author} {\bibfnamefont
  {Y.}~\bibnamefont {Umeda}},\ and\ \bibinfo {author} {\bibfnamefont
  {E.}~\bibnamefont {Yoshida}},\ }\bibfield  {title} {\bibinfo {title} {Towards
  accurate quantum chemical calculations on noisy quantum computers},\ }\href
  {https://arxiv.org/abs/2311.09634} {\bibfield  {journal} {\bibinfo  {journal}
  {arXiv preprint arXiv:2311.09634}\ } (\bibinfo {year} {2023})}\BibitemShut
  {NoStop}%
\bibitem [{\citenamefont {Rossmannek}\ \emph {et~al.}(2023)\citenamefont
  {Rossmannek}, \citenamefont {Pavosevic}, \citenamefont {Rubio},\ and\
  \citenamefont {Tavernelli}}]{rossmannek2023quantum}%
  \BibitemOpen
  \bibfield  {author} {\bibinfo {author} {\bibfnamefont {M.}~\bibnamefont
  {Rossmannek}}, \bibinfo {author} {\bibfnamefont {F.}~\bibnamefont
  {Pavosevic}}, \bibinfo {author} {\bibfnamefont {A.}~\bibnamefont {Rubio}},\
  and\ \bibinfo {author} {\bibfnamefont {I.}~\bibnamefont {Tavernelli}},\
  }\bibfield  {title} {\bibinfo {title} {Quantum embedding method for the
  simulation of strongly correlated systems on quantum computers},\ }\href
  {https://pubs.acs.org/doi/10.1021/acs.jpclett.3c00330} {\bibfield  {journal}
  {\bibinfo  {journal} {The Journal of Physical Chemistry Letters}\ }\textbf
  {\bibinfo {volume} {14}},\ \bibinfo {pages} {3491} (\bibinfo {year}
  {2023})}\BibitemShut {NoStop}%
\bibitem [{\citenamefont {Kirby}\ and\ \citenamefont
  {Love}(2019)}]{kirby2019contextuality}%
  \BibitemOpen
  \bibfield  {author} {\bibinfo {author} {\bibfnamefont {W.~M.}\ \bibnamefont
  {Kirby}}\ and\ \bibinfo {author} {\bibfnamefont {P.~J.}\ \bibnamefont
  {Love}},\ }\bibfield  {title} {\bibinfo {title} {Contextuality test of the
  nonclassicality of variational quantum eigensolvers},\ }\href
  {https://doi.org/10.1103/PhysRevLett.123.200501} {\bibfield  {journal}
  {\bibinfo  {journal} {Phys. Rev. Lett.}\ }\textbf {\bibinfo {volume} {123}},\
  \bibinfo {pages} {200501} (\bibinfo {year} {2019})}\BibitemShut {NoStop}%
\bibitem [{\citenamefont {Kirby}\ \emph {et~al.}(2021)\citenamefont {Kirby},
  \citenamefont {Tranter},\ and\ \citenamefont
  {Love}}]{kirby2021contextualsubspace}%
  \BibitemOpen
  \bibfield  {author} {\bibinfo {author} {\bibfnamefont {W.~M.}\ \bibnamefont
  {Kirby}}, \bibinfo {author} {\bibfnamefont {A.}~\bibnamefont {Tranter}},\
  and\ \bibinfo {author} {\bibfnamefont {P.~J.}\ \bibnamefont {Love}},\
  }\bibfield  {title} {\bibinfo {title} {Contextual {S}ubspace {V}ariational
  {Q}uantum {E}igensolver},\ }\href {https://doi.org/10.22331/q-2021-05-14-456}
  {\bibfield  {journal} {\bibinfo  {journal} {{Quantum}}\ }\textbf {\bibinfo
  {volume} {5}},\ \bibinfo {pages} {456} (\bibinfo {year} {2021})}\BibitemShut
  {NoStop}%
\bibitem [{\citenamefont {Weaving}\ \emph
  {et~al.}(2023{\natexlab{a}})\citenamefont {Weaving}, \citenamefont {Ralli},
  \citenamefont {Kirby}, \citenamefont {Tranter}, \citenamefont {Love},\ and\
  \citenamefont {Coveney}}]{weaving2023stabilizer}%
  \BibitemOpen
  \bibfield  {author} {\bibinfo {author} {\bibfnamefont {T.}~\bibnamefont
  {Weaving}}, \bibinfo {author} {\bibfnamefont {A.}~\bibnamefont {Ralli}},
  \bibinfo {author} {\bibfnamefont {W.~M.}\ \bibnamefont {Kirby}}, \bibinfo
  {author} {\bibfnamefont {A.}~\bibnamefont {Tranter}}, \bibinfo {author}
  {\bibfnamefont {P.~J.}\ \bibnamefont {Love}},\ and\ \bibinfo {author}
  {\bibfnamefont {P.~V.}\ \bibnamefont {Coveney}},\ }\bibfield  {title}
  {\bibinfo {title} {A stabilizer framework for the contextual subspace
  variational quantum eigensolver and the noncontextual projection ansatz},\
  }\href {https://doi.org/10.1021/acs.jctc.2c00910} {\bibfield  {journal}
  {\bibinfo  {journal} {Journal of Chemical Theory and Computation}\ }\textbf
  {\bibinfo {volume} {19}},\ \bibinfo {pages} {808–821} (\bibinfo {year}
  {2023}{\natexlab{a}})}\BibitemShut {NoStop}%
\bibitem [{\citenamefont {Ralli}\ \emph {et~al.}(2023)\citenamefont {Ralli},
  \citenamefont {Weaving}, \citenamefont {Tranter}, \citenamefont {Kirby},
  \citenamefont {Love},\ and\ \citenamefont {Coveney}}]{ralli2023unitary}%
  \BibitemOpen
  \bibfield  {author} {\bibinfo {author} {\bibfnamefont {A.}~\bibnamefont
  {Ralli}}, \bibinfo {author} {\bibfnamefont {T.}~\bibnamefont {Weaving}},
  \bibinfo {author} {\bibfnamefont {A.}~\bibnamefont {Tranter}}, \bibinfo
  {author} {\bibfnamefont {W.~M.}\ \bibnamefont {Kirby}}, \bibinfo {author}
  {\bibfnamefont {P.~J.}\ \bibnamefont {Love}},\ and\ \bibinfo {author}
  {\bibfnamefont {P.~V.}\ \bibnamefont {Coveney}},\ }\bibfield  {title}
  {\bibinfo {title} {Unitary partitioning and the contextual subspace
  variational quantum eigensolver},\ }\bibfield  {journal} {\bibinfo  {journal}
  {Physical Review Research}\ }\textbf {\bibinfo {volume} {5}},\ \href
  {https://doi.org/10.1103/physrevresearch.5.013095}
  {10.1103/physrevresearch.5.013095} (\bibinfo {year} {2023})\BibitemShut
  {NoStop}%
\bibitem [{\citenamefont {Weaving}\ \emph
  {et~al.}(2023{\natexlab{b}})\citenamefont {Weaving}, \citenamefont {Ralli},
  \citenamefont {Love}, \citenamefont {Succi},\ and\ \citenamefont
  {Coveney}}]{weaving2023contextual}%
  \BibitemOpen
  \bibfield  {author} {\bibinfo {author} {\bibfnamefont {T.}~\bibnamefont
  {Weaving}}, \bibinfo {author} {\bibfnamefont {A.}~\bibnamefont {Ralli}},
  \bibinfo {author} {\bibfnamefont {P.~J.}\ \bibnamefont {Love}}, \bibinfo
  {author} {\bibfnamefont {S.}~\bibnamefont {Succi}},\ and\ \bibinfo {author}
  {\bibfnamefont {P.~V.}\ \bibnamefont {Coveney}},\ }\bibfield  {title}
  {\bibinfo {title} {Contextual subspace variational quantum eigensolver
  calculation of the dissociation curve of molecular nitrogen on a
  superconducting quantum computer},\ }\href {https://arxiv.org/abs/2312.04392}
  {\bibfield  {journal} {\bibinfo  {journal} {arXiv preprint arXiv:2312.04392}\
  } (\bibinfo {year} {2023}{\natexlab{b}})}\BibitemShut {NoStop}%
\bibitem [{\citenamefont {Weaving}\ \emph
  {et~al.}(2023{\natexlab{c}})\citenamefont {Weaving}, \citenamefont {Ralli},
  \citenamefont {Kirby}, \citenamefont {Love}, \citenamefont {Succi},\ and\
  \citenamefont {Coveney}}]{weaving2023benchmarking}%
  \BibitemOpen
  \bibfield  {author} {\bibinfo {author} {\bibfnamefont {T.}~\bibnamefont
  {Weaving}}, \bibinfo {author} {\bibfnamefont {A.}~\bibnamefont {Ralli}},
  \bibinfo {author} {\bibfnamefont {W.~M.}\ \bibnamefont {Kirby}}, \bibinfo
  {author} {\bibfnamefont {P.~J.}\ \bibnamefont {Love}}, \bibinfo {author}
  {\bibfnamefont {S.}~\bibnamefont {Succi}},\ and\ \bibinfo {author}
  {\bibfnamefont {P.~V.}\ \bibnamefont {Coveney}},\ }\bibfield  {title}
  {\bibinfo {title} {Benchmarking noisy intermediate scale quantum error
  mitigation strategies for ground state preparation of the hcl molecule},\
  }\href {https://doi.org/10.1103/PhysRevResearch.5.043054} {\bibfield
  {journal} {\bibinfo  {journal} {Phys. Rev. Res.}\ }\textbf {\bibinfo {volume}
  {5}},\ \bibinfo {pages} {043054} (\bibinfo {year}
  {2023}{\natexlab{c}})}\BibitemShut {NoStop}%
\bibitem [{\citenamefont {Liang}\ \emph {et~al.}(2023)\citenamefont {Liang},
  \citenamefont {Song}, \citenamefont {Cheng}, \citenamefont {Ren},
  \citenamefont {Hao}, \citenamefont {Yang}, \citenamefont {Shi},\ and\
  \citenamefont {Li}}]{liang2023spacepulse}%
  \BibitemOpen
  \bibfield  {author} {\bibinfo {author} {\bibfnamefont {Z.}~\bibnamefont
  {Liang}}, \bibinfo {author} {\bibfnamefont {Z.}~\bibnamefont {Song}},
  \bibinfo {author} {\bibfnamefont {J.}~\bibnamefont {Cheng}}, \bibinfo
  {author} {\bibfnamefont {H.}~\bibnamefont {Ren}}, \bibinfo {author}
  {\bibfnamefont {T.}~\bibnamefont {Hao}}, \bibinfo {author} {\bibfnamefont
  {R.}~\bibnamefont {Yang}}, \bibinfo {author} {\bibfnamefont {Y.}~\bibnamefont
  {Shi}},\ and\ \bibinfo {author} {\bibfnamefont {T.}~\bibnamefont {Li}},\
  }\bibfield  {title} {\bibinfo {title} {Spacepulse: Combining parameterized
  pulses and contextual subspace for more practical vqe},\ }\href
  {https://arxiv.org/abs/2311.17423} {\bibfield  {journal} {\bibinfo  {journal}
  {arXiv preprint arXiv:2311.17423}\ } (\bibinfo {year} {2023})}\BibitemShut
  {NoStop}%
\bibitem [{\citenamefont {Zhang}\ and\ \citenamefont
  {Krakauer}(2003)}]{zhang2003quantum}%
  \BibitemOpen
  \bibfield  {author} {\bibinfo {author} {\bibfnamefont {S.}~\bibnamefont
  {Zhang}}\ and\ \bibinfo {author} {\bibfnamefont {H.}~\bibnamefont
  {Krakauer}},\ }\bibfield  {title} {\bibinfo {title} {Quantum {M}onte {C}arlo
  method using phase-free random walks with slater determinants},\ }\href
  {https://doi.org/10.1103/PhysRevLett.90.136401} {\bibfield  {journal}
  {\bibinfo  {journal} {Phys. Rev. Lett.}\ }\textbf {\bibinfo {volume} {90}},\
  \bibinfo {pages} {136401} (\bibinfo {year} {2003})}\BibitemShut {NoStop}%
\bibitem [{\citenamefont {Motta}\ and\ \citenamefont
  {Zhang}(2018)}]{motta2018ab}%
  \BibitemOpen
  \bibfield  {author} {\bibinfo {author} {\bibfnamefont {M.}~\bibnamefont
  {Motta}}\ and\ \bibinfo {author} {\bibfnamefont {S.}~\bibnamefont {Zhang}},\
  }\bibfield  {title} {\bibinfo {title} {Ab initio computations of molecular
  systems by the auxiliary-field quantum {M}onte {C}arlo method},\ }\href
  {https://arxiv.org/abs/1711.02242} {\bibfield  {journal} {\bibinfo  {journal}
  {Wiley Interdisciplinary Reviews: Computational Molecular Science}\ }\textbf
  {\bibinfo {volume} {8}},\ \bibinfo {pages} {e1364} (\bibinfo {year}
  {2018})}\BibitemShut {NoStop}%
\bibitem [{\citenamefont {Lee}\ \emph {et~al.}(2022{\natexlab{a}})\citenamefont
  {Lee}, \citenamefont {Pham},\ and\ \citenamefont {Reichman}}]{lee2022twenty}%
  \BibitemOpen
  \bibfield  {author} {\bibinfo {author} {\bibfnamefont {J.}~\bibnamefont
  {Lee}}, \bibinfo {author} {\bibfnamefont {H.~Q.}\ \bibnamefont {Pham}},\ and\
  \bibinfo {author} {\bibfnamefont {D.~R.}\ \bibnamefont {Reichman}},\
  }\bibfield  {title} {\bibinfo {title} {Twenty years of auxiliary-field
  quantum {M}onte {C}arlo in quantum chemistry: An overview and assessment on
  main group chemistry and bond-breaking},\ }\href
  {https://pubs.acs.org/doi/10.1021/acs.jctc.2c00802} {\bibfield  {journal}
  {\bibinfo  {journal} {Journal of Chemical Theory and Computation}\ }\textbf
  {\bibinfo {volume} {18}},\ \bibinfo {pages} {7024} (\bibinfo {year}
  {2022}{\natexlab{a}})}\BibitemShut {NoStop}%
\bibitem [{\citenamefont {Landinez~Borda}\ \emph {et~al.}(2019)\citenamefont
  {Landinez~Borda}, \citenamefont {Gomez},\ and\ \citenamefont
  {Morales}}]{landinez2019non}%
  \BibitemOpen
  \bibfield  {author} {\bibinfo {author} {\bibfnamefont {E.~J.}\ \bibnamefont
  {Landinez~Borda}}, \bibinfo {author} {\bibfnamefont {J.}~\bibnamefont
  {Gomez}},\ and\ \bibinfo {author} {\bibfnamefont {M.~A.}\ \bibnamefont
  {Morales}},\ }\bibfield  {title} {\bibinfo {title} {Non-orthogonal
  multi-slater determinant expansions in auxiliary field quantum {M}onte
  {C}arlo},\ }\href
  {https://pubs.aip.org/aip/jcp/article/150/7/074105/197704/Non-orthogonal-multi-Slater-determinant-expansions}
  {\bibfield  {journal} {\bibinfo  {journal} {The Journal of Chemical Physics}\
  }\textbf {\bibinfo {volume} {150}} (\bibinfo {year} {2019})}\BibitemShut
  {NoStop}%
\bibitem [{\citenamefont {Amsler}\ \emph
  {et~al.}(2023{\natexlab{a}})\citenamefont {Amsler}, \citenamefont {Deglmann},
  \citenamefont {Degroote}, \citenamefont {Kaicher}, \citenamefont {Kiser},
  \citenamefont {K{\"u}hn}, \citenamefont {Kumar}, \citenamefont {Maier},
  \citenamefont {Samsonidze}, \citenamefont {Schroeder} \emph
  {et~al.}}]{amsler2023quantum}%
  \BibitemOpen
  \bibfield  {author} {\bibinfo {author} {\bibfnamefont {M.}~\bibnamefont
  {Amsler}}, \bibinfo {author} {\bibfnamefont {P.}~\bibnamefont {Deglmann}},
  \bibinfo {author} {\bibfnamefont {M.}~\bibnamefont {Degroote}}, \bibinfo
  {author} {\bibfnamefont {M.~P.}\ \bibnamefont {Kaicher}}, \bibinfo {author}
  {\bibfnamefont {M.}~\bibnamefont {Kiser}}, \bibinfo {author} {\bibfnamefont
  {M.}~\bibnamefont {K{\"u}hn}}, \bibinfo {author} {\bibfnamefont
  {C.}~\bibnamefont {Kumar}}, \bibinfo {author} {\bibfnamefont
  {A.}~\bibnamefont {Maier}}, \bibinfo {author} {\bibfnamefont
  {G.}~\bibnamefont {Samsonidze}}, \bibinfo {author} {\bibfnamefont
  {A.}~\bibnamefont {Schroeder}}, \emph {et~al.},\ }\bibfield  {title}
  {\bibinfo {title} {Quantum-enhanced quantum {M}onte {C}arlo: an industrial
  view},\ }\href {https://arxiv.org/abs/2301.11838} {\bibfield  {journal}
  {\bibinfo  {journal} {arXiv preprint arXiv:2301.11838}\ } (\bibinfo {year}
  {2023}{\natexlab{a}})}\BibitemShut {NoStop}%
\bibitem [{\citenamefont {Pham}\ \emph {et~al.}(2024)\citenamefont {Pham},
  \citenamefont {Ouyang},\ and\ \citenamefont {Lv}}]{pham2024scalable}%
  \BibitemOpen
  \bibfield  {author} {\bibinfo {author} {\bibfnamefont {H.~Q.}\ \bibnamefont
  {Pham}}, \bibinfo {author} {\bibfnamefont {R.}~\bibnamefont {Ouyang}},\ and\
  \bibinfo {author} {\bibfnamefont {D.}~\bibnamefont {Lv}},\ }\bibfield
  {title} {\bibinfo {title} {Scalable quantum {M}onte {C}arlo with
  direct-product trial wave functions},\ }\href
  {https://pubs.acs.org/doi/10.1021/acs.jctc.3c00769} {\bibfield  {journal}
  {\bibinfo  {journal} {Journal of Chemical Theory and Computation}\ }\textbf
  {\bibinfo {volume} {20}},\ \bibinfo {pages} {3524} (\bibinfo {year}
  {2024})}\BibitemShut {NoStop}%
\bibitem [{\citenamefont {Huang}\ \emph
  {et~al.}(2024{\natexlab{a}})\citenamefont {Huang}, \citenamefont {Guo},
  \citenamefont {Pham},\ and\ \citenamefont {Lv}}]{huang2024gpu}%
  \BibitemOpen
  \bibfield  {author} {\bibinfo {author} {\bibfnamefont {Y.}~\bibnamefont
  {Huang}}, \bibinfo {author} {\bibfnamefont {Z.}~\bibnamefont {Guo}}, \bibinfo
  {author} {\bibfnamefont {H.~Q.}\ \bibnamefont {Pham}},\ and\ \bibinfo
  {author} {\bibfnamefont {D.}~\bibnamefont {Lv}},\ }\bibfield  {title}
  {\bibinfo {title} {Gpu-accelerated auxiliary-field quantum {M}onte {C}arlo
  with multi-slater determinant trial states},\ }\href
  {https://arxiv.org/abs/2406.08314} {\bibfield  {journal} {\bibinfo  {journal}
  {arXiv preprint arXiv:2406.08314}\ } (\bibinfo {year}
  {2024}{\natexlab{a}})}\BibitemShut {NoStop}%
\bibitem [{\citenamefont {Huggins}\ \emph
  {et~al.}(2022{\natexlab{a}})\citenamefont {Huggins}, \citenamefont
  {O'Gorman}, \citenamefont {Rubin}, \citenamefont {Reichman}, \citenamefont
  {Babbush},\ and\ \citenamefont {Lee}}]{huggin2022unbiasing}%
  \BibitemOpen
  \bibfield  {author} {\bibinfo {author} {\bibfnamefont {W.~J.}\ \bibnamefont
  {Huggins}}, \bibinfo {author} {\bibfnamefont {B.~A.}\ \bibnamefont
  {O'Gorman}}, \bibinfo {author} {\bibfnamefont {N.~C.}\ \bibnamefont {Rubin}},
  \bibinfo {author} {\bibfnamefont {D.~R.}\ \bibnamefont {Reichman}}, \bibinfo
  {author} {\bibfnamefont {R.}~\bibnamefont {Babbush}},\ and\ \bibinfo {author}
  {\bibfnamefont {J.}~\bibnamefont {Lee}},\ }\bibfield  {title} {\bibinfo
  {title} {Unbiasing fermionic quantum {M}onte {C}arlo with a quantum
  computer},\ }\href {https://doi.org/10.1038/s41586-021-04351-z} {\bibfield
  {journal} {\bibinfo  {journal} {Nature}\ }\textbf {\bibinfo {volume} {603}},\
  \bibinfo {pages} {416} (\bibinfo {year} {2022}{\natexlab{a}})}\BibitemShut
  {NoStop}%
\bibitem [{\citenamefont {Sinibaldi}\ \emph {et~al.}(2023)\citenamefont
  {Sinibaldi}, \citenamefont {Giuliani}, \citenamefont {Carleo},\ and\
  \citenamefont {Vicentini}}]{sinibaldi2023unbiasing}%
  \BibitemOpen
  \bibfield  {author} {\bibinfo {author} {\bibfnamefont {A.}~\bibnamefont
  {Sinibaldi}}, \bibinfo {author} {\bibfnamefont {C.}~\bibnamefont {Giuliani}},
  \bibinfo {author} {\bibfnamefont {G.}~\bibnamefont {Carleo}},\ and\ \bibinfo
  {author} {\bibfnamefont {F.}~\bibnamefont {Vicentini}},\ }\bibfield  {title}
  {\bibinfo {title} {Unbiasing time-dependent variational {M}onte {C}arlo by
  projected quantum evolution},\ }\href
  {https://quantum-journal.org/papers/q-2023-10-10-1131/} {\bibfield  {journal}
  {\bibinfo  {journal} {Quantum}\ }\textbf {\bibinfo {volume} {7}},\ \bibinfo
  {pages} {1131} (\bibinfo {year} {2023})}\BibitemShut {NoStop}%
\bibitem [{\citenamefont {Montanaro}\ and\ \citenamefont
  {Stanisic}(2023)}]{montanaro2023accelerating}%
  \BibitemOpen
  \bibfield  {author} {\bibinfo {author} {\bibfnamefont {A.}~\bibnamefont
  {Montanaro}}\ and\ \bibinfo {author} {\bibfnamefont {S.}~\bibnamefont
  {Stanisic}},\ }\bibfield  {title} {\bibinfo {title} {Accelerating variational
  quantum {M}onte {C}arlo using the variational quantum eigensolver},\ }\href
  {https://arxiv.org/abs/2307.07719} {\bibfield  {journal} {\bibinfo  {journal}
  {arXiv preprint arXiv:2307.07719}\ } (\bibinfo {year} {2023})}\BibitemShut
  {NoStop}%
\bibitem [{\citenamefont {Xu}\ and\ \citenamefont {Li}(2023)}]{xu2023quantum}%
  \BibitemOpen
  \bibfield  {author} {\bibinfo {author} {\bibfnamefont {X.}~\bibnamefont
  {Xu}}\ and\ \bibinfo {author} {\bibfnamefont {Y.}~\bibnamefont {Li}},\
  }\bibfield  {title} {\bibinfo {title} {Quantum-assisted {M}onte {C}arlo
  algorithms for fermions},\ }\href
  {https://quantum-journal.org/papers/q-2023-08-03-1072/} {\bibfield  {journal}
  {\bibinfo  {journal} {Quantum}\ }\textbf {\bibinfo {volume} {7}},\ \bibinfo
  {pages} {1072} (\bibinfo {year} {2023})}\BibitemShut {NoStop}%
\bibitem [{\citenamefont {Kanno}\ \emph {et~al.}(2024)\citenamefont {Kanno},
  \citenamefont {Nakamura}, \citenamefont {Kobayashi}, \citenamefont {Gocho},
  \citenamefont {Hatanaka}, \citenamefont {Yamamoto},\ and\ \citenamefont
  {Gao}}]{kanno2024quantum}%
  \BibitemOpen
  \bibfield  {author} {\bibinfo {author} {\bibfnamefont {S.}~\bibnamefont
  {Kanno}}, \bibinfo {author} {\bibfnamefont {H.}~\bibnamefont {Nakamura}},
  \bibinfo {author} {\bibfnamefont {T.}~\bibnamefont {Kobayashi}}, \bibinfo
  {author} {\bibfnamefont {S.}~\bibnamefont {Gocho}}, \bibinfo {author}
  {\bibfnamefont {M.}~\bibnamefont {Hatanaka}}, \bibinfo {author}
  {\bibfnamefont {N.}~\bibnamefont {Yamamoto}},\ and\ \bibinfo {author}
  {\bibfnamefont {Q.}~\bibnamefont {Gao}},\ }\bibfield  {title} {\bibinfo
  {title} {Quantum computing quantum {M}onte {C}arlo with hybrid tensor network
  for electronic structure calculations},\ }\href
  {https://www.nature.com/articles/s41534-024-00851-8} {\bibfield  {journal}
  {\bibinfo  {journal} {npj Quantum Information}\ }\textbf {\bibinfo {volume}
  {10}},\ \bibinfo {pages} {56} (\bibinfo {year} {2024})}\BibitemShut {NoStop}%
\bibitem [{\citenamefont {Mazzola}\ and\ \citenamefont
  {Carleo}(2022)}]{mazzola2022exponential}%
  \BibitemOpen
  \bibfield  {author} {\bibinfo {author} {\bibfnamefont {G.}~\bibnamefont
  {Mazzola}}\ and\ \bibinfo {author} {\bibfnamefont {G.}~\bibnamefont
  {Carleo}},\ }\bibfield  {title} {\bibinfo {title} {Exponential challenges in
  unbiasing quantum {M}onte {C}arlo algorithms with quantum computers},\ }\href
  {https://arxiv.org/abs/2205.09203} {\bibfield  {journal} {\bibinfo  {journal}
  {arXiv preprint arXiv:2205.09203}\ } (\bibinfo {year} {2022})}\BibitemShut
  {NoStop}%
\bibitem [{\citenamefont {Lee}\ \emph {et~al.}(2022{\natexlab{b}})\citenamefont
  {Lee}, \citenamefont {Reichman}, \citenamefont {Babbush}, \citenamefont
  {Rubin}, \citenamefont {Malone}, \citenamefont {O'Gorman},\ and\
  \citenamefont {Huggins}}]{lee2022response}%
  \BibitemOpen
  \bibfield  {author} {\bibinfo {author} {\bibfnamefont {J.}~\bibnamefont
  {Lee}}, \bibinfo {author} {\bibfnamefont {D.~R.}\ \bibnamefont {Reichman}},
  \bibinfo {author} {\bibfnamefont {R.}~\bibnamefont {Babbush}}, \bibinfo
  {author} {\bibfnamefont {N.~C.}\ \bibnamefont {Rubin}}, \bibinfo {author}
  {\bibfnamefont {F.~D.}\ \bibnamefont {Malone}}, \bibinfo {author}
  {\bibfnamefont {B.}~\bibnamefont {O'Gorman}},\ and\ \bibinfo {author}
  {\bibfnamefont {W.~J.}\ \bibnamefont {Huggins}},\ }\bibfield  {title}
  {\bibinfo {title} {Response to ``{E}xponential challenges in unbiasing
  quantum {M}onte {C}arlo algorithms with quantum computers"},\ }\href
  {https://arxiv.org/abs/2207.13776} {\bibfield  {journal} {\bibinfo  {journal}
  {arXiv preprint arXiv:2207.13776}\ } (\bibinfo {year}
  {2022}{\natexlab{b}})}\BibitemShut {NoStop}%
\bibitem [{\citenamefont {Aaronson}(2017)}]{aaronson2017shadow}%
  \BibitemOpen
  \bibfield  {author} {\bibinfo {author} {\bibfnamefont {S.}~\bibnamefont
  {Aaronson}},\ }\bibfield  {title} {\bibinfo {title} {{Shadow Tomography of
  Quantum States}},\ }\bibfield  {journal} {\bibinfo  {journal} {arXiv}\ }\href
  {https://doi.org/10.48550/arXiv.1711.01053} {10.48550/arXiv.1711.01053}
  (\bibinfo {year} {2017}),\ \Eprint {https://arxiv.org/abs/1711.01053}
  {1711.01053} \BibitemShut {NoStop}%
\bibitem [{\citenamefont {Huang}\ \emph {et~al.}(2020)\citenamefont {Huang},
  \citenamefont {Kueng},\ and\ \citenamefont {Preskill}}]{huang2020predicting}%
  \BibitemOpen
  \bibfield  {author} {\bibinfo {author} {\bibfnamefont {H.-Y.}\ \bibnamefont
  {Huang}}, \bibinfo {author} {\bibfnamefont {R.}~\bibnamefont {Kueng}},\ and\
  \bibinfo {author} {\bibfnamefont {J.}~\bibnamefont {Preskill}},\ }\bibfield
  {title} {\bibinfo {title} {{Predicting Many Properties of a Quantum System
  from Very Few Measurements}},\ }\bibfield  {journal} {\bibinfo  {journal}
  {arXiv}\ }\href {https://doi.org/10.1038/s41567-020-0932-7}
  {10.1038/s41567-020-0932-7} (\bibinfo {year} {2020}),\ \Eprint
  {https://arxiv.org/abs/2002.08953} {2002.08953} \BibitemShut {NoStop}%
\bibitem [{\citenamefont {Wan}\ \emph {et~al.}(2022)\citenamefont {Wan},
  \citenamefont {Huggins}, \citenamefont {Lee},\ and\ \citenamefont
  {Babbush}}]{wan2022matchgate}%
  \BibitemOpen
  \bibfield  {author} {\bibinfo {author} {\bibfnamefont {K.}~\bibnamefont
  {Wan}}, \bibinfo {author} {\bibfnamefont {W.~J.}\ \bibnamefont {Huggins}},
  \bibinfo {author} {\bibfnamefont {J.}~\bibnamefont {Lee}},\ and\ \bibinfo
  {author} {\bibfnamefont {R.}~\bibnamefont {Babbush}},\ }\bibfield  {title}
  {\bibinfo {title} {{Matchgate Shadows for Fermionic Quantum Simulation}},\
  }\bibfield  {journal} {\bibinfo  {journal} {arXiv}\ }\href
  {https://doi.org/10.48550/arXiv.2207.13723} {10.48550/arXiv.2207.13723}
  (\bibinfo {year} {2022}),\ \Eprint {https://arxiv.org/abs/2207.13723}
  {2207.13723} \BibitemShut {NoStop}%
\bibitem [{\citenamefont {Low}(2022)}]{low2022classical}%
  \BibitemOpen
  \bibfield  {author} {\bibinfo {author} {\bibfnamefont {G.~H.}\ \bibnamefont
  {Low}},\ }\bibfield  {title} {\bibinfo {title} {{Classical shadows of
  fermions with particle number symmetry}},\ }\bibfield  {journal} {\bibinfo
  {journal} {arXiv}\ }\href {https://doi.org/10.48550/arXiv.2208.08964}
  {10.48550/arXiv.2208.08964} (\bibinfo {year} {2022}),\ \Eprint
  {https://arxiv.org/abs/2208.08964} {2208.08964} \BibitemShut {NoStop}%
\bibitem [{\citenamefont {Wu}\ and\ \citenamefont {Koh}(2024)}]{wu2024error}%
  \BibitemOpen
  \bibfield  {author} {\bibinfo {author} {\bibfnamefont {B.}~\bibnamefont
  {Wu}}\ and\ \bibinfo {author} {\bibfnamefont {D.~E.}\ \bibnamefont {Koh}},\
  }\bibfield  {title} {\bibinfo {title} {Error-mitigated fermionic classical
  shadows on noisy quantum devices},\ }\href
  {https://www.nature.com/articles/s41534-024-00836-7} {\bibfield  {journal}
  {\bibinfo  {journal} {npj Quantum Information}\ }\textbf {\bibinfo {volume}
  {10}},\ \bibinfo {pages} {39} (\bibinfo {year} {2024})}\BibitemShut {NoStop}%
\bibitem [{\citenamefont {Zhao}\ and\ \citenamefont
  {Miyake}(2024)}]{zhao2024group}%
  \BibitemOpen
  \bibfield  {author} {\bibinfo {author} {\bibfnamefont {A.}~\bibnamefont
  {Zhao}}\ and\ \bibinfo {author} {\bibfnamefont {A.}~\bibnamefont {Miyake}},\
  }\bibfield  {title} {\bibinfo {title} {Group-theoretic error mitigation
  enabled by classical shadows and symmetries},\ }\href
  {https://www.nature.com/articles/s41534-024-00854-5} {\bibfield  {journal}
  {\bibinfo  {journal} {npj Quantum Information}\ }\textbf {\bibinfo {volume}
  {10}},\ \bibinfo {pages} {57} (\bibinfo {year} {2024})}\BibitemShut {NoStop}%
\bibitem [{\citenamefont {Kiser}\ \emph {et~al.}(2024)\citenamefont {Kiser},
  \citenamefont {Schroeder}, \citenamefont {Anselmetti}, \citenamefont {Kumar},
  \citenamefont {Moll}, \citenamefont {Streif},\ and\ \citenamefont
  {Vodola}}]{kiser2024classical}%
  \BibitemOpen
  \bibfield  {author} {\bibinfo {author} {\bibfnamefont {M.}~\bibnamefont
  {Kiser}}, \bibinfo {author} {\bibfnamefont {A.}~\bibnamefont {Schroeder}},
  \bibinfo {author} {\bibfnamefont {G.-L.~R.}\ \bibnamefont {Anselmetti}},
  \bibinfo {author} {\bibfnamefont {C.}~\bibnamefont {Kumar}}, \bibinfo
  {author} {\bibfnamefont {N.}~\bibnamefont {Moll}}, \bibinfo {author}
  {\bibfnamefont {M.}~\bibnamefont {Streif}},\ and\ \bibinfo {author}
  {\bibfnamefont {D.}~\bibnamefont {Vodola}},\ }\bibfield  {title} {\bibinfo
  {title} {Classical and quantum cost of measurement strategies for
  quantum-enhanced auxiliary field quantum monte carlo},\ }\href
  {https://iopscience.iop.org/article/10.1088/1367-2630/ad2f67} {\bibfield
  {journal} {\bibinfo  {journal} {New Journal of Physics}\ }\textbf {\bibinfo
  {volume} {26}},\ \bibinfo {pages} {033022} (\bibinfo {year}
  {2024})}\BibitemShut {NoStop}%
\bibitem [{\citenamefont {Jiang}\ \emph {et~al.}(2024)\citenamefont {Jiang},
  \citenamefont {O'Gorman}, \citenamefont {Mahajan},\ and\ \citenamefont
  {Lee}}]{jiang2024unbiasing}%
  \BibitemOpen
  \bibfield  {author} {\bibinfo {author} {\bibfnamefont {T.}~\bibnamefont
  {Jiang}}, \bibinfo {author} {\bibfnamefont {B.}~\bibnamefont {O'Gorman}},
  \bibinfo {author} {\bibfnamefont {A.}~\bibnamefont {Mahajan}},\ and\ \bibinfo
  {author} {\bibfnamefont {J.}~\bibnamefont {Lee}},\ }\href@noop {} {\bibinfo
  {title} {Unbiasing fermionic auxiliary-field quantum {M}onte {C}arlo with
  matrix product state trial wavefunctions}} (\bibinfo {year} {2024}),\ \Eprint
  {https://arxiv.org/abs/2405.05440} {arXiv:2405.05440} \BibitemShut {NoStop}%
\bibitem [{\citenamefont {Huang}\ \emph
  {et~al.}(2024{\natexlab{b}})\citenamefont {Huang}, \citenamefont {Chen},
  \citenamefont {Gupt}, \citenamefont {Suchara}, \citenamefont {Tran},
  \citenamefont {McArdle},\ and\ \citenamefont {Galli}}]{huang2024evaluating}%
  \BibitemOpen
  \bibfield  {author} {\bibinfo {author} {\bibfnamefont {B.}~\bibnamefont
  {Huang}}, \bibinfo {author} {\bibfnamefont {Y.-T.}\ \bibnamefont {Chen}},
  \bibinfo {author} {\bibfnamefont {B.}~\bibnamefont {Gupt}}, \bibinfo {author}
  {\bibfnamefont {M.}~\bibnamefont {Suchara}}, \bibinfo {author} {\bibfnamefont
  {A.}~\bibnamefont {Tran}}, \bibinfo {author} {\bibfnamefont {S.}~\bibnamefont
  {McArdle}},\ and\ \bibinfo {author} {\bibfnamefont {G.}~\bibnamefont
  {Galli}},\ }\href@noop {} {\bibinfo {title} {Evaluating a quantum-classical
  quantum {M}onte {C}arlo algorithm with matchgate shadows}} (\bibinfo {year}
  {2024}{\natexlab{b}}),\ \Eprint {https://arxiv.org/abs/2404.18303}
  {arXiv:2404.18303} \BibitemShut {NoStop}%
\bibitem [{\citenamefont {Bartlett}\ and\ \citenamefont
  {Musia{\l}}(2007)}]{bartlett2007coupled}%
  \BibitemOpen
  \bibfield  {author} {\bibinfo {author} {\bibfnamefont {R.~J.}\ \bibnamefont
  {Bartlett}}\ and\ \bibinfo {author} {\bibfnamefont {M.}~\bibnamefont
  {Musia{\l}}},\ }\bibfield  {title} {\bibinfo {title} {Coupled-cluster theory
  in quantum chemistry},\ }\href
  {https://journals.aps.org/rmp/abstract/10.1103/RevModPhys.79.291} {\bibfield
  {journal} {\bibinfo  {journal} {Reviews of Modern Physics}\ }\textbf
  {\bibinfo {volume} {79}},\ \bibinfo {pages} {291} (\bibinfo {year}
  {2007})}\BibitemShut {NoStop}%
\bibitem [{\citenamefont {Van~der Ven}\ \emph {et~al.}(2020)\citenamefont
  {Van~der Ven}, \citenamefont {Deng}, \citenamefont {Banerjee},\ and\
  \citenamefont {Ong}}]{vanderven2020batterytheory}%
  \BibitemOpen
  \bibfield  {author} {\bibinfo {author} {\bibfnamefont {A.}~\bibnamefont
  {Van~der Ven}}, \bibinfo {author} {\bibfnamefont {Z.}~\bibnamefont {Deng}},
  \bibinfo {author} {\bibfnamefont {S.}~\bibnamefont {Banerjee}},\ and\
  \bibinfo {author} {\bibfnamefont {S.~P.}\ \bibnamefont {Ong}},\ }\bibfield
  {title} {\bibinfo {title} {Rechargeable alkali-ion battery materials: Theory
  and computation},\ }\href {https://doi.org/10.1021/acs.chemrev.9b00601}
  {\bibfield  {journal} {\bibinfo  {journal} {Chemical Reviews}\ }\textbf
  {\bibinfo {volume} {120}},\ \bibinfo {pages} {6977} (\bibinfo {year}
  {2020})},\ \bibinfo {note} {pMID: 32022553}\BibitemShut {NoStop}%
\bibitem [{\citenamefont {Wang}\ \emph {et~al.}(2001)\citenamefont {Wang},
  \citenamefont {Nakamura}, \citenamefont {Ue},\ and\ \citenamefont
  {Balbuena}}]{wang2001ecdecomposition}%
  \BibitemOpen
  \bibfield  {author} {\bibinfo {author} {\bibfnamefont {Y.}~\bibnamefont
  {Wang}}, \bibinfo {author} {\bibfnamefont {S.}~\bibnamefont {Nakamura}},
  \bibinfo {author} {\bibfnamefont {M.}~\bibnamefont {Ue}},\ and\ \bibinfo
  {author} {\bibfnamefont {P.~B.}\ \bibnamefont {Balbuena}},\ }\bibfield
  {title} {\bibinfo {title} {Theoretical studies to understand surface
  chemistry on carbon anodes for lithium-ion batteries: Reduction mechanisms of
  ethylene carbonate},\ }\href {https://doi.org/10.1021/ja0164529} {\bibfield
  {journal} {\bibinfo  {journal} {Journal of the American Chemical Society}\
  }\textbf {\bibinfo {volume} {123}},\ \bibinfo {pages} {11708} (\bibinfo
  {year} {2001})},\ \bibinfo {note} {pMID: 11716728}\BibitemShut {NoStop}%
\bibitem [{\citenamefont {Debnath}\ \emph {et~al.}(2023)\citenamefont
  {Debnath}, \citenamefont {Neufeld}, \citenamefont {Jacobson}, \citenamefont
  {Rudshteyn}, \citenamefont {Weber}, \citenamefont {Berkelbach},\ and\
  \citenamefont {Friesner}}]{debnath2023afqmcec}%
  \BibitemOpen
  \bibfield  {author} {\bibinfo {author} {\bibfnamefont {S.}~\bibnamefont
  {Debnath}}, \bibinfo {author} {\bibfnamefont {V.~A.}\ \bibnamefont
  {Neufeld}}, \bibinfo {author} {\bibfnamefont {L.~D.}\ \bibnamefont
  {Jacobson}}, \bibinfo {author} {\bibfnamefont {B.}~\bibnamefont {Rudshteyn}},
  \bibinfo {author} {\bibfnamefont {J.~L.}\ \bibnamefont {Weber}}, \bibinfo
  {author} {\bibfnamefont {T.~C.}\ \bibnamefont {Berkelbach}},\ and\ \bibinfo
  {author} {\bibfnamefont {R.~A.}\ \bibnamefont {Friesner}},\ }\bibfield
  {title} {\bibinfo {title} {Accurate quantum chemical reaction energies for
  lithium-mediated electrolyte decomposition and evaluation of density
  functional approximations},\ }\href
  {https://doi.org/10.1021/acs.jpca.3c04369} {\bibfield  {journal} {\bibinfo
  {journal} {The Journal of Physical Chemistry A}\ }\textbf {\bibinfo {volume}
  {127}},\ \bibinfo {pages} {9178} (\bibinfo {year} {2023})},\ \bibinfo {note}
  {pMID: 37878768}\BibitemShut {NoStop}%
\bibitem [{\citenamefont {Bravyi}\ \emph {et~al.}(2017)\citenamefont {Bravyi},
  \citenamefont {Gambetta}, \citenamefont {Mezzacapo},\ and\ \citenamefont
  {Temme}}]{bravyi2017tapering}%
  \BibitemOpen
  \bibfield  {author} {\bibinfo {author} {\bibfnamefont {S.}~\bibnamefont
  {Bravyi}}, \bibinfo {author} {\bibfnamefont {J.~M.}\ \bibnamefont
  {Gambetta}}, \bibinfo {author} {\bibfnamefont {A.}~\bibnamefont
  {Mezzacapo}},\ and\ \bibinfo {author} {\bibfnamefont {K.}~\bibnamefont
  {Temme}},\ }\href {https://arxiv.org/abs/1701.08213} {\bibinfo {title}
  {Tapering off qubits to simulate fermionic hamiltonians}} (\bibinfo {year}
  {2017}),\ \Eprint {https://arxiv.org/abs/1701.08213} {arXiv:1701.08213
  [quant-ph]} \BibitemShut {NoStop}%
\bibitem [{\citenamefont {Raussendorf}\ \emph {et~al.}(2020)\citenamefont
  {Raussendorf}, \citenamefont {Bermejo-Vega}, \citenamefont {Tyhurst},
  \citenamefont {Okay},\ and\ \citenamefont
  {Zurel}}]{raussendorf2020phasespace}%
  \BibitemOpen
  \bibfield  {author} {\bibinfo {author} {\bibfnamefont {R.}~\bibnamefont
  {Raussendorf}}, \bibinfo {author} {\bibfnamefont {J.}~\bibnamefont
  {Bermejo-Vega}}, \bibinfo {author} {\bibfnamefont {E.}~\bibnamefont
  {Tyhurst}}, \bibinfo {author} {\bibfnamefont {C.}~\bibnamefont {Okay}},\ and\
  \bibinfo {author} {\bibfnamefont {M.}~\bibnamefont {Zurel}},\ }\bibfield
  {title} {\bibinfo {title} {Phase-space-simulation method for quantum
  computation with magic states on qubits},\ }\href
  {https://doi.org/10.1103/PhysRevA.101.012350} {\bibfield  {journal} {\bibinfo
   {journal} {Phys. Rev. A}\ }\textbf {\bibinfo {volume} {101}},\ \bibinfo
  {pages} {012350} (\bibinfo {year} {2020})}\BibitemShut {NoStop}%
\bibitem [{\citenamefont {Mermin}(1993)}]{mermin1993hidden}%
  \BibitemOpen
  \bibfield  {author} {\bibinfo {author} {\bibfnamefont {N.~D.}\ \bibnamefont
  {Mermin}},\ }\bibfield  {title} {\bibinfo {title} {Hidden variables and the
  two theorems of john bell},\ }\href
  {https://doi.org/10.1103/RevModPhys.65.803} {\bibfield  {journal} {\bibinfo
  {journal} {Rev. Mod. Phys.}\ }\textbf {\bibinfo {volume} {65}},\ \bibinfo
  {pages} {803} (\bibinfo {year} {1993})}\BibitemShut {NoStop}%
\bibitem [{\citenamefont {Spekkens}(2007)}]{spekkens2007evidence}%
  \BibitemOpen
  \bibfield  {author} {\bibinfo {author} {\bibfnamefont {R.~W.}\ \bibnamefont
  {Spekkens}},\ }\bibfield  {title} {\bibinfo {title} {Evidence for the
  epistemic view of quantum states: A toy theory},\ }\href
  {https://doi.org/10.1103/PhysRevA.75.032110} {\bibfield  {journal} {\bibinfo
  {journal} {Phys. Rev. A}\ }\textbf {\bibinfo {volume} {75}},\ \bibinfo
  {pages} {032110} (\bibinfo {year} {2007})}\BibitemShut {NoStop}%
\bibitem [{\citenamefont {Spekkens}(2008)}]{spekkens2008negativity}%
  \BibitemOpen
  \bibfield  {author} {\bibinfo {author} {\bibfnamefont {R.~W.}\ \bibnamefont
  {Spekkens}},\ }\bibfield  {title} {\bibinfo {title} {Negativity and
  contextuality are equivalent notions of nonclassicality},\ }\href
  {https://doi.org/10.1103/PhysRevLett.101.020401} {\bibfield  {journal}
  {\bibinfo  {journal} {Phys. Rev. Lett.}\ }\textbf {\bibinfo {volume} {101}},\
  \bibinfo {pages} {020401} (\bibinfo {year} {2008})}\BibitemShut {NoStop}%
\bibitem [{\citenamefont {Kirby}\ and\ \citenamefont
  {Love}(2020)}]{kirby2020classical}%
  \BibitemOpen
  \bibfield  {author} {\bibinfo {author} {\bibfnamefont {W.~M.}\ \bibnamefont
  {Kirby}}\ and\ \bibinfo {author} {\bibfnamefont {P.~J.}\ \bibnamefont
  {Love}},\ }\bibfield  {title} {\bibinfo {title} {Classical simulation of
  noncontextual pauli hamiltonians},\ }\href
  {https://doi.org/10.1103/PhysRevA.102.032418} {\bibfield  {journal} {\bibinfo
   {journal} {Phys. Rev. A}\ }\textbf {\bibinfo {volume} {102}},\ \bibinfo
  {pages} {032418} (\bibinfo {year} {2020})}\BibitemShut {NoStop}%
\bibitem [{\citenamefont {Iosue}(2019)}]{qubovert}%
  \BibitemOpen
  \bibfield  {author} {\bibinfo {author} {\bibfnamefont {J.~T.}\ \bibnamefont
  {Iosue}},\ }\href@noop {} {\bibinfo {title} {Qubovert documentation}},\
  \bibinfo {howpublished} {\url{https://qubovert.readthedocs.io/en/stable/}}
  (\bibinfo {year} {2019})\BibitemShut {NoStop}%
\bibitem [{\citenamefont {Zhang}(2013)}]{zhang2013auxiliary}%
  \BibitemOpen
  \bibfield  {author} {\bibinfo {author} {\bibfnamefont {S.}~\bibnamefont
  {Zhang}},\ }\bibfield  {title} {\bibinfo {title} {Auxiliary-field quantum
  {Monte} {Carlo} for correlated electron systems},\ }in\ \href
  {https://juser.fz-juelich.de/record/137827} {\emph {\bibinfo {booktitle}
  {{E}mergent {P}henomena in {C}orrelated {M}atter}}},\ Vol.~\bibinfo {volume}
  {3}\ (\bibinfo  {publisher} {Forschungszentrum},\ \bibinfo {address}
  {Jülich},\ \bibinfo {year} {2013})\ p.~\bibinfo {pages} {15}\BibitemShut
  {NoStop}%
\bibitem [{\citenamefont {Malone}\ \emph {et~al.}(2022)\citenamefont {Malone},
  \citenamefont {Mahajan}, \citenamefont {Spencer},\ and\ \citenamefont
  {Lee}}]{malone2022ipie}%
  \BibitemOpen
  \bibfield  {author} {\bibinfo {author} {\bibfnamefont {F.~D.}\ \bibnamefont
  {Malone}}, \bibinfo {author} {\bibfnamefont {A.}~\bibnamefont {Mahajan}},
  \bibinfo {author} {\bibfnamefont {J.~S.}\ \bibnamefont {Spencer}},\ and\
  \bibinfo {author} {\bibfnamefont {J.}~\bibnamefont {Lee}},\ }\bibfield
  {title} {\bibinfo {title} {ipie: A python-based auxiliary-field quantum
  {M}onte {C}arlo program with flexibility and efficiency on cpus and gpus},\
  }\href {https://pubs.acs.org/doi/10.1021/acs.jctc.2c00934} {\bibfield
  {journal} {\bibinfo  {journal} {Journal of Chemical Theory and Computation}\
  }\textbf {\bibinfo {volume} {19}},\ \bibinfo {pages} {109} (\bibinfo {year}
  {2022})}\BibitemShut {NoStop}%
\bibitem [{\citenamefont {Negele}(2018)}]{negele2018quantum}%
  \BibitemOpen
  \bibfield  {author} {\bibinfo {author} {\bibfnamefont {J.~W.}\ \bibnamefont
  {Negele}},\ }\href {https://doi.org/10.1201/9780429497926} {\emph {\bibinfo
  {title} {Quantum many-particle systems}}}\ (\bibinfo  {publisher} {CRC
  Press},\ \bibinfo {year} {2018})\BibitemShut {NoStop}%
\bibitem [{\citenamefont {Huggins}\ \emph
  {et~al.}(2022{\natexlab{b}})\citenamefont {Huggins}, \citenamefont
  {O{'}Gorman}, \citenamefont {Rubin}, \citenamefont {Reichman}, \citenamefont
  {Babbush},\ and\ \citenamefont {Lee}}]{huggins2022unbiasing}%
  \BibitemOpen
  \bibfield  {author} {\bibinfo {author} {\bibfnamefont {W.~J.}\ \bibnamefont
  {Huggins}}, \bibinfo {author} {\bibfnamefont {B.~A.}\ \bibnamefont
  {O{'}Gorman}}, \bibinfo {author} {\bibfnamefont {N.~C.}\ \bibnamefont
  {Rubin}}, \bibinfo {author} {\bibfnamefont {D.~R.}\ \bibnamefont {Reichman}},
  \bibinfo {author} {\bibfnamefont {R.}~\bibnamefont {Babbush}},\ and\ \bibinfo
  {author} {\bibfnamefont {J.}~\bibnamefont {Lee}},\ }\bibfield  {title}
  {\bibinfo {title} {{Unbiasing fermionic quantum {M}onte {C}arlo with a
  quantum computer}},\ }\href {https://doi.org/10.1038/s41586-021-04351-z}
  {\bibfield  {journal} {\bibinfo  {journal} {Nature}\ }\textbf {\bibinfo
  {volume} {603}},\ \bibinfo {pages} {416} (\bibinfo {year}
  {2022}{\natexlab{b}})}\BibitemShut {NoStop}%
\bibitem [{\citenamefont {Amsler}\ \emph
  {et~al.}(2023{\natexlab{b}})\citenamefont {Amsler}, \citenamefont {Deglmann},
  \citenamefont {Degroote}, \citenamefont {Kaicher}, \citenamefont {Kiser},
  \citenamefont {K{\"u}hn}, \citenamefont {Kumar}, \citenamefont {Maier},
  \citenamefont {Samsonidze}, \citenamefont {Schroeder} \emph
  {et~al.}}]{amsler2023classical}%
  \BibitemOpen
  \bibfield  {author} {\bibinfo {author} {\bibfnamefont {M.}~\bibnamefont
  {Amsler}}, \bibinfo {author} {\bibfnamefont {P.}~\bibnamefont {Deglmann}},
  \bibinfo {author} {\bibfnamefont {M.}~\bibnamefont {Degroote}}, \bibinfo
  {author} {\bibfnamefont {M.~P.}\ \bibnamefont {Kaicher}}, \bibinfo {author}
  {\bibfnamefont {M.}~\bibnamefont {Kiser}}, \bibinfo {author} {\bibfnamefont
  {M.}~\bibnamefont {K{\"u}hn}}, \bibinfo {author} {\bibfnamefont
  {C.}~\bibnamefont {Kumar}}, \bibinfo {author} {\bibfnamefont
  {A.}~\bibnamefont {Maier}}, \bibinfo {author} {\bibfnamefont
  {G.}~\bibnamefont {Samsonidze}}, \bibinfo {author} {\bibfnamefont
  {A.}~\bibnamefont {Schroeder}}, \emph {et~al.},\ }\bibfield  {title}
  {\bibinfo {title} {Classical and quantum trial wave functions in
  auxiliary-field quantum {M}onte {C}arlo applied to oxygen allotropes and a
  {CuBr}$_2$ model system},\ }\href {https://doi.org/10.1063/5.0146934}
  {\bibfield  {journal} {\bibinfo  {journal} {The Journal of Chemical Physics}\
  }\textbf {\bibinfo {volume} {159}} (\bibinfo {year}
  {2023}{\natexlab{b}})}\BibitemShut {NoStop}%
\bibitem [{\citenamefont {Cleve}\ \emph {et~al.}(1998)\citenamefont {Cleve},
  \citenamefont {Ekert}, \citenamefont {Macchiavello},\ and\ \citenamefont
  {Mosca}}]{cleve1998quantum}%
  \BibitemOpen
  \bibfield  {author} {\bibinfo {author} {\bibfnamefont {R.}~\bibnamefont
  {Cleve}}, \bibinfo {author} {\bibfnamefont {A.}~\bibnamefont {Ekert}},
  \bibinfo {author} {\bibfnamefont {C.}~\bibnamefont {Macchiavello}},\ and\
  \bibinfo {author} {\bibfnamefont {M.}~\bibnamefont {Mosca}},\ }\bibfield
  {title} {\bibinfo {title} {{Quantum algorithms revisited}},\ }\href
  {https://doi.org/10.1098/rspa.1998.0164} {\bibfield  {journal} {\bibinfo
  {journal} {Proc. R. Soc. Lond. A.}\ }\textbf {\bibinfo {volume} {454}},\
  \bibinfo {pages} {339} (\bibinfo {year} {1998})}\BibitemShut {NoStop}%
\bibitem [{\citenamefont {Aharonov}\ \emph {et~al.}(2005)\citenamefont
  {Aharonov}, \citenamefont {Jones},\ and\ \citenamefont
  {Landau}}]{aharonov2005a}%
  \BibitemOpen
  \bibfield  {author} {\bibinfo {author} {\bibfnamefont {D.}~\bibnamefont
  {Aharonov}}, \bibinfo {author} {\bibfnamefont {V.}~\bibnamefont {Jones}},\
  and\ \bibinfo {author} {\bibfnamefont {Z.}~\bibnamefont {Landau}},\
  }\bibfield  {title} {\bibinfo {title} {{A Polynomial Quantum Algorithm for
  Approximating the Jones Polynomial}},\ }\bibfield  {journal} {\bibinfo
  {journal} {arXiv}\ }\href {https://doi.org/10.48550/arXiv.quant-ph/0511096}
  {10.48550/arXiv.quant-ph/0511096} (\bibinfo {year} {2005}),\ \Eprint
  {https://arxiv.org/abs/quant-ph/0511096} {quant-ph/0511096} \BibitemShut
  {NoStop}%
\bibitem [{\citenamefont {Kitaev}(1995)}]{kitaev1995quantum}%
  \BibitemOpen
  \bibfield  {author} {\bibinfo {author} {\bibfnamefont {A.~{\relax Yu}.}\
  \bibnamefont {Kitaev}},\ }\bibfield  {title} {\bibinfo {title} {{Quantum
  measurements and the Abelian Stabilizer Problem}},\ }\bibfield  {journal}
  {\bibinfo  {journal} {arXiv}\ }\href
  {https://doi.org/10.48550/arXiv.quant-ph/9511026}
  {10.48550/arXiv.quant-ph/9511026} (\bibinfo {year} {1995}),\ \Eprint
  {https://arxiv.org/abs/quant-ph/9511026} {quant-ph/9511026} \BibitemShut
  {NoStop}%
\bibitem [{\citenamefont {Zhao}\ \emph {et~al.}(2020)\citenamefont {Zhao},
  \citenamefont {Rubin},\ and\ \citenamefont {Miyake}}]{zhao2020fermionic}%
  \BibitemOpen
  \bibfield  {author} {\bibinfo {author} {\bibfnamefont {A.}~\bibnamefont
  {Zhao}}, \bibinfo {author} {\bibfnamefont {N.~C.}\ \bibnamefont {Rubin}},\
  and\ \bibinfo {author} {\bibfnamefont {A.}~\bibnamefont {Miyake}},\
  }\bibfield  {title} {\bibinfo {title} {{Fermionic partial tomography via
  classical shadows}},\ }\bibfield  {journal} {\bibinfo  {journal} {arXiv}\
  }\href {https://doi.org/10.1103/PhysRevLett.127.110504}
  {10.1103/PhysRevLett.127.110504} (\bibinfo {year} {2020}),\ \Eprint
  {https://arxiv.org/abs/2010.16094} {2010.16094} \BibitemShut {NoStop}%
\bibitem [{\citenamefont {Helgaker}\ \emph {et~al.}(2013)\citenamefont
  {Helgaker}, \citenamefont {Jorgensen},\ and\ \citenamefont
  {Olsen}}]{helgaker2013molecular}%
  \BibitemOpen
  \bibfield  {author} {\bibinfo {author} {\bibfnamefont {T.}~\bibnamefont
  {Helgaker}}, \bibinfo {author} {\bibfnamefont {P.}~\bibnamefont
  {Jorgensen}},\ and\ \bibinfo {author} {\bibfnamefont {J.}~\bibnamefont
  {Olsen}},\ }\href@noop {} {\emph {\bibinfo {title} {Molecular
  electronic-structure theory}}}\ (\bibinfo  {publisher} {John Wiley \& Sons},\
  \bibinfo {year} {2013})\BibitemShut {NoStop}%
\bibitem [{\citenamefont {Peled}\ and\ \citenamefont
  {Menkin}(2017)}]{peled2017sei}%
  \BibitemOpen
  \bibfield  {author} {\bibinfo {author} {\bibfnamefont {E.}~\bibnamefont
  {Peled}}\ and\ \bibinfo {author} {\bibfnamefont {S.}~\bibnamefont {Menkin}},\
  }\bibfield  {title} {\bibinfo {title} {Review—sei: Past, present and
  future},\ }\href {https://doi.org/10.1149/2.1441707jes} {\bibfield  {journal}
  {\bibinfo  {journal} {Journal of The Electrochemical Society}\ }\textbf
  {\bibinfo {volume} {164}},\ \bibinfo {pages} {A1703} (\bibinfo {year}
  {2017})}\BibitemShut {NoStop}%
\bibitem [{\citenamefont {Metzger}\ \emph {et~al.}(2016)\citenamefont
  {Metzger}, \citenamefont {Strehle}, \citenamefont {Solchenbach},\ and\
  \citenamefont {Gasteiger}}]{metzger2016h2evolution}%
  \BibitemOpen
  \bibfield  {author} {\bibinfo {author} {\bibfnamefont {M.}~\bibnamefont
  {Metzger}}, \bibinfo {author} {\bibfnamefont {B.}~\bibnamefont {Strehle}},
  \bibinfo {author} {\bibfnamefont {S.}~\bibnamefont {Solchenbach}},\ and\
  \bibinfo {author} {\bibfnamefont {H.~A.}\ \bibnamefont {Gasteiger}},\
  }\bibfield  {title} {\bibinfo {title} {Origin of {H}2 evolution in {LIB}s:
  {H}2{O} reduction vs. electrolyte oxidation},\ }\href
  {https://doi.org/10.1149/2.1151605jes} {\bibfield  {journal} {\bibinfo
  {journal} {Journal of The Electrochemical Society}\ }\textbf {\bibinfo
  {volume} {163}},\ \bibinfo {pages} {A798} (\bibinfo {year}
  {2016})}\BibitemShut {NoStop}%
\bibitem [{\citenamefont {Shkrob}\ \emph {et~al.}(2013)\citenamefont {Shkrob},
  \citenamefont {Zhu}, \citenamefont {Marin},\ and\ \citenamefont
  {Abraham}}]{shkrob2013eprec}%
  \BibitemOpen
  \bibfield  {author} {\bibinfo {author} {\bibfnamefont {I.}~\bibnamefont
  {Shkrob}}, \bibinfo {author} {\bibfnamefont {Y.}~\bibnamefont {Zhu}},
  \bibinfo {author} {\bibfnamefont {T.}~\bibnamefont {Marin}},\ and\ \bibinfo
  {author} {\bibfnamefont {D.}~\bibnamefont {Abraham}},\ }\bibfield  {title}
  {\bibinfo {title} {Reduction of carbonate electrolytes and the formation of
  solid-electrolyte interface (sei) in lithium-ion batteries. 1. spectroscopic
  observations of radical intermediates generated in one-electron reduction of
  carbonates},\ }\href {https://doi.org/10.1021/jp406274e} {\bibfield
  {journal} {\bibinfo  {journal} {J. Phys. Chem. C}\ }\textbf {\bibinfo
  {volume} {117}},\ \bibinfo {pages} {19255} (\bibinfo {year}
  {2013})}\BibitemShut {NoStop}%
\bibitem [{\citenamefont {Dunning}(1989)}]{dunning1989ccpvnz}%
  \BibitemOpen
  \bibfield  {author} {\bibinfo {author} {\bibfnamefont {J.}~\bibnamefont
  {Dunning}, \bibfnamefont {Thom~H.}},\ }\bibfield  {title} {\bibinfo {title}
  {{Gaussian basis sets for use in correlated molecular calculations. I. The
  atoms boron through neon and hydrogen}},\ }\href
  {https://doi.org/10.1063/1.456153} {\bibfield  {journal} {\bibinfo  {journal}
  {The Journal of Chemical Physics}\ }\textbf {\bibinfo {volume} {90}},\
  \bibinfo {pages} {1007} (\bibinfo {year} {1989})}\BibitemShut {NoStop}%
\bibitem [{\citenamefont {Prascher}\ \emph {et~al.}(2011)\citenamefont
  {Prascher}, \citenamefont {Woon}, \citenamefont {Peterson}, \citenamefont
  {Dunning},\ and\ \citenamefont {Wilson}}]{prascher2011ccpvnz}%
  \BibitemOpen
  \bibfield  {author} {\bibinfo {author} {\bibfnamefont {B.}~\bibnamefont
  {Prascher}}, \bibinfo {author} {\bibfnamefont {D.}~\bibnamefont {Woon}},
  \bibinfo {author} {\bibfnamefont {K.}~\bibnamefont {Peterson}}, \bibinfo
  {author} {\bibfnamefont {T.}~\bibnamefont {Dunning}},\ and\ \bibinfo {author}
  {\bibfnamefont {A.}~\bibnamefont {Wilson}},\ }\bibfield  {title} {\bibinfo
  {title} {Gaussian basis sets for use in correlated molecular calculations.
  vii. valence, core-valence, and scalar relativistic basis sets for {L}i,
  {B}e, {N}a, and {M}g},\ }\href {https://doi.org/10.1007/s00214-010-0764-0}
  {\bibfield  {journal} {\bibinfo  {journal} {Theor. Chem. Acc.}\ }\textbf
  {\bibinfo {volume} {128}},\ \bibinfo {pages} {69} (\bibinfo {year}
  {2011})}\BibitemShut {NoStop}%
\bibitem [{\citenamefont {HQS{\ }Quantum{\ }Simulations{\
  }GmbH}(2024)}]{hqs_asf}%
  \BibitemOpen
  \bibfield  {author} {\bibinfo {author} {\bibnamefont {HQS{\ }Quantum{\
  }Simulations{\ }GmbH}},\ }\href@noop {} {\bibinfo {title}
  {Activespacefinder}},\ \bibinfo {howpublished}
  {\url{https://github.com/HQSquantumsimulations/ActiveSpaceFinder.git}}
  (\bibinfo {year} {2024})\BibitemShut {NoStop}%
\bibitem [{\citenamefont {Sun}\ \emph {et~al.}(2018)\citenamefont {Sun},
  \citenamefont {Berkelbach}, \citenamefont {Blunt}, \citenamefont {Booth},
  \citenamefont {Guo}, \citenamefont {Li}, \citenamefont {Liu}, \citenamefont
  {McClain}, \citenamefont {Sayfutyarova}, \citenamefont {Sharma},
  \citenamefont {Wouters},\ and\ \citenamefont {Chan}}]{PySCF:2018}%
  \BibitemOpen
  \bibfield  {author} {\bibinfo {author} {\bibfnamefont {Q.}~\bibnamefont
  {Sun}}, \bibinfo {author} {\bibfnamefont {T.~C.}\ \bibnamefont {Berkelbach}},
  \bibinfo {author} {\bibfnamefont {N.~S.}\ \bibnamefont {Blunt}}, \bibinfo
  {author} {\bibfnamefont {G.~H.}\ \bibnamefont {Booth}}, \bibinfo {author}
  {\bibfnamefont {S.}~\bibnamefont {Guo}}, \bibinfo {author} {\bibfnamefont
  {Z.}~\bibnamefont {Li}}, \bibinfo {author} {\bibfnamefont {J.}~\bibnamefont
  {Liu}}, \bibinfo {author} {\bibfnamefont {J.~D.}\ \bibnamefont {McClain}},
  \bibinfo {author} {\bibfnamefont {E.~R.}\ \bibnamefont {Sayfutyarova}},
  \bibinfo {author} {\bibfnamefont {S.}~\bibnamefont {Sharma}}, \bibinfo
  {author} {\bibfnamefont {S.}~\bibnamefont {Wouters}},\ and\ \bibinfo {author}
  {\bibfnamefont {G.~K.-L.}\ \bibnamefont {Chan}},\ }\bibfield  {title}
  {\bibinfo {title} {Pyscf: the python-based simulations of chemistry
  framework},\ }\href {https://doi.org/https://doi.org/10.1002/wcms.1340}
  {\bibfield  {journal} {\bibinfo  {journal} {WIREs Computational Molecular
  Science}\ }\textbf {\bibinfo {volume} {8}},\ \bibinfo {pages} {e1340}
  (\bibinfo {year} {2018})}\BibitemShut {NoStop}%
\bibitem [{\citenamefont {Ralli}\ and\ \citenamefont {Weaving}(2022)}]{symmer}%
  \BibitemOpen
  \bibfield  {author} {\bibinfo {author} {\bibfnamefont {A.}~\bibnamefont
  {Ralli}}\ and\ \bibinfo {author} {\bibfnamefont {T.}~\bibnamefont
  {Weaving}},\ }\href@noop {} {\bibinfo {title} {Symmer}},\ \bibinfo
  {howpublished} {\url{https://github.com/UCL-CCS/symmer}} (\bibinfo {year}
  {2022})\BibitemShut {NoStop}%
\end{thebibliography}%

\clearpage
\onecolumngrid
\appendix

\section{Application of Shadow Tomography for CS-AFQMC}

\subsection{Derivation from \cref{eq:gen_channel} to \cref{eq:derived_channel}}\label{app:derivation}

\subsection{Commutation of the Contextual Rotation with Matchgate Circuits}\label{app:cont_rot_commutation}

To investigate the commutation relations between the unitaries for the rotation into the contextual subspace and the unitaries related to matchgate shadow circuits, we compare the terms in the contextual rotations $R_k$ given by \cref{eq:stab_rotations_paulis} and the gates in the matchgate circuit $U_Q$ (\cref{eq:matchgates}). There are multiple cases to consider depending on the intersection of the qubits of the two operators. The rotations trivially commute if they act on separate qubits. In the case of two common qubits, the matchgate will have the form $\exp(i\theta X_jX_{j+1})$, and will thus also commute since the anticommutations have even parity. In the case of a single qubit overlap, the intersection qubit corresponds to a Pauli $Y$ term in the stabilizer rotation or the matchgate is acting on the qubit with a Pauli $X$ term. When these cases occur, the terms anticommute, and the prefactor can be absorbed into the stabilizer rotation and updated as one iterates through every matchgate in the matchgate circuit. The updated rotation operator $\tilde U_{\mathcal{W}}$ can be computed iteratively in this way such that $U_Q U_\mathcal{W}^\dagger = \tilde U^{\dagger}_\mathcal{W} U_Q$ is satisfied. Further, we note that the resulting operator will only differ from the original operator by a global sign of $\pm1$.

Here we show the derivation from \cref{eq:gen_channel} to \cref{eq:derived_channel} that describes the channel for the shadow procedure. 
\begin{align}
    \mathcal{M} (\rho) & = \underset{U\sim D}{\mathbb{E}}\sum_{b\in\{0,1\}^n}\bra{b}\left(U\otimes \mathbb{I}_{\overline{\text{cs}}}\right)\rho\left(U^\dagger\otimes \mathbb{I}_{\overline{\text{cs}}}\right)\ket{b}\left(U^\dagger\otimes \mathbb{I}_{\overline{\text{cs}}}\right)\ket{b}\bra{b}\left(U\otimes \mathbb{I}_{\overline{\text{cs}}}\right)\\
                       & = \underset{U\sim D}{\mathbb{E}}\sum_{b\in\{0,1\}^n}\bra{b}\tilde{U}_\mathcal{W}^\dagger\left(U\otimes \mathbb{I}_{\overline{\text{cs}}}\right)\rho_{\text{unrot}}\left(U^\dagger\otimes \mathbb{I}_{\overline{\text{cs}}}\right)\tilde{U}_\mathcal{W}\ket{b}\left(U^\dagger\otimes \mathbb{I}_{\overline{\text{cs}}}\right)\ket{b}\bra{b}\left(U\otimes \mathbb{I}_{\overline{\text{cs}}}\right)\\
                       & = \underset{U\sim D}{\mathbb{E}}\sum_{b\in\{0,1\}^n}\bra{b'}\left(U\otimes \mathbb{I}_{\overline{\text{cs}}}\right)\rho_{\text{unrot}}\left(U^\dagger\otimes \mathbb{I}_{\overline{\text{cs}}}\right)\ket{b'}\left(U^\dagger\otimes \mathbb{I}_{\overline{\text{cs}}}\right)\ket{b}\bra{b}\left(U\otimes \mathbb{I}_{\overline{\text{cs}}}\right)\\
                       & = \underset{U\sim D}{\mathbb{E}}\sum_{b\in\{0,1\}^n}\bra{b_\text{cs},b_{\overline{\text{cs}}}'; b_\text{cs}}\left(U\otimes \mathbb{I}_{\overline{\text{cs}}}\right)\rho_{\text{unrot}}\left(U^\dagger\otimes \mathbb{I}_{\overline{\text{cs}}}\right)\ket{b_\text{cs},b_{\overline{\text{cs}}}'; b_\text{cs}}\left(U^\dagger\otimes \mathbb{I}_{\overline{\text{cs}}}\right)\ket{b}\bra{b}\left(U\otimes \mathbb{I}_{\overline{\text{cs}}}\right)\\
                       & = \underset{U\sim D}{\mathbb{E}}\sum_{b\in\{0,1\}^n}\bra{b_\text{cs}}\bra{b_{\overline{\text{cs}}}'; b_\text{cs}}\left(U\otimes \mathbb{I}_{\overline{\text{cs}}}\right)\rho_\text{cs}\otimes\ket{\text{nc}'}\bra{\text{nc}'}\left(U^\dagger\otimes \mathbb{I}_{\overline{\text{cs}}}\right)\ket{b_\text{cs}}\ket{b_{\overline{\text{cs}}}'; b_\text{cs}}\left(U^\dagger\otimes \mathbb{I}_{\overline{\text{cs}}}\right)\ket{b}\bra{b}\left(U\otimes \mathbb{I}_{\overline{\text{cs}}}\right)\\
                       & = \underset{U\sim D}{\mathbb{E}}\sum_{b\in\{0,1\}^n}\bra{b_\text{cs}}U\rho_\text{cs}U^\dagger\ket{b_\text{cs}}
                       \norm{\bra{b_{\overline{\text{cs}}}'; b_\text{cs}}\ket{\text{nc}'}}
                       \left[U^\dagger\ket{b_\text{cs}}\bra{b_\text{cs}}U\right]\otimes\ket{b_{\overline{\text{cs}}}}\bra{b_{\overline{\text{cs}}}}\\
                       & = \underset{U\sim D}{\mathbb{E}}\sum_{b_{\text{cs}}\in\{0,1\}^{n_{\text{cs}}}}\bra{b_\text{cs}}U\rho_\text{cs}U^\dagger\ket{b_\text{cs}}
                       \left[U^\dagger\ket{b_\text{cs}}\bra{b_\text{cs}}U\right]\otimes\ket{b_{\overline{\text{cs}}};\text{nc}'; b_\text{cs}}\bra{b_{\overline{\text{cs}}};\text{nc}'; b_\text{cs}}\,,
\end{align}
where $|b'\rangle$ is the computational basis state $|b\rangle$ rotated by $\tilde{U}_\mathcal{W}$ and the state 
\begin{equation}
\ket{b_{\overline{\text{cs}}}'; b_\text{cs}}=\left(\bra{b_\text{cs}}\otimes\mathbb{I}_{\overline{\text{cs}}}\right)\tilde{U}_\mathcal{W}^\dagger\left[\ket{b_{\text{cs}}}\otimes\ket{b_{\overline{\text{cs}}}}\right]\,,\end{equation}  
where the state $\ket{b_{\overline{\text{cs}}};\text{nc}'; b_\text{cs}}$ is the state with the bitstring $b_{\overline{\text{cs}}}$ such that when rotated with the state with bitstring $b_\text{cs}$ it results in the state $\ket{\text{nc}'; b_\text{cs}}$, or in other words the state $\ket{b_{\overline{\text{cs}}}}$ given $\ket{b_\text{cs}}$ such that $\tilde{U}_\mathcal{W}\left[\ket{b_\text{cs}}\otimes\ket{b_{\overline{\text{cs}}}}\right]= \ket{b_\text{cs}}\otimes\ket{\text{nc}'}$.

\subsection{Variance}\label{app:variance}
The number of snapshots required to estimate an expectation value is based on the variance of the unbiased estimates. This is given by
\begin{align}
    \text{Var}[\hat{o}] = &\mathbb{E}\left[|\hat{o}|^2\right] - |\mathbb{E}\left[\hat{o}\right]|^2\\
    =& \underset{U\sim D}{\mathbb{E}}\sum_{b_{\text{cs}}\in\{0,1\}^{n_{\text{cs}}}}\bra{b_\text{cs}}U\rho_\text{cs}U^\dagger\ket{b_\text{cs}}
                       \left|\left[U^\dagger\ket{b_\text{cs}}\bra{b_\text{cs}}U\right]\otimes\ket{b_{\overline{\text{cs}}};\text{nc}'; b_\text{cs}}\bra{b_{\overline{\text{cs}}};\text{nc}'; b_\text{cs}}\right|^2 - |\Tr(O\rho)|^2\\
                       = &\Tr\left[\sum_{b_{\text{cs}}\in\{0,1\}^{n_{\text{cs}}}}\underset{U\sim D}{\mathbb{E}}\left(\mathcal{U}\otimes\mathbb{I}\right)^{\otimes3}\left(\left(\ket{b_\text{cs}}\bra{b_\text{cs}}\otimes\ket{b_{\overline{\text{cs}}};\text{nc}'; b_\text{cs}}\bra{b_{\overline{\text{cs}}};\text{nc}'; b_\text{cs}}\right)^{\otimes3}\right) \left(\rho\otimes\mathcal{M}^{-1}(O)\otimes\mathcal{M}^{-1}(O^\dagger)\right)\right]\\
                       & - |\Tr(O\rho)|^2\,,\nonumber
\end{align}
where $\mathcal{U}(\rho)=U\rho U^\dagger$, $O, O^\dagger\in \mathcal{X}$ for a subspace of linear operators on the vector space  of the Hilbert space of $n$-qubits $\mathcal{X}\in\mathcal{L}(\mathcal{H}_n)$. It then follows that the variance is bounded under the same characteristics as the bound proven in Ref.~\cite{wan2022matchgate}. 

\section{Nitrogen dissociation}\label{app:nitrogen_dissoc}

Here we show a comparison of the different energy estimations for each distance used to construct the dissociation curve in \cref{fig:n2_triptych}.
\begin{figure}[h]
    \centering
    \includegraphics[width=0.97\textwidth]{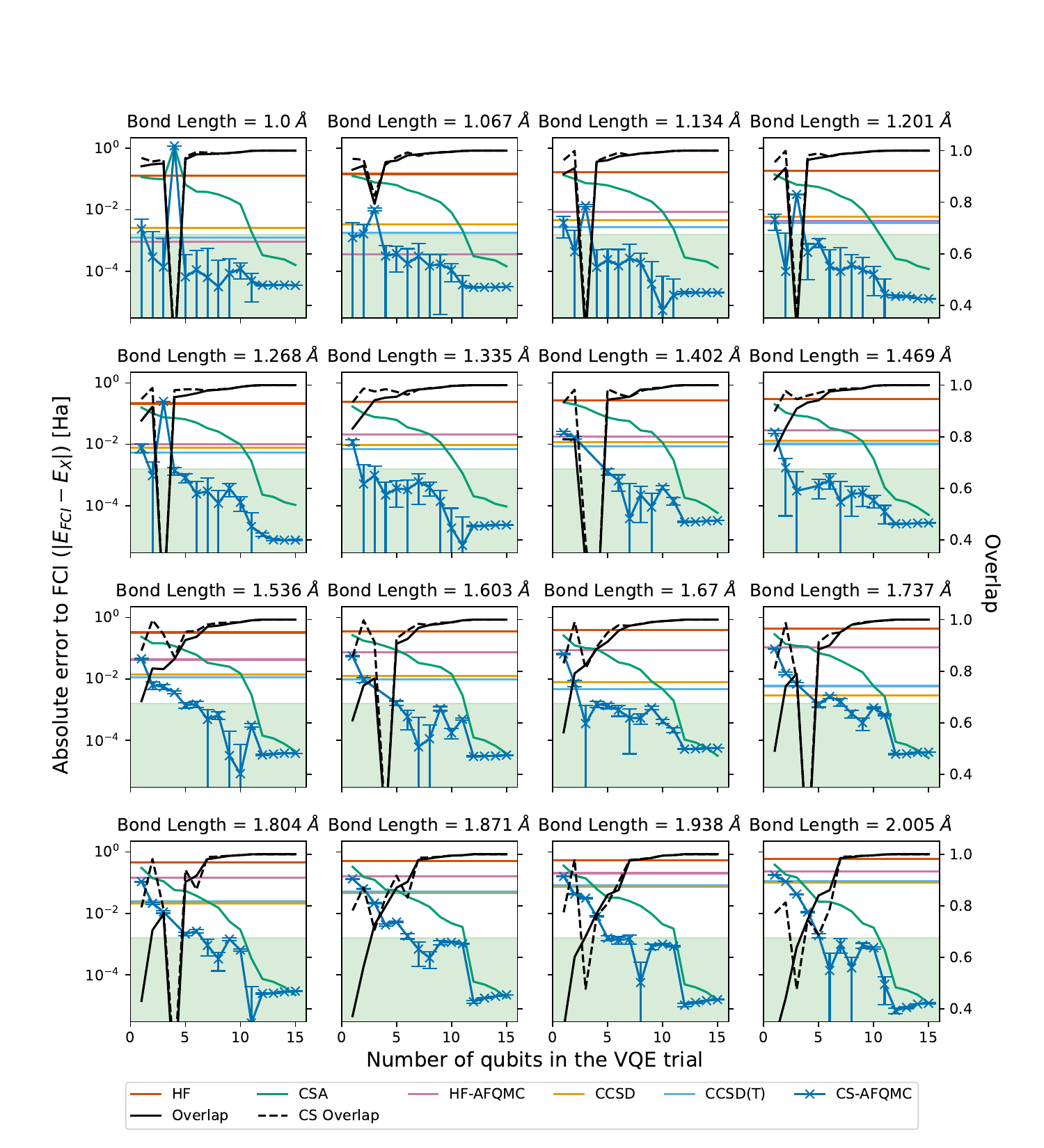}
    \caption{Calculated energies for inter-atomic distances between 1 and 2 \AA. Compares HF, CSA, AFQMC, CCSD, CCSD(T) and CS-AFQMC for increasing size of the contextual subspace and green shading representing chemical accuracy. The error bars are calculated based on the mean squared error (MSE) of multiple CS-AFQMC runs at each CS level.}
    \label{fig:toy_system_results}
\end{figure}

\end{document}